\documentclass[12pt]{article}

\addtocounter{footnote}{1}

\usepackage{graphicx}

\usepackage{epsfig}  

\def\A{{\bf{A}}}

\def\bra#1{ \langle{#1}| }
\def\braket#1#2{ \langle{#1}|{#2}\rangle }
\def\opket#1#2 {  {#1} |{#2}\rangle }
\def\braopket#1#2#3{ \langle{#1}| {#2} |{#3}\rangle }

\def\C{{\bf C}}

\def\d{{\rm d}}

\def\det{{\rm det}\,}

\def\ds{\displaystyle}

\def\expect#1{\left\langle{#1}\right\rangle}
\def\F{{\bf F}}

\def\Hh{\hat{H}}

\def\K2{{\cal K}}
\def\ket#1{ |{#1}\rangle }

\def\lin{{\rm lin}}

\def\mod{{\rm mod}}

\def\phi{\varphi}

\def\s{{\bf s}}
\def\sc{{\rm sc}}

\def\Tr{{\rm Tr}}

\def\Uh{\hat{U}}

\def\w{{\bf w}}

\def\x{{\bf x}}

\begin{document}
\baselineskip 15pt

\title{Wavefunction Statistics using Scar States}
\author{Soo-Young Lee\footnote{
Present Address: NCRIC for Controlling Optical Chaos,
Pai-Chai University, Daejeon 302-735, South Korea}
$\,$ and Stephen C. Creagh}
\date{}
\maketitle

\vspace{-25pt}
\begin{center}
{\it

School of Mathematical Sciences, University of Nottingham,
Nottingham NG7 2RD, UK.
\newline

}
\end{center}

\vspace*{10pt}

\begin{abstract}
We describe the statistics of chaotic wavefunctions near periodic
orbits using a basis of states which optimise the effect of
scarring. These states reflect the underlying structure of
stable and unstable manifolds in phase space and provide a
natural means of characterising scarring effects in individual
wavefunctions as well as their collective statistical properties.
In particular, these states may be used to find scarring
in regions of the spectrum normally associated with
antiscarring and suggest a characterisation of templates for scarred
wavefunctions which vary over the spectrum. The results are applied
to quantum maps and billiard systems.
\end{abstract}

%\pacs{PACS numbers: 03.65.Sq, 05.45.+b}

%\newpage
\section{Introduction}
\label{introduction}
\noindent The observation that random matrix theory (RMT) describes the
statistical properties of the eigensolutions of a classically chaotic
quantum system \cite{BGS} has formed a pivotal role in our understanding
of such systems. The most basic prediction for eigenfunctions is that they
form a Gaussian distribution when projected on a generic state.
More explicitly, the overlap between a generic probe state
$\ket{\phi}$ and the eigenstates $\ket{\psi_n}$ of the chaotic system
has the probability distribution function
\begin{equation}
P(x) = \frac{1}{\sqrt{2\pi}}\; e^{- x^2/2},
\label{rmt}
\end{equation}
where we normalise the states $\ket{\psi_n}$ so that the overlap
$x=\braket{\phi}{\psi_n}$ has unit variance and assume
time-reversal invariance for ease of notation. This distribution
has been shown to work very accurately for billiard systems, for example
\cite{Robnik}.

In any such comparison with RMT statistics, deviations due to short
time dynamics and other system-specific effects are particularly
interesting. In systems such as quantum dots \cite{Qdots,Kapqdot},
for example, such deviations may help us distinguish between RMT
statistics arising from a simple underlying chaotic dynamics and RMT
statistics which arise simply because the system is complex, perhaps
dominated by interactions etc. In the context of wavefunction statistics,
a remarkably strong example of such deviation has been demonstrated by
Kaplan, Heller and coworkers \cite{KA98,KAHE,BI01}. They find that
wavefunction statistics around a periodic orbit of the classical limit
can deviate strongly from (\ref{rmt}), an effect attributed
to the phenomenon of scarring \cite{MC83,HE84,BO88,BE89} (whereby
wavefunctions of classically chaotic systems are seen frequently to
behave atypically near periodic orbits).
These deviations can be robust and
survive the semiclassical limit if the probe state $\ket{\phi}$
is sufficiently localised in phase space around a periodic orbit.
In this paper we describe the {\it collective} statistics of many
components of the wavefunction using a basis of probe states which
are locally complete near a periodic orbit.
We find that a particular set of probe states can be defined
in terms of which these deviations may be described particularly simply
and these are related to states which have previously been used to
characterise scarring under different guises (\cite{KA98,KAHE,BI01}
and \cite{earlyscarf,NoVo,VE01,Wis01,FNdB}), known variously under
the names of universal test states \cite{KAHE},
scarfunctions \cite{VE01} or simply quasimodes \cite{FNdB}.

The first theoretical explanation for scarring \cite{HE84} was
based on the evolution of a Gaussian wavepacket launched on a
classical periodic orbit. Fourier analysis of the evolving wavepacket
shows that typical overlaps of the wavepacket with chaotic eigenstates
$\psi_n$ fluctuate quasiperiodically as a function of energy. Peaks
occur in {\it scarred} parts of the spectrum associated
with enhanced wavefunction probabilities near the periodic orbit
and troughs are found  in the {\it antiscarred} parts
where wavefunctions are expected to be suppressed near the
periodic orbit. More recently this basic idea has been developed
further by Kaplan et al \cite{KA98,KAHE,BI01} to produce a
quantitative theory of wavefunction statistics near a periodic orbit.

Kaplan et al consider the statistics of overlap with a single
probe state $\ket{\phi}$, which might typically be Gaussian. By
considering quantum recurrence for
such a state launched on a periodic orbit, one can show that
the variance of the distribution $p(x)$ must deviate from the
constant value predicted by RMT and is a function of the spectral
parameter --- energy $E$ in the case of time-independent systems and
eigenangle $0\leq\theta< 2\pi$ in the case of quantum maps. The overlap
distribution in (\ref{rmt}) is then naturally replaced by
\begin{equation}
P(x;\theta) = \frac{1}{\left[2\pi C(\theta)\right]^{1/2}}\;
e^{-x^2 /2C(\theta)},
\label{scaledg}
\end{equation}
where we adopt the notation of eigenangle for the spectral parameter
and where the variance $C(\theta)$, which may be calculated from
the linearised autocorrelation function of the probe state, is
independent of $\hbar$.

This was generalised in \cite{CR02} to  provide a joint-probability
distribution describing the statistics of the wavefunction in a basis
of probe states which are complete in a neighbourhood in phase space
of the periodic orbit. A minimum-uncertainty Gaussian wavepacket
can be thought of as the ground state of a harmonic-oscillator
Hamiltonian. We can also use as probe states the excited states
$\ket{\phi_k}$ of the same Hamiltonian and thereby obtain a series of overlaps
$\x=(x_0,x_1,\cdots,x_k,\cdots)$ representing the components of
a chaotic eigenfunction in the harmonic-oscillator basis. (We adopt
notation here appropriate to a single degree of freedom transverse to
the periodic orbit, but it should be clear how to generalise these
statements to higher dimensions.) One can then derive a matrix of
correlations $\C(\theta)$ with components
\[
C_{kl}(\theta) = \expect{x_k x_l}
\]
which depends as before on the spectral parameter $\theta$ and which is
independent of $\hbar$ (and therefore survives the semiclassical limit).
The corresponding normal joint-probability distribution for the overlap
amplitudes
\begin{equation}\label{gausshyp}
P(\x;\theta) = \frac{1}{\left[(2\pi)^{M+1}\det{\bf C}(\theta)\right]^{1/2}}\;
e^{-\x^T {\bf C}^{-1}(\theta)\,\x/2}
\label{JPD}
\end{equation}
has been shown in  \cite{CR02} to describe well the statistics of
eigenstates of perturbed cat maps and to lead to a quantitative
understanding of the influence of scarring on the statistics of
tunnelling from chaotic potential wells. The distribution has been
written for a finite number of components $(x_0,x_1,\cdots,x_M)$.
The statistics of probe states sufficiently localised near the
periodic orbit (but otherwise arbitrary) can straightforwardly
be obtained from the limit $M\to\infty$. More generally though,
this limit needs to be treated carefully, as discussed in Section~2.

In this paper we investigate the local structure of scarred eigenfunctions
by using the envelope matrix ${\bf C}(\theta)$ to define a series of probe
states adapted to a given periodic orbit and a given part of the spectrum.
These {\it scar states} are defined so that the joint-probability distribution
$P(\x ; \theta)$ predicts statistical
independence for their overlaps with chaotic wavefunctions. In a
phase-space representation, these probe states are concentrated along the
stable and unstable manifolds of the periodic orbit. In a region of the
spectrum associated with maximal scarring, the first of these states
corresponds to the scarfunction introduced by Vergini and Carlo in \cite{VE01}.
Elsewhere they correspond to linearised eigenstates such as considered by
Nonnenmacher and Voros \cite{NoVo} and Kaplan and Heller \cite{KAHE}.
The analysis here suggests an extension to a series of such states,
however, forming a local basis near the periodic orbit. Furthermore
these scar states vary as a function of the spectral parameter so that
the local basis varies quasiperiodically as the spectrum is traversed.
A surprising outcome, which has previously been noted in \cite{KAHE},
is that in antiscarred regions of the spectrum, where conventionally
we expect the chaotic wavefunction to be suppressed in a neighbourhood
of the periodic orbit, we can define probe states for which
chaotic wavefunctions have larger overlaps than predicted by RMT.
That is, we find positive scarring in the antiscarred regions.
We see in this paper that this observation generalises to
a basis of such states, in terms of which a statistical description
of wavefunction overlaps can be achieved using a simple product of
independent Gaussians.

The organisation of the paper is as follows.
In Section 2 we start by recalling the definition of the envelope
matrix ${\bf C}(\theta)$ and then we define optimised probe states
which give decoupled overlap distributions with
scarred eigenfunctions. We also discuss the local pattern of the
scarred eigenfunctions and how these vary with the spectral parameter
$\theta$. We see how this works for a specific quantum map model in
Section~3 and we consider billiard systems in Section 4, where we show
that the statistics of the boundary eigenfunctions can also be described by
the joint-probability distribution $P(\x ; \theta)$.

\section{The envelope matrix and scar states }

In this section we begin with a brief description of the envelope
matrix ${\bf C}(\theta)$ which determines the joint-probability
distribution $P(\x; \theta)$. The eigenvectors of ${\bf C}(\theta)$
can be interpreted as providing the statistically independent basis
for the joint-probability distribution. The eigenvectors with the
largest eigenvalues provide us with probe states which give overlap
distributions in which the effect of scarring is optimised. Borrowing
the terminology of Vergini and Carlo, we refer to these probe states
as {\it scar states}. We show in particular
that the first of these scar states coincides in the semiclassical limit
with the scarfunction introduced by Vergini and Carlo \cite{VE01}
and that in general they form a subset of the hyperbolic stationary
states whose coherent-state representations have been described in
detail by Nonnenmacher and Voros \cite{NoVo}.

The notation in this section is adapted to the case of a quantum map
$\hat{U}$ with eigensolutions $\hat{U}\ket{\psi_n}
=e^{-i\theta_n}\ket{\psi_n}$. We consider statistics of ensembles
which are constructed by restricting the eigenangle $\theta_n$ to a
narrow window centred at the value $\theta$. The corresponding construction
in the case of time-independent systems is to consider states for
which the action $S(E_n)$ of the periodic orbit under consideration
is such that $\theta_n=S(E_n)/\hbar$ is similarly
restricted.

\subsection{Envelope matrix ${\bf C}(\theta)$}

The joint-probability distribution $P(\x; \theta)$ completely characterises
the influence of scarring on  wavefunction statistics near the periodic
orbit and our assertion is that this is in turn completely determined 
(to a good approximation at least) by 
the envelope matrix ${\bf C}(\theta)$. For a given periodic orbit, the 
matrix ${\bf C}(\theta)$ is constructed in practice from a Fourier transform 
of a correlation matrix ${\bf A}(t)$, formed by writing linearised
quantum evolution in a harmonic-oscillator 
basis $\{\ket{\phi_k}\}$. For simplicity of notation, let us assume that 
the periodic orbit is in fact a fixed point $p_0$ of a chaotic 
map and that this map is quantised by the unitary operator $\Uh$, while
$\Uh_\lin$ quantises the linearised dynamics around $p_0$. The matrix
${\bf A}(t)$ is simply $\Uh_\lin^t$ written in the basis 
$\{\ket{\phi_k}\}$, with elements
\begin{equation}
A_{kl} (t) = \bra{ \phi_k } \hat{U}^t_{\lin} \ket{\phi_l}.
\label{Amat}
\end{equation}
We might of course use any basis of states localised in phase space
around $p_0$ but we use a harmonic oscillator basis because we can 
then write closed-form expressions for $A_{kl} (t)$.
These are given explicitly in Appendix~\ref{appgetA}, but here we 
simply note the essential features.

The linearised evolution underlying  $\Uh_\lin$ has inversion 
through $p_0$ as a symmetry (even if the full map $\hat{U}$ does not)
and this leads to a decoupling of $\A(t)$ into blocks 
corresponding to even and odd values of $k$. That is, 
$A_{kl}(t)=0$ unless $k$ and $l$ are both 
even or both odd. In any discussion of statistics it is therefore 
natural to consider separately the components $x_k$ for even and odd $k$
--- any explicit calculation we give will be for the even case.
The linearised evolution also has a time-reversal symmetry (again, even
if the full map $\hat{U}$ does not). This means that we can choose
phases of $\ket{\phi_k}$ so that $\A(t)$ is symmetric and 
$\left(\A(t)\right)^* = \A(-t)$.

For a map with one degree of freedom, two classical parameters which 
characterise the classical situation are sufficient to determine 
$\A(t)$. The first of these is the stability eigenvalue 
$\lambda^t=\pm e^{\rho t}$ of the unstable dynamics around $p_0$. This 
sets the rate at which  $\A(t)$ decays with $t$ and 
subsequently the strength of deviations due to scarring. The second 
parameter effectively characterises the harmonic-oscillator basis. 
Any elliptic linear evolution which has $p_0$ as a fixed point can be used 
to generate a harmonic-oscillator basis $\ket{\phi_k}$, and we should expect 
$\A(t)$ to depend on the eccentricity and orientation of
this elliptic evolution (relative to the hyperbolic structure
of the unstable dynamics around $p_0$). It turns out that a single 
parameter, which we denote $Q$ following Kaplan and Heller, is sufficient
to completely characterise this dependence, and is easily interpreted
geometrically. Let  $(q,p)$ be canonical coordinates for which the 
harmonic oscillator defining $\ket{\phi_k}$ is a multiple of $q^2+p^2$.
Then $Q=\cot\alpha$, where $\alpha$ is the angle in coordinates
$(q,p)$ between the stable and unstable manifolds of the chaotic
dynamics about $p_0$.

The expressions for ${\bf A}(t)$ given in Appendix~\ref{appgetA} 
are simplest for the case $Q=0$, which corresponds to orthogonal stable 
and unstable manifolds (relative to the coordinates $(q,p)$ suggested
by the normal form for the harmonic oscillator). If we are 
interested in the wavefunction statistics in isolation, then the basis 
$\ket{\phi_k}$ may be chosen at our convenience and it seems natural in
that case to ensure that $Q=0$. In applications to tunnelling statistics
\cite{CR02}, however, the basis $\ket{\phi_k}$ is imposed by a secondary 
calculation and it is therefore useful to be able to treat the case 
$Q\neq 0$. We will also find this option convenient when we apply the 
results to an explicit model in the next section.

Given ${\bf A}(t)$, we define the envelope matrix by
\[
\C(\theta) = \sum_{t=-\infty}^\infty e^{i\theta t}\A(t).
\]
Given the properties of $\A(t)$ described above, this is a real 
symmetric matrix which decouples into even and odd blocks.
If we construct an ensemble of chaotic eigenstates $\ket{\psi_n}$
from a small window around the spectral parameter $\theta$, modulo
$2\pi$, then it can be shown that the components $C_{lk}(\theta)$ 
give the averages $\expect{x_l^*x_k}$ \cite{CR02}. Given such an 
RMT-violating constraint on variance and correlation, a 
common procedure in the analysis of wavefunction statistics has 
been to assert that the statistical distribution remains normal 
but with an appropriately adjusted covariance matrix. In particular,
this has been done by Antonsen et al \cite{OT95} for averages of the
wavefunction along the periodic orbit and by Narimanov et al {\cite{Qdots}
for the wavefunction itself. In our case this amounts to
assuming the joint probability distribution (\ref{JPD}) in the
case of GOE statistics, with an obvious generalisation for the GUE 
case. While we cannot prove the result, the distribution was shown 
in \cite{CR02} to successfully describe the statistics of low-lying
components of $\x$ and to explain deviations due to scarring of 
tunnelling-rate statistics. We posit the result simply as a 
conjecture and refer to the numerical evidence presented here
and in \cite{CR02} as justification for it.

The joint-probability distribution obtained from $\C(\theta)$ 
is written in (\ref{gausshyp}) for a truncated set of 
components $(\x=x_0,x_1,\cdots,x_M)$. To obtain a complete characterisation
of wavefunction statistics, we should take a limit $M\to\infty$. 
However, because $\C(\theta)$
is derived from a linearised hyperbolic system whose quantisation
does not have normalised eigenstates, the elements of $\C(\theta)$ 
decay quite slowly and the limit must be treated with care. We find a 
straightforward limit if we use (\ref{gausshyp}) to obtain the 
statistics of a measure such as 
\begin{equation}\label{defy}
	\xi_n = \braopket{\psi_n}{\hat{P}}{\psi_n}
\end{equation}
which samples the chaotic state $\ket{\psi_n}$ in a localised region 
of phase space around $p_0$. Localisation in the context of (\ref{defy})
means that $\hat{P}$ is an operator whose matrix representation 
${\bf P}$ in the harmonic-oscillator basis decays sufficiently rapidly that 
$\Tr\,{\bf P}\C(\theta)$ exists --- it has been
shown in \cite{CR02} that the statistics of $\xi_n$ are then governed by a 
distribution of the form (assuming GOE statistics and Hermitian $\hat{P}$)
\begin{equation}\label{pofy}
p(\xi;\theta) = \frac{1}{2\pi}\int
	\frac{e^{-iqy}}
     {\sqrt{\det\!\!\left(1+2iq{\bf P}\C(\theta)\right)}}
							\;\d q,
\end{equation}
that is well-defined in the limit $M\to\infty$. Alternatively,
any finite subset of the components $\x=(x_0,x_1,\cdots)$ can be 
described by a joint-probability distribution written
analogously to (\ref{JPD}) with an appropriate submatrix of 
$\C(\theta)$. The limit $M\to\infty$ is not simple however,
if we try to treat $P(\x;\theta)$ as a full distribution.

\subsection{The construction of scar states}

Consider the distribution for a finite set of coordinates 
$\x=(x_0,x_1,\cdots,x_M)$. As for any normal distribution it is 
natural to rotate these coordinates so as to produce 
a statistically independent basis
${\bf y}=(y_0,y_1,\cdots,y_M)$ for which
\begin{equation}
 \expect{ y_i y_j } =  s_i (\theta )\delta_{ij}
\label{indep} 
\end{equation}
and for which the  joint-probability distribution is a simple product of 
Gaussian distributions, 
\begin{equation}
P({\bf y};\theta) = \prod_{i=0}^{M} P(y_i;\theta)= 
\prod_{i=0}^{M} \frac{1}{ \sqrt{2\pi s_i(\theta)}} \;
e^{-y_i^2/2s_i(\theta)}.
\label{gaudis}
\end{equation}
These are obtained as the eigenvectors and eigenvalues of the
corresponding $(M+1)\times (M+1)$ submatrix of $\C(\theta)$.
It should be noted, however, that because the infinite-dimensional 
limit of $\C(\theta)$ does not have proper eigenvectors, the coordinates 
$y_i$ and the eigenvalues
\[
{\bf s}(\theta) \equiv (s_0(\theta),s_1(\theta),\cdots,s_M(\theta)),
\]
do not converge in a simple way as $M\to\infty$. We will see in
particular that the leading eigenvalues diverge logarithmically
with increasing $M$. This means that we can construct linear 
combinations of the harmonic-oscillator basis states $\ket{\phi_k}$
for which the overlaps with the chaotic states $\ket{\psi_n}$ can be 
made arbitrarily large, on average, compared with the expectation from
RMT. It should be noted however that for any finite value of $\hbar$
there is a limit to how large we can let $M$ become without
the linear approximation underlying the construction of 
$\C(\theta)$ becoming invalid and the limit $M\to\infty$ can
therefore only sensibly be taken in conjunction with the semiclassical 
limit \cite{note}.

We begin by deriving formally the eigenvectors of $\C(\theta)$ 
in an infinite basis and later indicate how these solutions are
regularised by truncation. For this purpose we use the fact that
$\A(t)$ and $\C(\theta)$ decompose into even and odd blocks and 
consider these blocks, denoted by $\A^\pm(t)$ and $\C^\pm(\theta)$,
separately. Using Poisson resummation we write
\begin{eqnarray}\label{decompC}
\C^\pm(\theta) &=& \sum_{m=-\infty}^\infty  \int_{-\infty}^\infty 
			e^{i(\theta+2\pi m) t}\A^\pm(t)\,\d t \nonumber\\[3pt]
&=& \sum_{m=-\infty}^\infty \F^\pm(\theta+2\pi m)
\end{eqnarray}
and we claim that 
\[
\F^\pm(\Omega) =  \int_{-\infty}^\infty e^{i\Omega t}\A^\pm(t) \,\d t
\]
is an unnormalised projection onto an eigenvector of $\C^\pm(\theta)$
if $\Omega=\theta+2\pi m$ for some integer $m$. We see this by 
writing
\[
F_{kl}^\pm(\Omega) = \braopket{\phi_k}{\hat{S}(\Omega)}{\phi_l},
\]
where the superscript $\pm$ indicates whether $\ket{\phi_k}$ and
$\ket{\phi_l}$ are restricted to the even or odd subspaces
and $\hat{S}(\Omega)$ is the operator
\[
\hat{S}(\Omega) = \int_{-\infty}^\infty e^{i\Omega t}\, {\Uh_\lin^t} 
							\,\d t.
\]
Let us now substitute 
\[
\Uh_\lin^t = e^{-i\Hh t/\hbar}
\]
where $\Hh$ is a quadratic Hamiltonian linearising the hyberbolic
dynamics about $p_0$. We expand 
\[
\Uh_\lin^t =\sum_\pm \int \d E \, 
			e^{-i E t/\hbar} \, \ket{E^\pm}\bra{E^\pm}
\]
in eigenbasis $\Hh\ket{E^\pm}=E\ket{E^\pm}$ of $\Hh$, with the usual 
normalisation $\braket{E^\pm}{{E'}^\pm}=\delta(E-E')$. The sum over 
$\pm$ here arises because the eigenstates for each $E$ are 
doubly-degenerate. In a representation where the problem amounts
to scattering from a parabolic barrier, the conventional description of
this degeneracy is in terms of scattering states with left or right 
incoming waves, but in our case we choose eigenstates that have even or 
odd parity. This decomposition gives in turn that
\[
\hat{S}(\Omega) = 2\pi\delta(\Omega-\Hh/\hbar)
= 2\pi\hbar \sum_\pm \ket{E(\Omega)^\pm}\bra{E(\Omega)^\pm}  
\]
where $E(\Omega)=\hbar\Omega$. We note that phase space representations
of the states $\ket{E^\pm}$ have been analysed in detail by Nonnenmacher 
and Voros \cite{NoVo} (beware however that in their notation the label 
$\pm$ sometimes refers to the direction of an incoming wave whereas here 
it always refers to parity) and we will return to their phase-space 
description later. This formal result states simply that the 
eigenvectors of $\C^\pm(\theta)$
are representations in the harmonic-oscillator basis of the subset of
these states for which $\Omega=\theta+2\pi m$ and which have a definite
parity. In particular we may 
decompose $\C^\pm(\theta)$ as promised in (\ref{decompC}), with
\begin{equation}\label{decompF}
\F^\pm(\Omega) = 2\pi\hbar \, \w^\pm(\Omega) \, \w^\pm(\Omega)^{\dagger} 
\end{equation}
and $w_k^\pm(\Omega) = \braket{\phi_k}{E(\Omega)^\pm}$.

Equation (\ref{decompC}) therefore provides us formally with a spectral 
decomposition of $\C^\pm(\theta)$ with eigenvectors $\w^\pm(\theta+2\pi m)$
labelled by the integers $m$. The corresponding eigenvalues
diverge, however, because the hyperbolic stationary states 
$\ket{E^\pm}$ are unnormalisable. This supports our earlier assertion 
that the eigenvalues of a finite submatrix of $\C(\theta)$ diverge as 
$M\to\infty$. To regularise the situation we must consider truncations
of $\C(\theta)$. This corresponds physically to restricting the probe
states we use to be sufficiently localised near $p_0$ in phase space,
which is in any case a necessity if the linearised evolution used in 
deriving the form of $\A(t)$ is to be valid.

We will consider truncations of the form
\[
\tilde{C}_{kl}(\theta) =  \sqrt{P_k} C_{kl}(\theta) \sqrt{P_l}
\]
where the coefficients $P_k$ decay from unity sufficiently rapidly that
$\tilde{\C}(\theta)$ has a series of nonvanishing eigenvalues
with normalisable eigenvectors, but sufficiently slowly that 
$\tilde{\C}(\theta)$ can be regarded as an approximation for $\C(\theta)$
in an appropriate neighbourhood of $p_0$. In the simplest case we use a 
step function $P_l=\Theta(M-l)$, which replaces $\C(\theta)$ with a 
finite submatrix describing the statistics of $(x_0,x_1,\cdots,x_M)$, 
but we may also let $P_k$ decay smoothly. This truncation is 
useful in the statistics of (\ref{defy}) above in the case 
$\hat{P} = \sum_k P_k \ket{\phi_k}\bra{\phi_k}$ (which is the form seen in
applications to tunnelling, where $P_k$ is a tunnelling rate associated
with the probe state $\ket{\phi_k}$). In particular we will use
truncating functions of the form
\begin{equation}\label{fermi}
 { P}_k=\frac{1}{{1+e^{\beta(k-M)}}},
\end{equation}
where $M$ behaves as a cutoff in the dimension of $\tilde{\C}(\theta)$
and $\beta$ provides a decay rate after cutoff. The sharp truncation
above corresponds to the limit $\beta\to\infty$.

\begin{figure}[h]
\vspace*{-1.2in}\hspace*{1.3in}\includegraphics[width=4.5in] {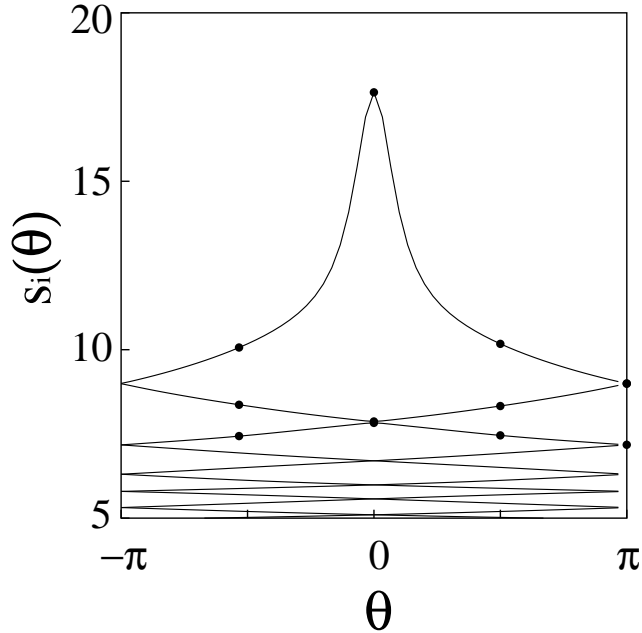}
\vspace*{-2.5cm}
\caption{The leading eigenvalues ${\bf s}(\theta) =
(s_0(\theta),s_1(\theta),\cdots)$
of the envelope matrix $\tilde{\C}(\theta)$. Here we use
truncation parameters $\beta=0.025$ and $M =300$. The bullets
indicate $s_i(\theta)$ for $i=0,1$ and $2$
and for $\theta =-\pi /2, 0, \pi/2$ and $\pi$. The Husimi distributions
corresponding to these eigenvalues are shown in Figure~\ref{Husplot}.}
\label{eigS}
\end{figure}

The eigenvectors of the truncated matrix $\tilde{\C}(\theta)$
are not as straightforwardly calculated as those of the full matrix
$\C(\theta)$. The leading eigenvectors, however, corresponding
to smaller absolute values of $m$, should approximately coincide
with those of $\C(\theta)$.
In detail, if $E(\Omega)$ with $\Omega=\theta+2\pi m$ is small
enough and the components of $\hat{P}$ decay slowly enough from unity
then
\[
\left(\sqrt{{\bf P}}\w^\pm (\Omega)\right)_k
					= \sqrt{P_k} w_k^\pm (\Omega)
\]
is approximately an eigenvector of $\tilde{\C}(\theta)$ with eigenvalue
\begin{eqnarray*}
s(\Omega) &=& 2\pi\hbar\, (\sqrt{{\bf P}}\w^\pm (\Omega))^\dagger
		 \sqrt{{\bf P}}\w^\pm (\Omega)\\[6pt]
	&=&  2\pi\hbar \, \braopket {E(\Omega)^\pm}{\hat{P}}{E(\Omega)^\pm}.
\end{eqnarray*}
Since $\ket{E(\Omega)^\pm}$ is unnormalisable, these eigenvalues
diverge as promised as the truncation dimension increases and
$\hat{P}\to I$.
The eigensolution is only approximate because, while (\ref{decompF})
still applies to the truncated matrix if $\w^\pm(\Omega)$ is replaced by
$\sqrt{{\bf P}}\w^\pm(\Omega)$, the vectors $\sqrt{{\bf P}}\w^\pm(\Omega)$
for two different values of $\Omega$ are only approximately orthogonal.

We present in Figure~\ref{eigS} the leading eigenvalues, relabelled
$(s_0(\theta),s_1(\theta),\cdots)$ in order of decreasing value,
for a fixed point $p_0$ with stable and unstable manifolds along
$p=\pm q$, a harmonic oscillator basis with $Q=0$ and using a smooth
truncation of the form given in (\ref{fermi}). As expected from the
preceding discussion, these eigenvalues unfold to give a smooth
function of $\Omega=\theta+2\pi m$, decreasing from the centre.
The leading eigenvalue is peaked in the
so-called {\it scarred region} of the spectrum --- we have chosen phases
so that this occurs at $\theta=0$. In terms of wavefunction
statistics, this means that the corresponding eigenvector defines
a probe state for which the variance is much larger than predicted
by RMT (about 18 times larger for the parameters illustrated in
Figure~\ref{eigS}). The scarred region is also where overlap statistics
of chaotic wavefunctions with a simple Gaussian probe state
have the greatest variance. In the case of a Gaussian probe state, however,
the variance falls between these peaks to a value significantly
smaller than predicted by RMT in the {\it anti-scarred region},
corresponding here to the region around $\theta=\pi$ --- this contrasts
with the case where we track a probe state corresponding to the
leading eigenvector, in which case the overlap statistics have large
variances even in the anti-scarred region (albeit not as large as in the
scarred region).

The leading eigenvalue $s_0(\theta)$ can be tracked through
crossings in the antiscarred region to give the next eigenvalue
$s_1(\theta)$ and so on. Although less dramatically deviant than
the leading eigenvalue, these subsequent eigenvalues define probe
states which also have overlaps that are significantly larger than
predicted by RMT. As with the leading probe state, the overlaps
they define are larger than RMT even in the antiscarred region.
For any finite truncation dimension and increasing index, the
eigenvalues will eventually become very sensitive to the truncation
and will not define useful probe states. For a fixed index
and increasing truncation dimension, however, the probes states
provide us with a useful characterisation of wavefunctions near
the periodic orbit and we will refer to these probe states
as {\it scar states}.

We will see later that the leading eigenvalues increase
logarithmically with the effective truncation dimension $M$.
The leading eigenvectors, on the other hand, converge to give
a meaningful limit in the region of phase space surrounding the
periodic orbit. A detailed discussion for the Husimi distributions
of such hyperbolic stationary states has been given in \cite{NoVo} and
we illustrate these with some explicit cases here. Given an
eigenvalue $s_i(\theta)$, we denote the corresponding scar state
by $\ket{\chi_i(\theta)}$
which in the limit $M\to\infty$ approaches
\[
\ket{\chi_i(\theta)} \to \ket{E(\theta+2\pi m_i)^\pm}
\]
for some integer $m_i$. We denote the corresponding Husimi distribution
by
\begin{equation}\label{calcHi}
H_i(q,p;\theta) = |\braket{z}{\chi_i(\theta)}|^2.
\end{equation}
Figure~\ref{Husplot} illustrates this Husimi distribution for some of the
leading scar states, whose eigenvalues $s_i(\theta)$ are indicated by
bullets in Figure \ref{eigS}. The scar states shown are appropriate to
a fixed point $p_0$ at the origin of phase space with stable and unstable
manifolds defined by $p=\pm q$.
Each of these Husimi functions is strongly localised around the
stable and unstable manifolds of the fixed point, although the details
vary with the spectral parameter $\theta$ and index $i$. In
distinguishing between different cases it is useful to note that
the zeros of the Husimi functions are visible as dark spots
--- these enable us to distinguish between states which are concentrated
in similar regions of phase space but which have different phase
structure.

\begin{figure}[t]

\vspace*{-1.0in}\hspace*{-0.35in}\includegraphics[width=2.5in] {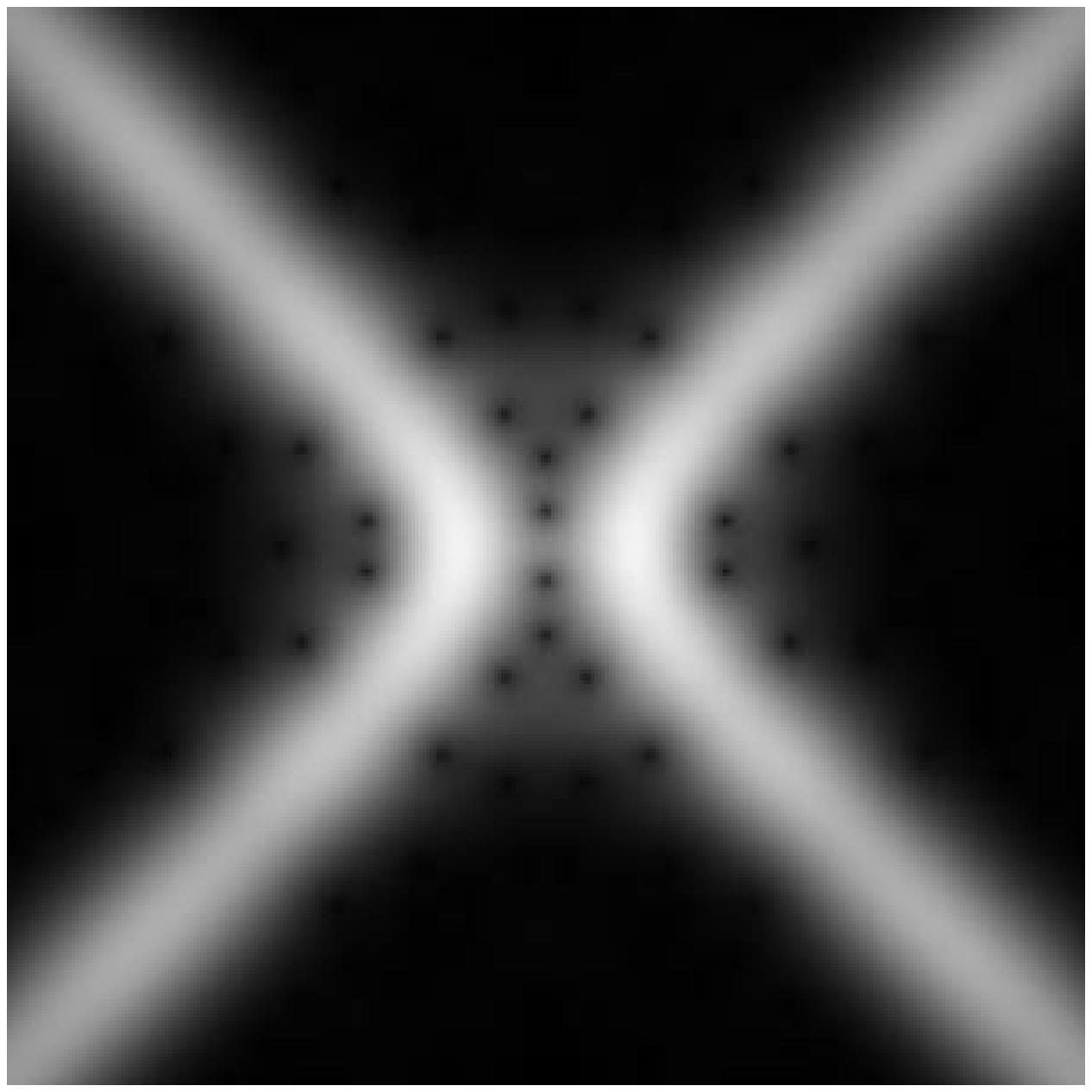}

\vspace*{-3.25in} \hspace*{1.3in}\includegraphics[width=2.5in] {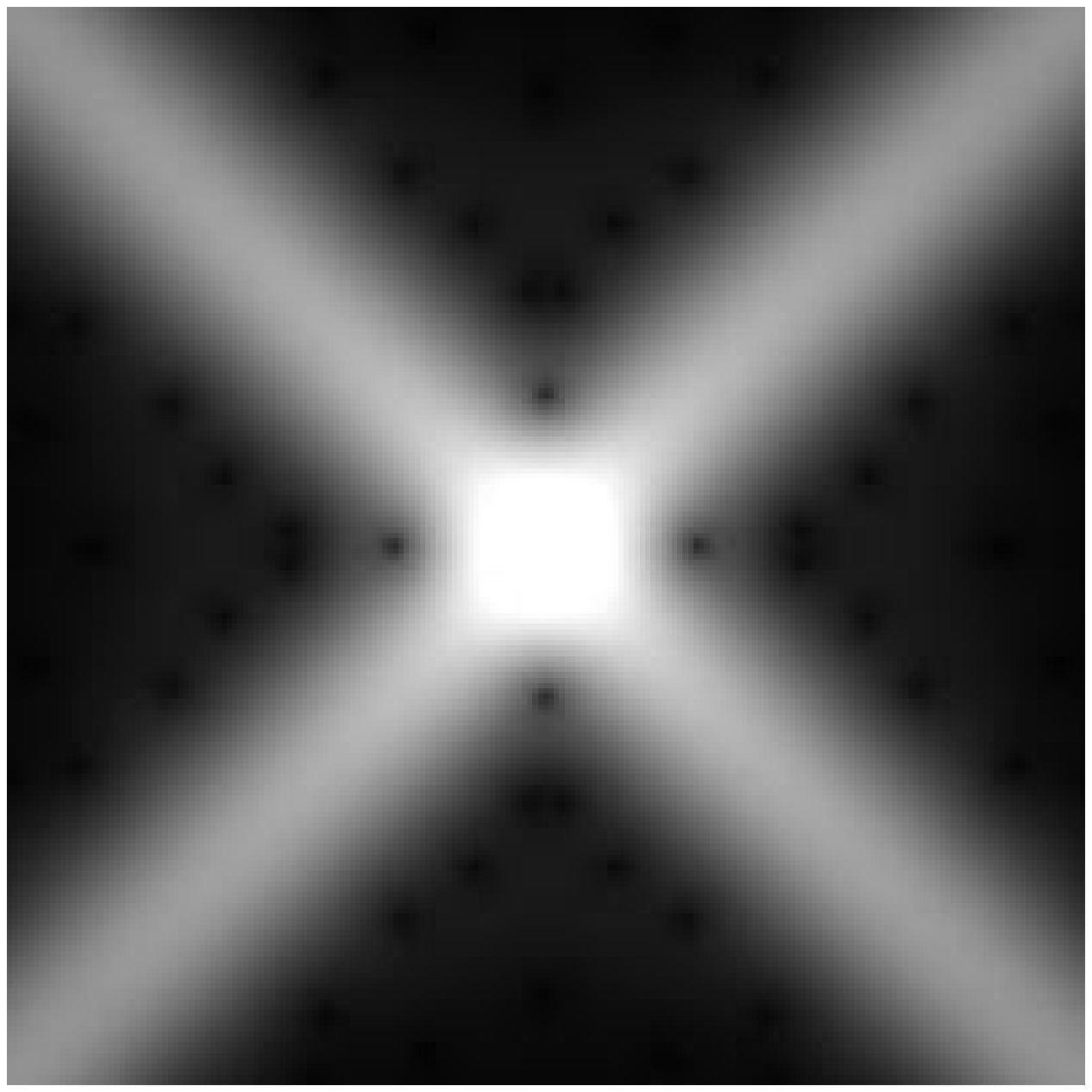}

\vspace*{-3.25in} \hspace*{2.95in}\includegraphics[width=2.5in] {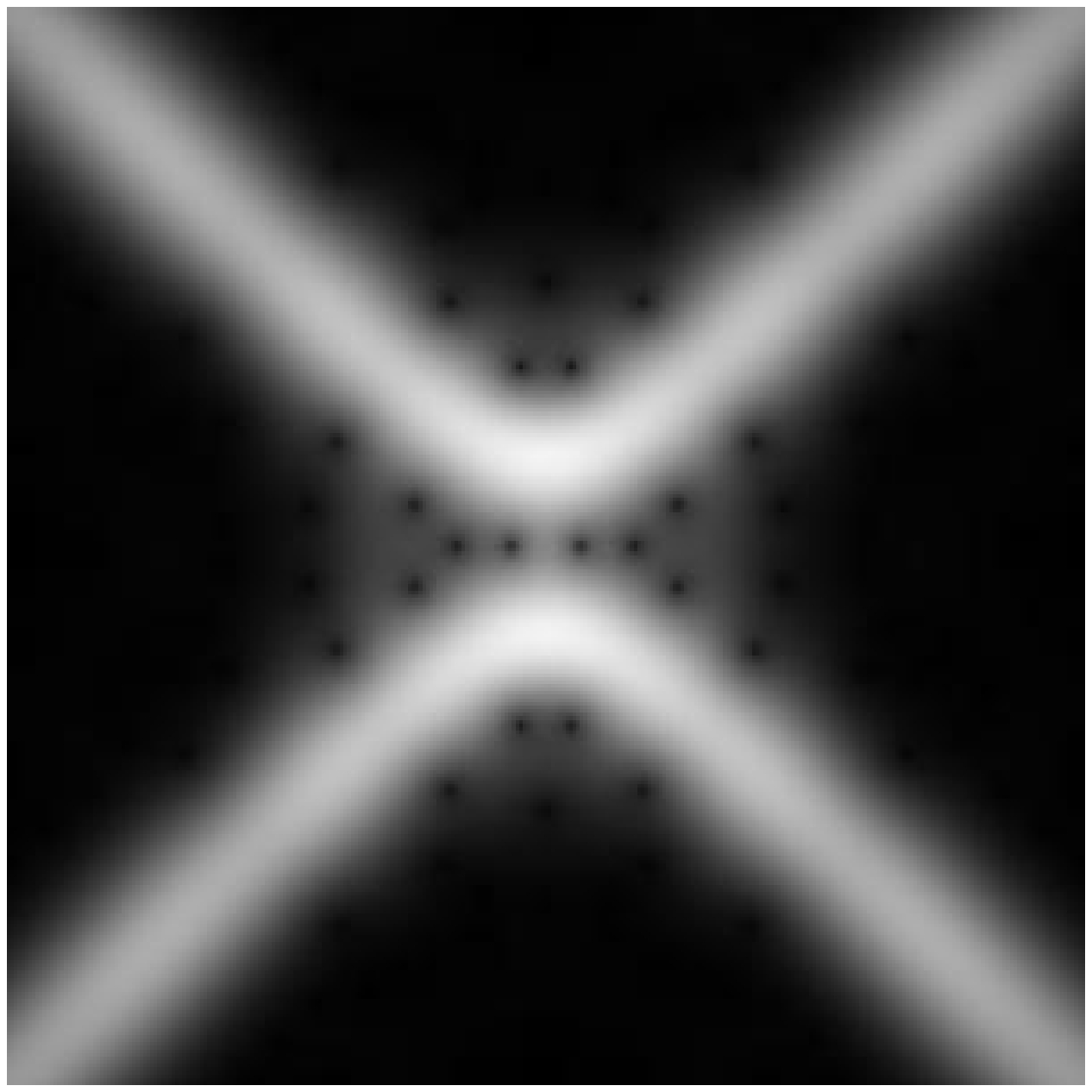}

\vspace*{-3.25in} \hspace*{4.6in}\includegraphics[width=2.5in] {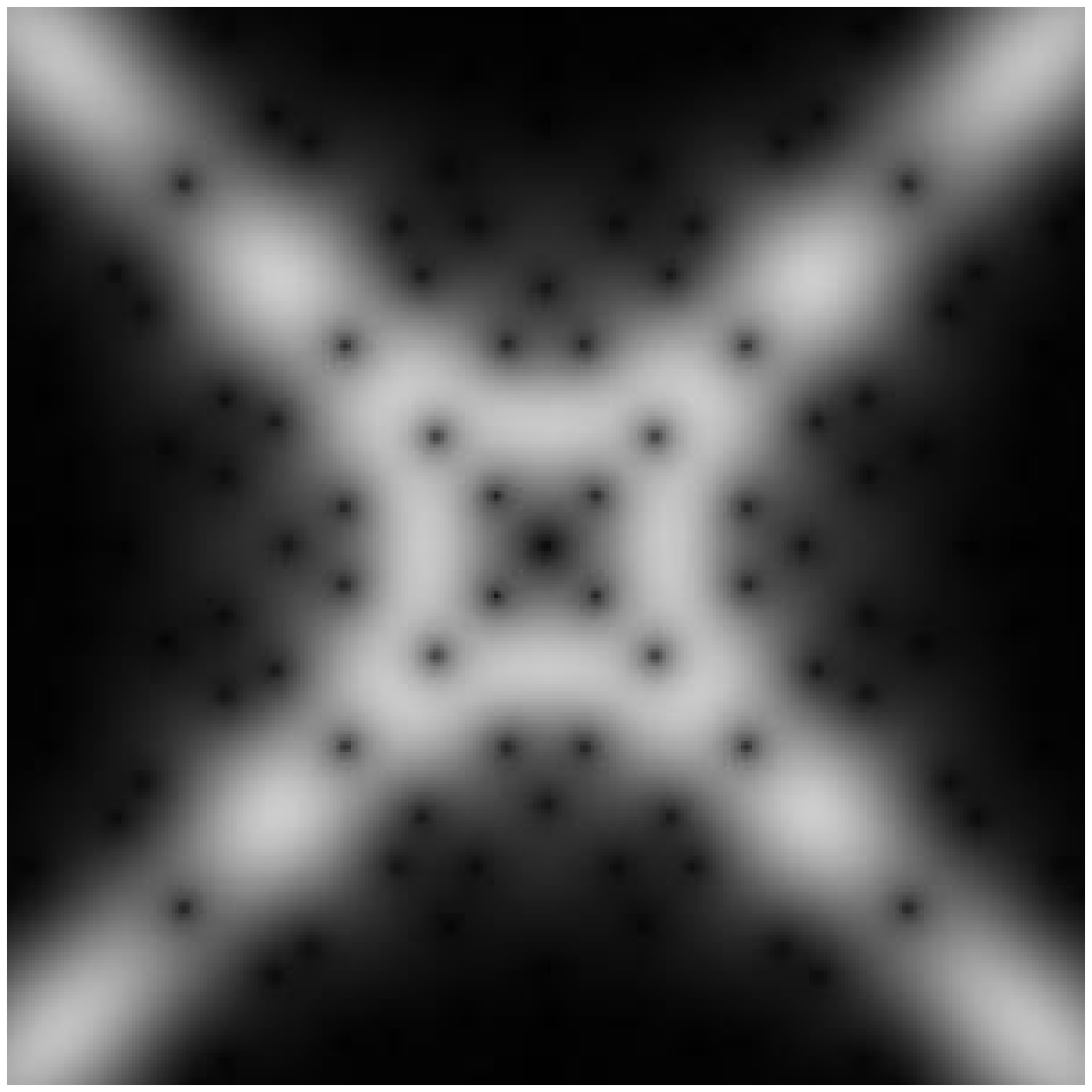}

\vspace*{-1.63in}\hspace*{-0.35in}\includegraphics[width=2.5in] {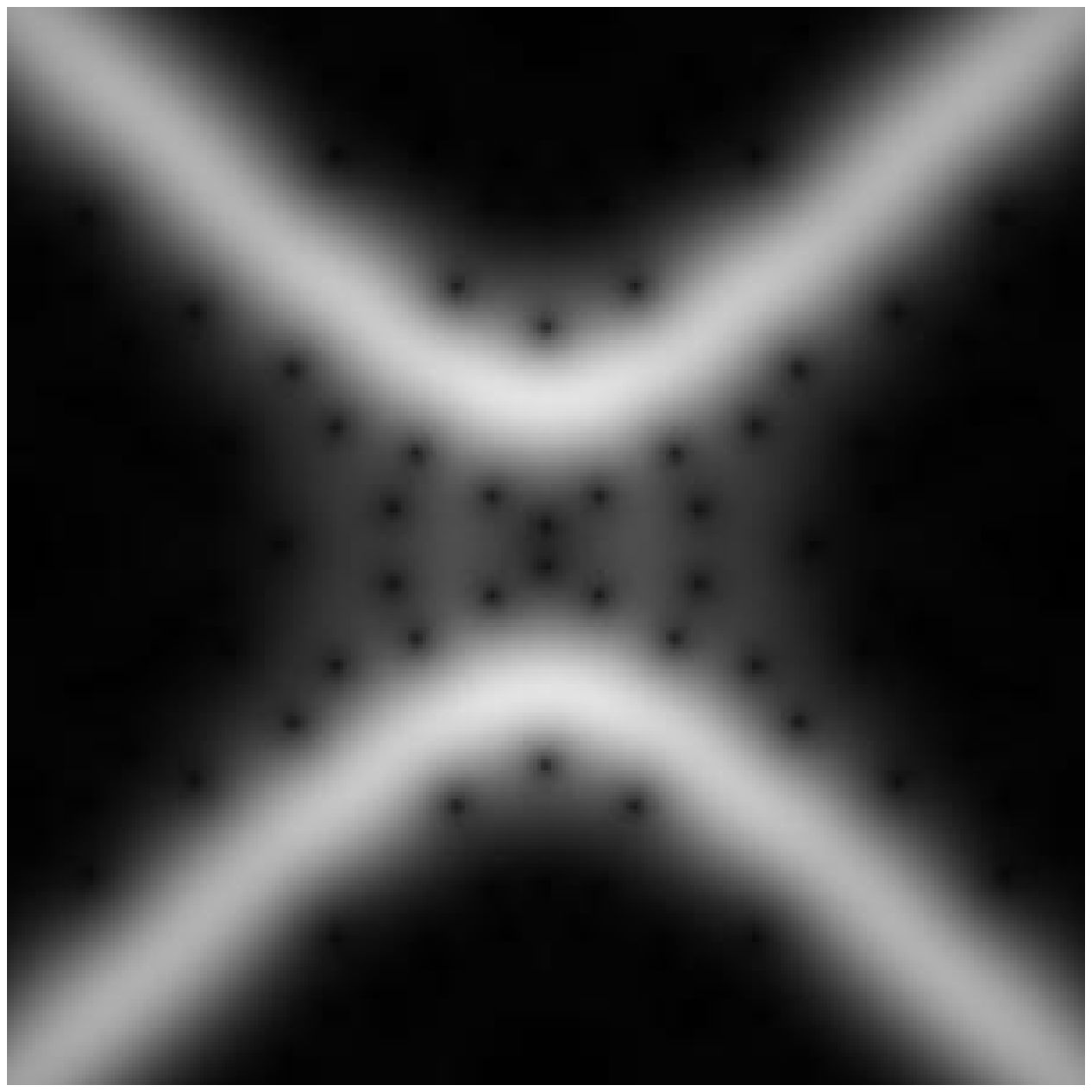}

\vspace*{-3.25in} \hspace*{1.3in}\includegraphics[width=2.5in] {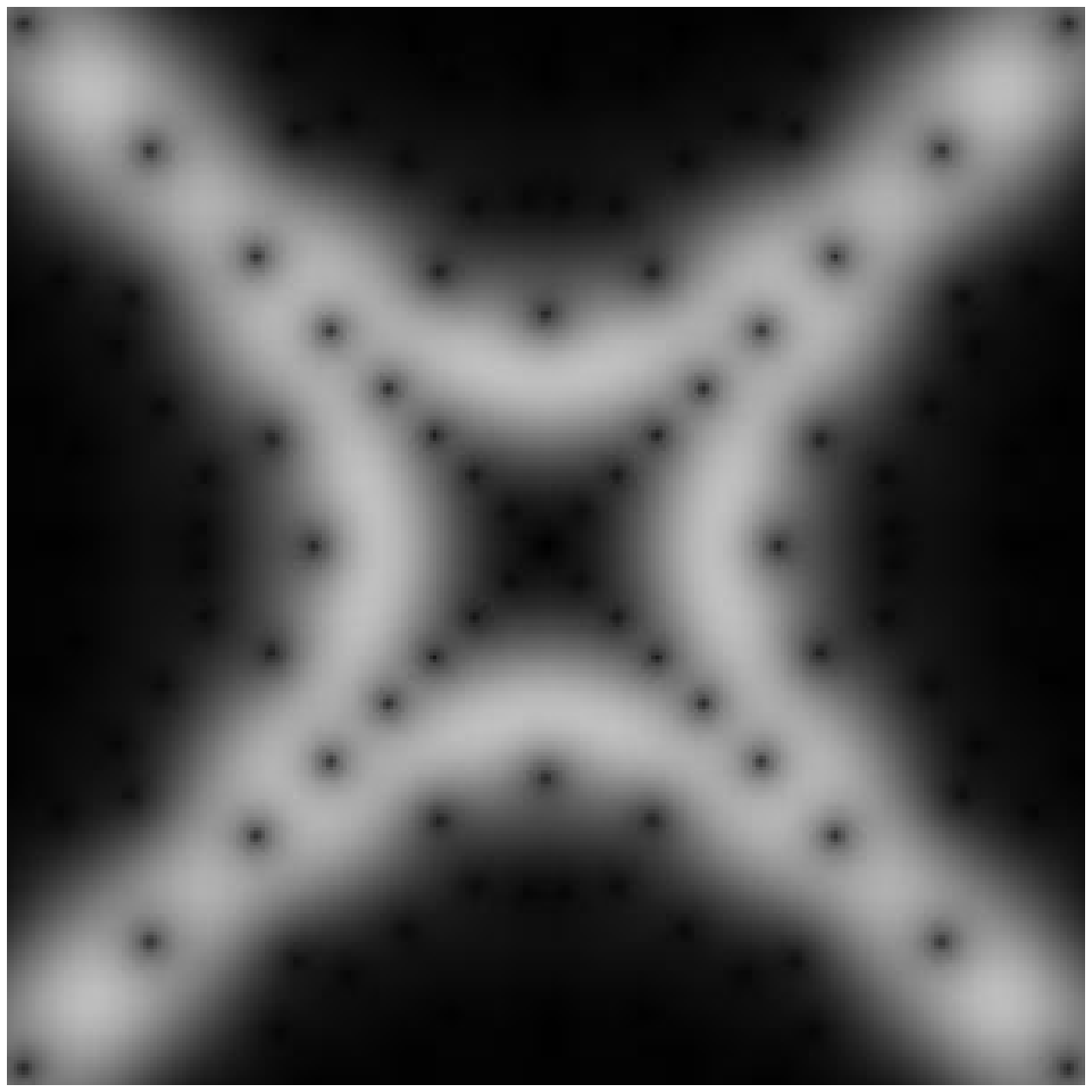}

\vspace*{-3.25in} \hspace*{2.95in}\includegraphics[width=2.5in] {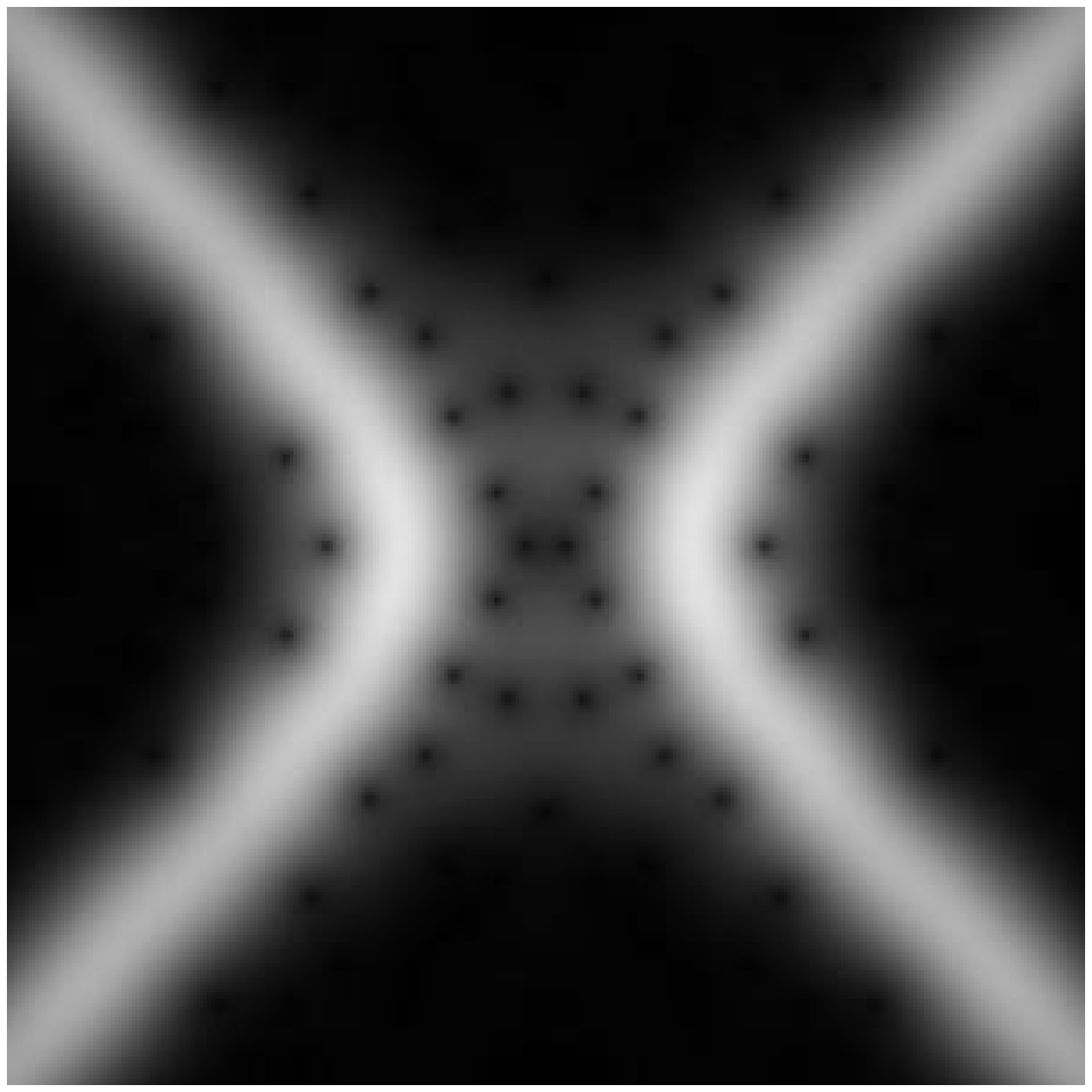}

\vspace*{-3.25in} \hspace*{4.6in}\includegraphics[width=2.5in] {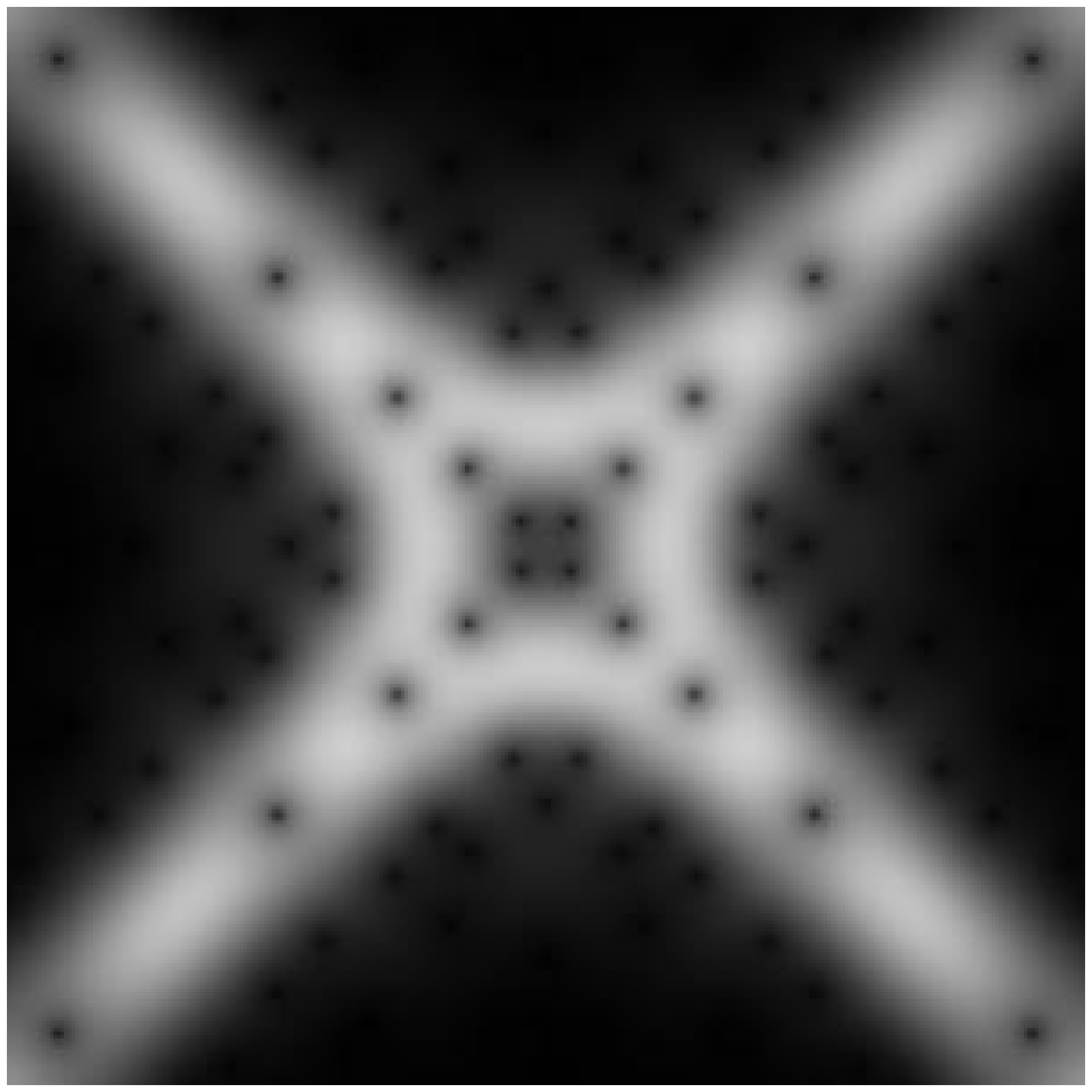}

\vspace*{-1.4in}\hspace*{-0.35in}\includegraphics[width=2.5in] {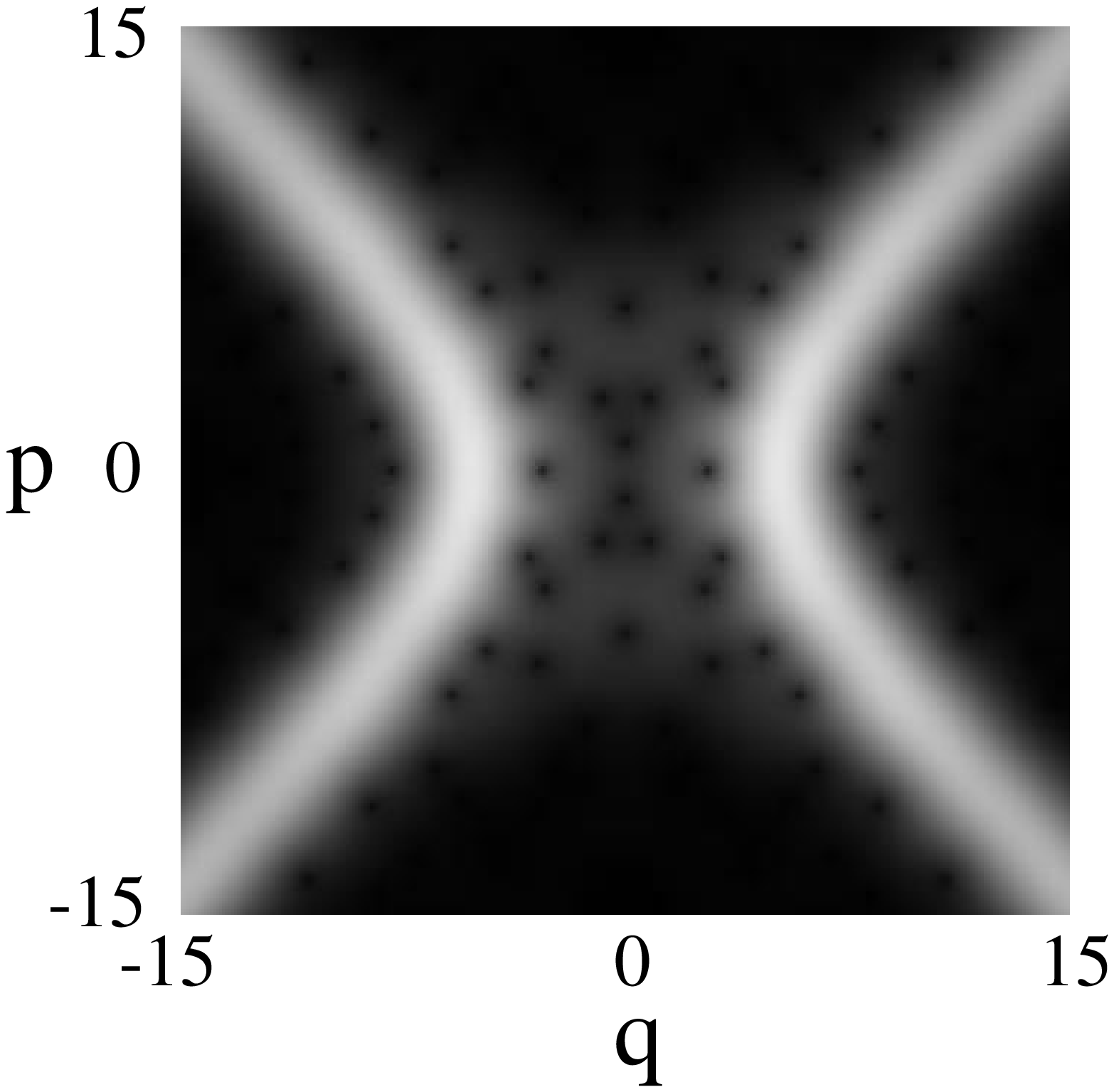}

\vspace*{-3.25in} \hspace*{1.3in}\includegraphics[width=2.5in] {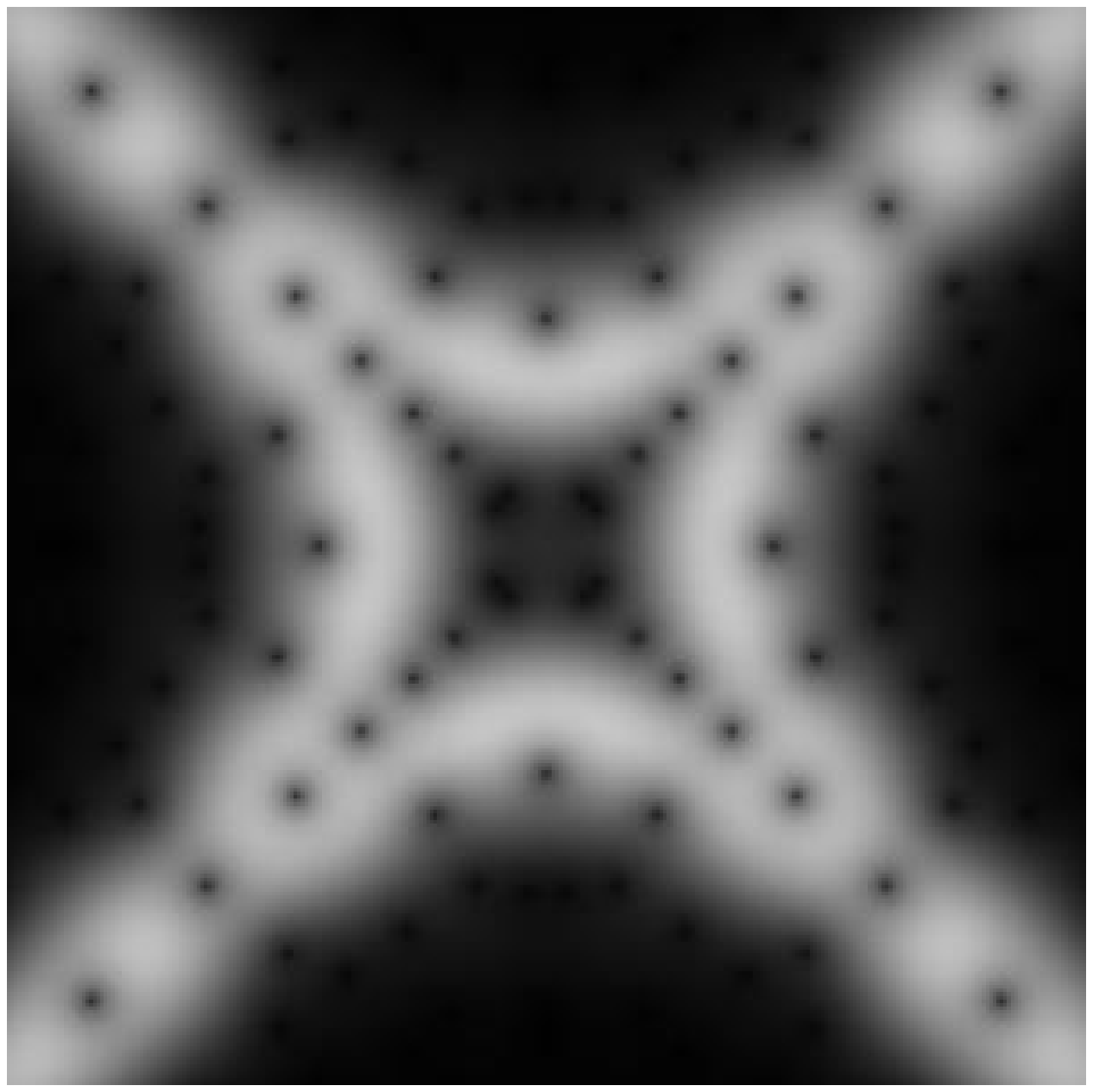}

\vspace*{-3.25in} \hspace*{2.95in}\includegraphics[width=2.5in] {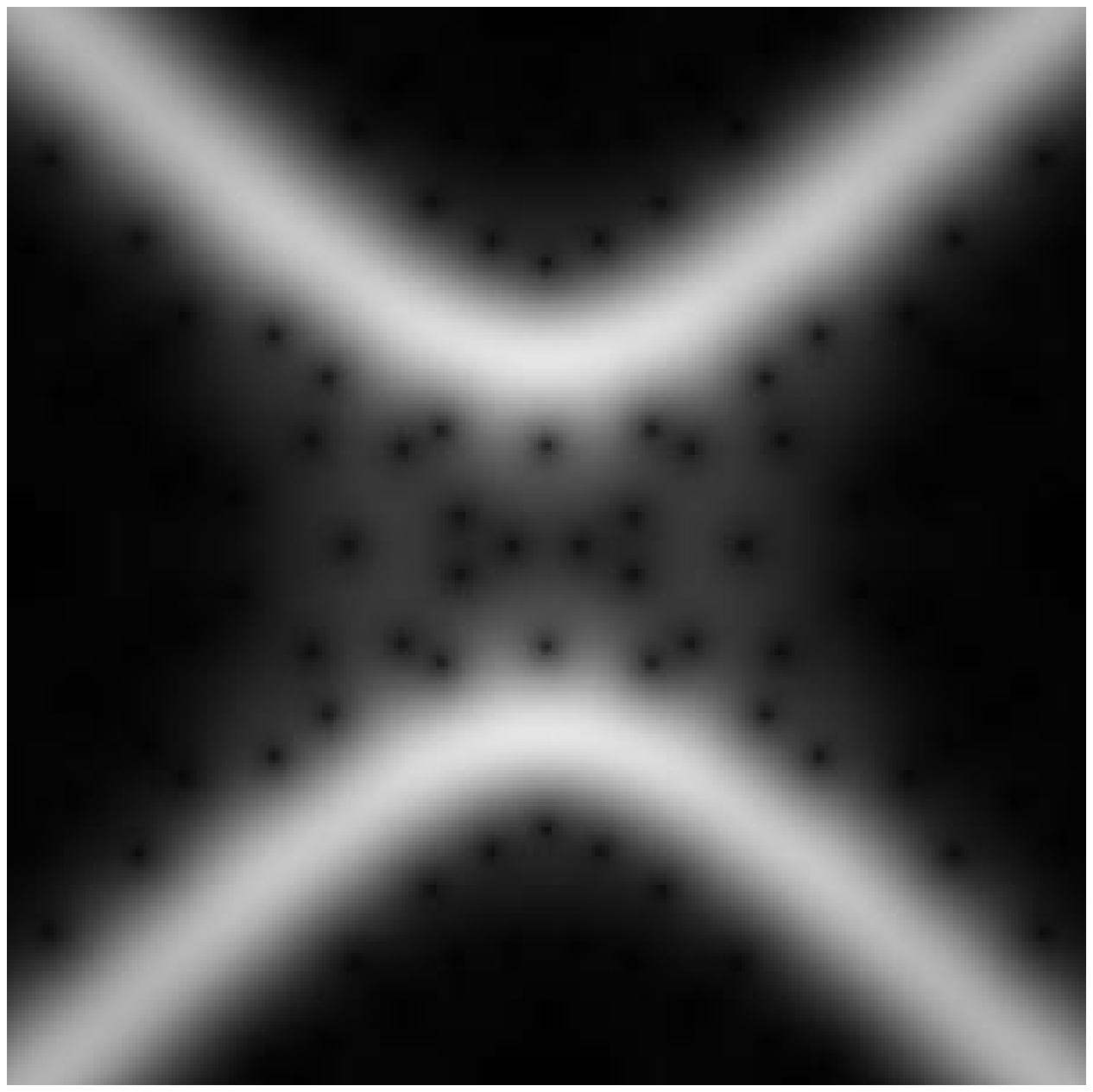}

\vspace*{-3.25in} \hspace*{4.6in}\includegraphics[width=2.5in] {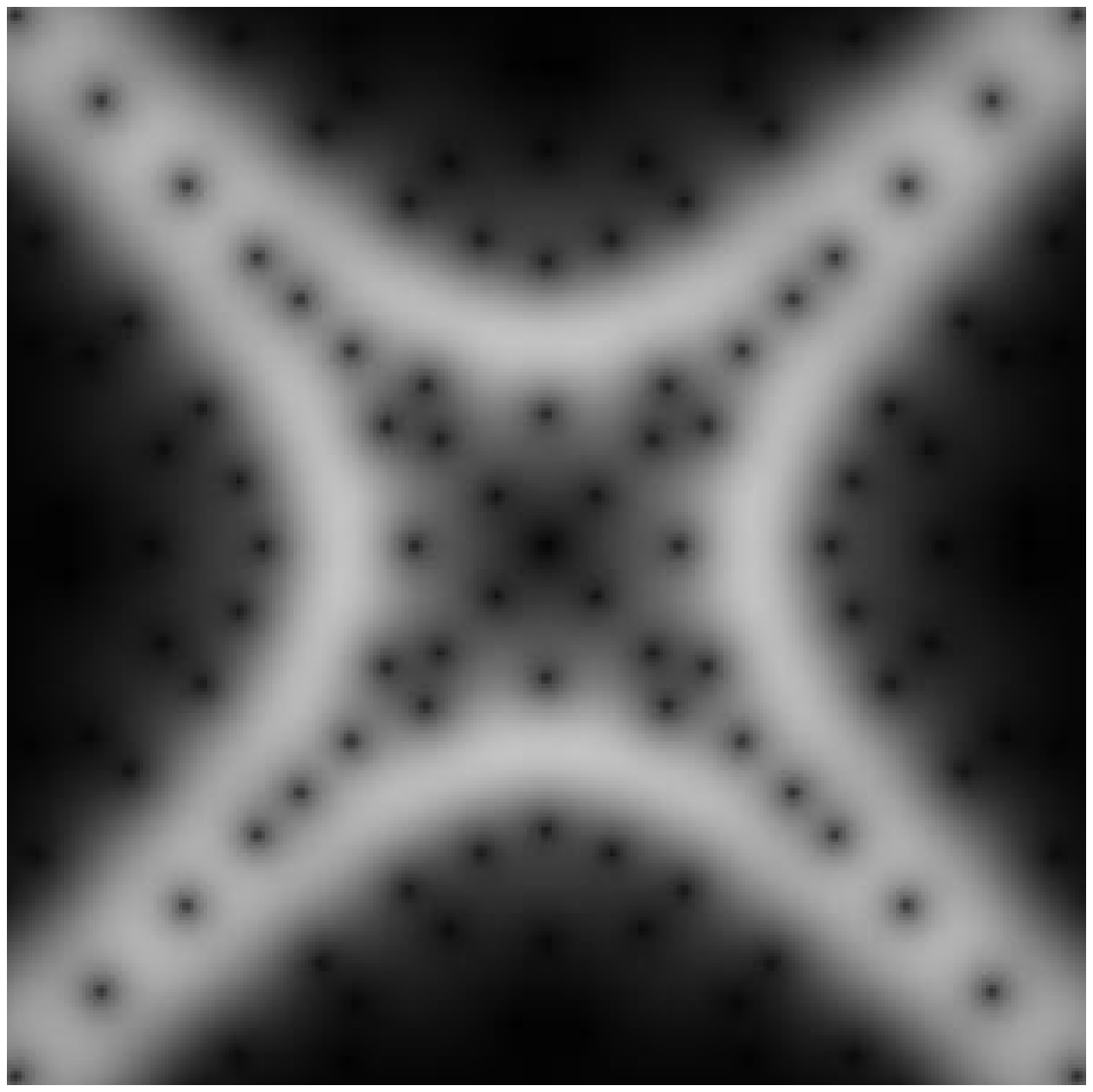}

\vspace*{-1.3cm}
\caption{The Husimi functions of the scar states corresponding to the
bullets in Figure~\ref{eigS}. The underlying hyperbolic system has
stable and unstable manifolds along the diagonals $p=\pm q$.
The first row gives $H_0(q,p,\theta)$ for $\theta=-\pi/2, 0,\pi/2$ and
$\pi$. The second and third rows give $H_1(q,p,\theta)$ and 
$H_2(q,p,\theta)$ for the same values of $\theta$.
Zeros of the Husimi functions appear as dark points.}
\label{Husplot}
\end{figure}

The leading scar states $\ket{\chi_i(\theta)}$ are potentially 
useful not only for understanding the statistics of wavefunctions 
but also for characterising the structure of individual wavefunctions. The 
variance of the overlap $\braket{\chi_i(\theta)}{\psi_n}$ is quite 
large and in fact can diverge in the semiclassical limit. In a given part of
the spectrum, chaotic eigenfunctions will therefore tend to have 
especially large components along the corresponding scar states.
One aspect of this is that we might therefore recognise the 
characteristics of the scar states in individual states, and the 
changing form of the scar states across the spectrum will provide a 
template for recognising such structures. We will
see explicit examples of this in the next section. A second aspect
is that the scar states should provide a powerful 
guide to choosing bases for the explicit calculation of chaotic 
eigenstates. This sort of construction has recently been proposed
by Vergini and Carlo \cite{VE00}. They have in particular defined
for each periodic orbit a {\it scarfunction} which is concentrated
around the stable and unstable manifolds much like the 
scar states $\ket{\chi_i(\theta)}$ are. In fact we will see in the 
next subsection that the scarfunction coincides with the leading state
$\ket{\chi_0(0)}$ in the semiclassical limit. The property of 
the states $\ket{\chi_0(\theta)}$ that they change as a function of 
$\theta$ and adapt to a given part of the spectrum, as well as the ability
to generate additional orthogonal states (labelled by $i$),
might therefore provide a useful generalisation of that approach.

\subsection{The leading scar state}

We show now how the decomposition outlined in the previous subsection
emerges in the harmonic-oscillator basis. We cannot give closed-form 
expressions for general scar states $\ket{\chi_i(\theta)}$ in this basis,
but we can calculate the basic properties of the leading state 
$\ket{\chi_0(0)}$ in the scarred part of the spectrum, corresponding 
to $\theta=0$ in our convention. We will see in particular that this 
reproduces the scarfunction of Vergini and Carlo \cite{VE01}.

We have noted that the envelope matrix $\C(\theta)$ depends on the 
stability exponent $\rho$ of the underlying hyperbolic dynamics and 
a parameter $Q$ depending on the basis. We are free to choose a 
basis for which $Q=0$ and we assume that to be the case in this 
subsection. The stability exponent $\rho$ determines how quickly 
the correlation matrix $\A(t)$ decays with $t$ and therefore 
how far the envelope matrix $\C(\theta)$ deviates from the RMT 
case $\C(\theta)=I$.
The eigenvalues $\s(\theta)$ in particular depend strongly on $\rho$.
The scar states, however, depend only on the underlying 
geometry of the stable and unstable manifolds and not on the rate of decay.
We therefore consider 
the limit $\rho\to 0$, which simplifies because we can then replace the 
Fourier series defining $\C(\theta)$ by an integral
\begin{equation}
\sum_t {\bf A}(t) e^{i\theta t} \longrightarrow 
\int \, {\bf A}(t) e^{i\theta t}\;\d t.
\end{equation}
In other words, the sum in (\ref{decompC}) is dominated by the $m=0$ term.
We note that this formulation is similar to the construction of the 
universal scar measure in \cite{KAHE} or the continuous quasimode
in \cite{FNdB}, except here we also incorporate the propagation of
excited harmonic-oscillator eigenstates as well as the Gaussian
ground state.

Using the form for ${\bf A}(t)$ given in 
Appendix~\ref{appgetA} we can in particular evaluate this integral 
explicitly for the case $\theta=0$. We find a decomposition
of the form
\begin{equation}
{\bf C}(0) \simeq \frac{1}{\rho} {\bf v} {\bf v}^T,
\end{equation}
where the components of the vector ${\bf v}$ are  
\begin{equation}\label{givew}
v_k  =   \cases {\ds
(-1)^{n}  (2\pi)^{-1/4}
  \frac{\sqrt{(4n)!}}{2^{2n}(2n)!}
\frac{\Gamma(n+\frac{1}{2})}{\Gamma(n +\frac{3}{4})}
& \quad for $k=4n$,    \cr
			\noalign{\vskip12pt}
\qquad\qquad 0
& \quad otherwise. 
\cr &			
}
\end{equation}
This means that ${\bf v}$ is the leading eigenvector of $\C(0)$
and that the corresponding eigenvalue is
\begin{equation}\label{higheig}
 s_0(0) \simeq \frac{1}{\rho} {\bf v}^T {\bf v}.
\end{equation}
The vanishing of components $v_k$ for odd $k$ is a reflection of the 
fact that linearised evolution has inversion through the origin as a 
symmetry --- we have already see that this symmetry leads in the general
case to a decoupling of $\C(\theta)$ into odd and even blocks. When
we choose a basis for which $Q=0$, the stable and unstable manifolds are
orthogonal and we have in addition a symmetry
of interchange $(q,p)\to(p,-q)$ of the corresponding canonical coordinates.
this is responsible for a further decoupling of the index $k$ mod $4$.
This four-fold symmetry is also reflected in the general case, in 
particular through the symmetry $H_i(q,p,\theta)=H_i(p,q,-\theta)$ 
visible in the Husimi functions of Figure~\ref{Husplot}.

We remark that this eigensolution depends on the stability exponent $\rho$
only through the prefactor in (\ref{higheig}). The calculation has 
been performed formally for the untruncated matrix $\C(\theta)$ and
we note that (\ref{higheig}) is divergent and gives a meaningful 
eigenvalue only if we truncate the matrix at a finite dimension $M$. 
We can obtain an estimate for the divergence of  $s_0(0)$ with truncation 
dimension by noting that the components of ${\bf v}$ have the
asymptotic behaviour (as $k=4n\to\infty$)
\[
v_{4n}^2 \simeq  \frac{1}{n}.
\]
The leading eigenvalue for a matrix sharply truncated at dimension $M$
therefore asymptotes to (as $\rho\to 0$ and $M\to\infty$)
\begin{equation}\label{sislog}
s_0(0) \simeq  \frac{1}{\rho}\sum_{n=1}^{M/4} v_{4n}^2 
			\simeq \frac{1}{\rho}
				\log (M/4) + {\rm const}.
\end{equation}
Although our derivation has been restricted to the limit $\rho\to 0$,
we remark that this describes quite well the divergence of $s_0(0)$
even for moderate values of $\rho$, as indicated in Figure~\ref{assymps}
where we compare it with a numerical calculation for $\rho=0.6$.

\begin{figure}[h]
\vspace*{-1.in}\hspace*{1.3in}\includegraphics[width=4.5in] {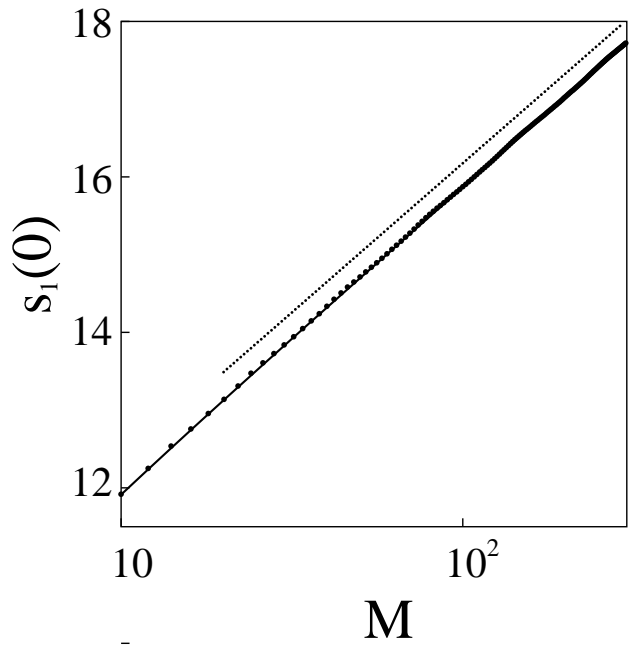}
\vspace*{-2.5cm}
\caption{ The leading eigenvalue $s_0(0)$ as a function of 
the sharp cut-off dimension $M$ when $\rho =0.6$ and $Q=0$, showing
a logarithmic divergence. The solid curve is the approximation in 
(\ref{sislog}).
 }
\label{assymps}
\end{figure}

We expect similar divergence to hold for the next-to-leading 
eigenvalues. Since $s_i(\theta)$ diverges we might then expect to 
be able to make the statistical deviation from RMT arbitrarily 
large by choosing appropriate probe states. It should be noted, 
however, that our results are meaningful only if we take the
limit $M\to\infty$ in conjunction with the semiclassical limit
$\hbar\to 0$. Our calculation of the correlation $C_{kl}(\theta)$
between two basis states was obtained by assuming evolution 
that is linearised about the periodic orbit. This is always valid for 
fixed $k$ and $l$ in the semiclassical limit. If we fix $\hbar$ 
and take the limit $k,l\to\infty$, however, the basis states
$\ket{\phi_k}$ will eventually fall outside the region of phase 
space where the linearisation is appropriate and the correlations
$C_{kl}(\theta)$ will then not accurately describe the statistics.
For any given quantum system then, we should restrict ourselves
to truncations for which linearised evolution is accurate and 
this effectively limits the value of $s_i(\theta)$ and therefore
the deviation from RMT.

Consistent with the preceding subsection, the vector  ${\bf v}$ can 
also be expressed formally as an eigenstate of the Hamiltonian
\begin{equation}
\Hh = \frac{\rho}{2} (p^2 -q^2)
\end{equation}
giving the underlying hyperbolic motion in appropriate canonical 
coordinates $(q,p)$. We can express this Hamiltonian in terms of creation and
annihilation operators  of a harmonic oscillator basis for which $Q=0$ as 
\[
\Hh = -\frac{\rho}{2}({a^\dagger}^2+ a^2).
\]
The vector ${\bf v}$ is then easily seen formally to be a null eigenvector 
\begin{equation}
{\bf H} {\bf v} = 0
\end{equation}
of the Hamiltonian matrix
\[
       ({\bf H})_{kl} \equiv \braopket{\phi_k}{\Hh}{\phi_l}.
\]
In particular the  leading scar state is also one of the simpler cases
for which Husimi functions were constructed in \cite{NoVo}.
This can be seen by using the representation
\[
\ket{z} = e^{-|z|^2/2} \sum_{k=0}^\infty
				\frac{1}{\sqrt{k!}}{z^*}^k\ket{\phi_k}
\]
of the coherent states, where $z=(q-ip)/\sqrt{2\hbar}$, and substituting
the coefficients of (\ref{givew}) in (\ref{calcHi}), 
giving
\begin{eqnarray*}
H_0(z,0) &=& \frac{1}{\sqrt{2\pi}} \,
			e^{-|z|^2}\left| \sum_{n=0}^\infty (-1)^n   
\frac{\Gamma(n+\frac{1}{2})}{\Gamma(n +\frac{3}{4})} 
\frac{z^{4n}}{2^{2n}(2n)!} \right|^2 \\[3pt]
&=& \frac{1}{2}\sqrt{\frac{\pi}{2}} e^{-|z|^2}\left|
\sqrt{z}J_{-1/4}\left(\frac{z^2}{2}\right)
\right|^2
\end{eqnarray*}
for the Husimi function of this state. This coincides with the 
expression in \cite{NoVo} for the $E=0$ eigenstate of the 
hyperbolic system as expected from the discussion of the previous 
subsection.

The preceding discussion also helps us to relate $\ket{\chi_0(0)}$ 
to the scarfunction of Vergini and Carlo \cite{VE01}. They define the
scar function $\ket{\chi}$ to be a finite linear combination of the states 
$\ket{\phi_k}$ which minimises
\[
\braopket {\chi}{\Hh^2}{\chi}.
\]
In the limit $M\to 0$ the eigenvector ${\bf v}$ provides a solution 
to just this problem and the corresponding state $\ket{\chi_0(0)}$ 
should therefore coincide with the scarfunction of \cite{VE01} in 
that limit.

Away from the scarred region, that is for $\theta\neq 0$, the scar 
states should correspond to universal test states \cite{KAHE,Wis01},
or quasimodes \cite{FNdB}, although we have not written explicit forms
for them in a harmonic-oscillator representation. Note however
that one can find explicit expression for the Husimi functions 
in \cite{NoVo}. We also note that phase space representations of operators 
with the structure of $\tilde{\C}(\theta)$ have also been given by Rivas
and Ozorio de Almeida \cite{ROdA}, and these might also be useful
in deriving explicit expressions for the scar states in phase space.

\section{Application to a quantum map model}

We now investigate how well the scar states describe wavefunction 
structure and statistics for a specific quantum map model. 
The details of this model are unimportant and we
merely remark that the classical system is a hyperbolic mapping of the 
torus with time-reversal symmetry and a fixed point at the origin --- 
the detailed parameters for the system are the same as those given 
in \cite{CR02} and the reader should look there for further details. 
As in \cite{CR02}, we choose as basis states $\ket{\phi_k}$ the 
eigenstates of a Harper Hamiltonian which is harmonic near the 
fixed point. This problem is characterised by a stability exponent 
$\rho=0.6006$ and a basis-parameter $Q=0.5658$. The semiclassical
limit is controlled by the dimension of Hilbert space, which we 
denote by $N$. Because of numerical limitations we use relatively 
modest truncations of $\C(\theta)$ in this investigation ---
typically we use a sharp truncation with $M=28$ or less.

%We perform numerical calculations for a quantisation of the cat map
%\[
%\spinor{q}{p} \mapsto \mat{1}{1}{1}{2} \spinor{q}{p}
%				\qquad \mbox{mod $1$},
%\]
%which we perturb by composing  with a near-identity map
%\[
%\spinor{q}{p} \mapsto \spinor{q-\eps\sin{2\pi p}}{p}
%		\qquad \mbox{mod $1$}
%\]
%in order to eliminate the degeneracies of the simple cat map.

\begin{figure}[h]

\vspace*{-0.5in}\hspace*{-0.35in}\includegraphics[width=2.5in] {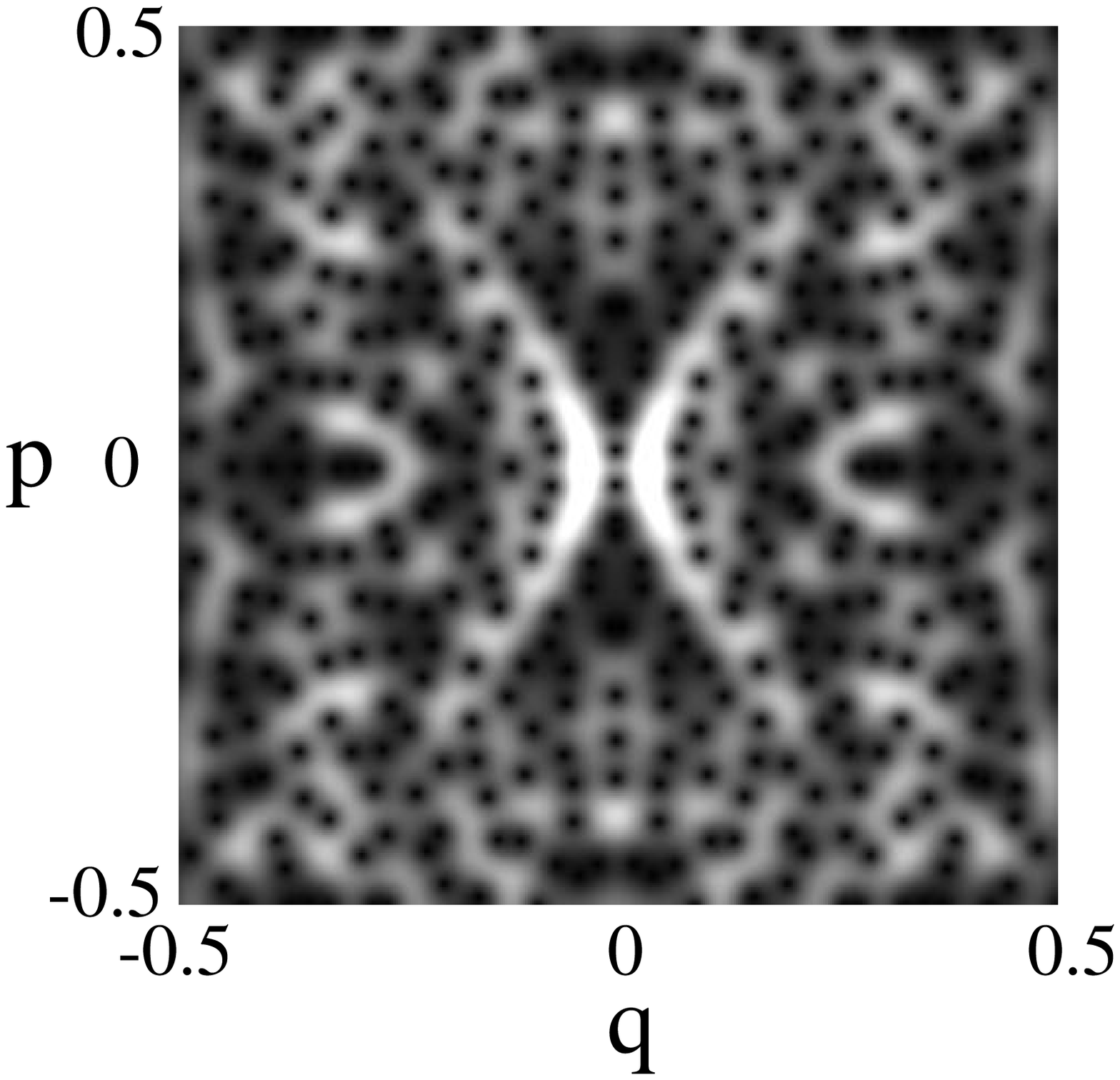}

\vspace*{-3.25in} \hspace*{1.3in}\includegraphics[width=2.5in] {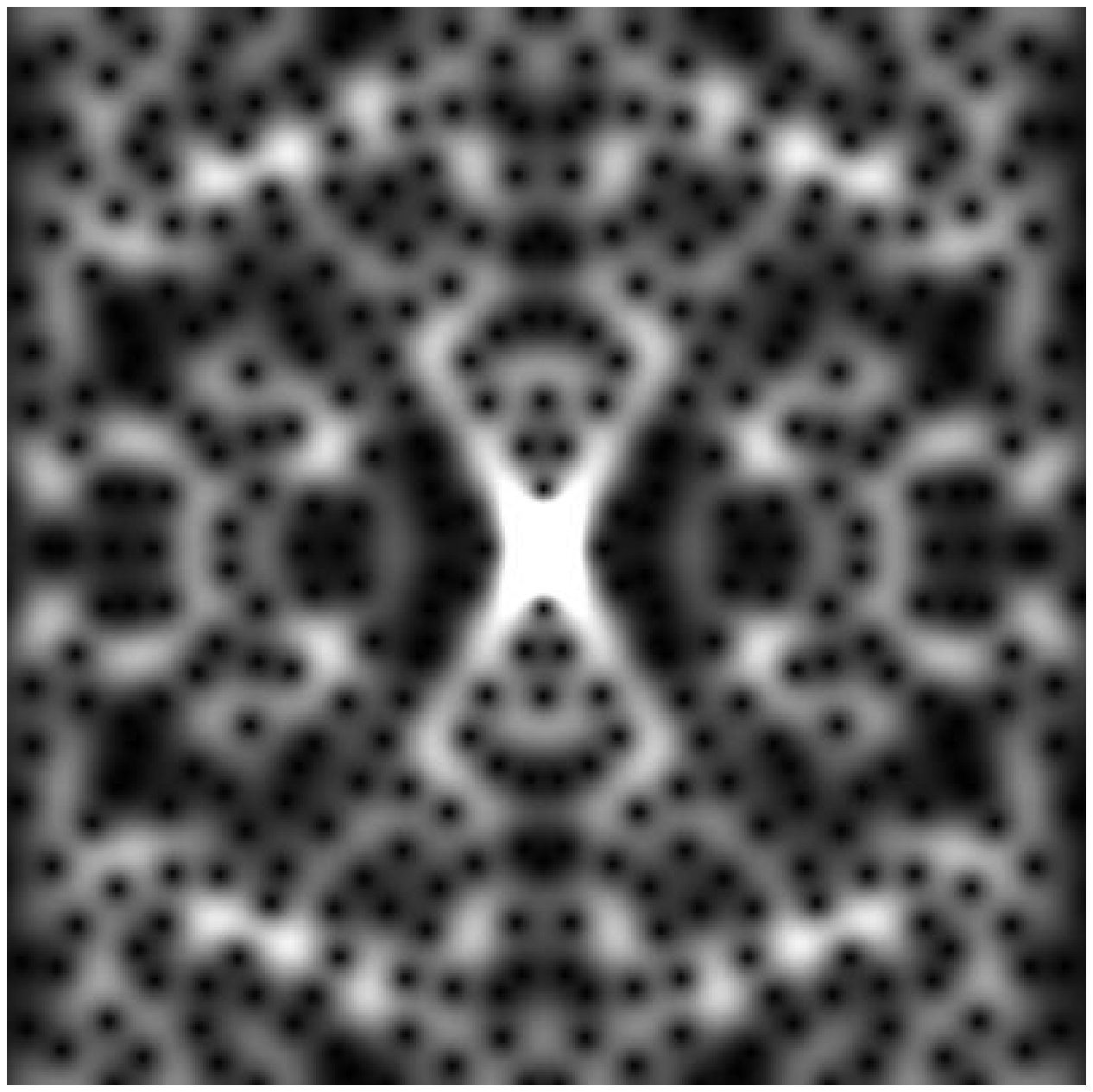}

\vspace*{-3.25in} \hspace*{2.95in}\includegraphics[width=2.5in] {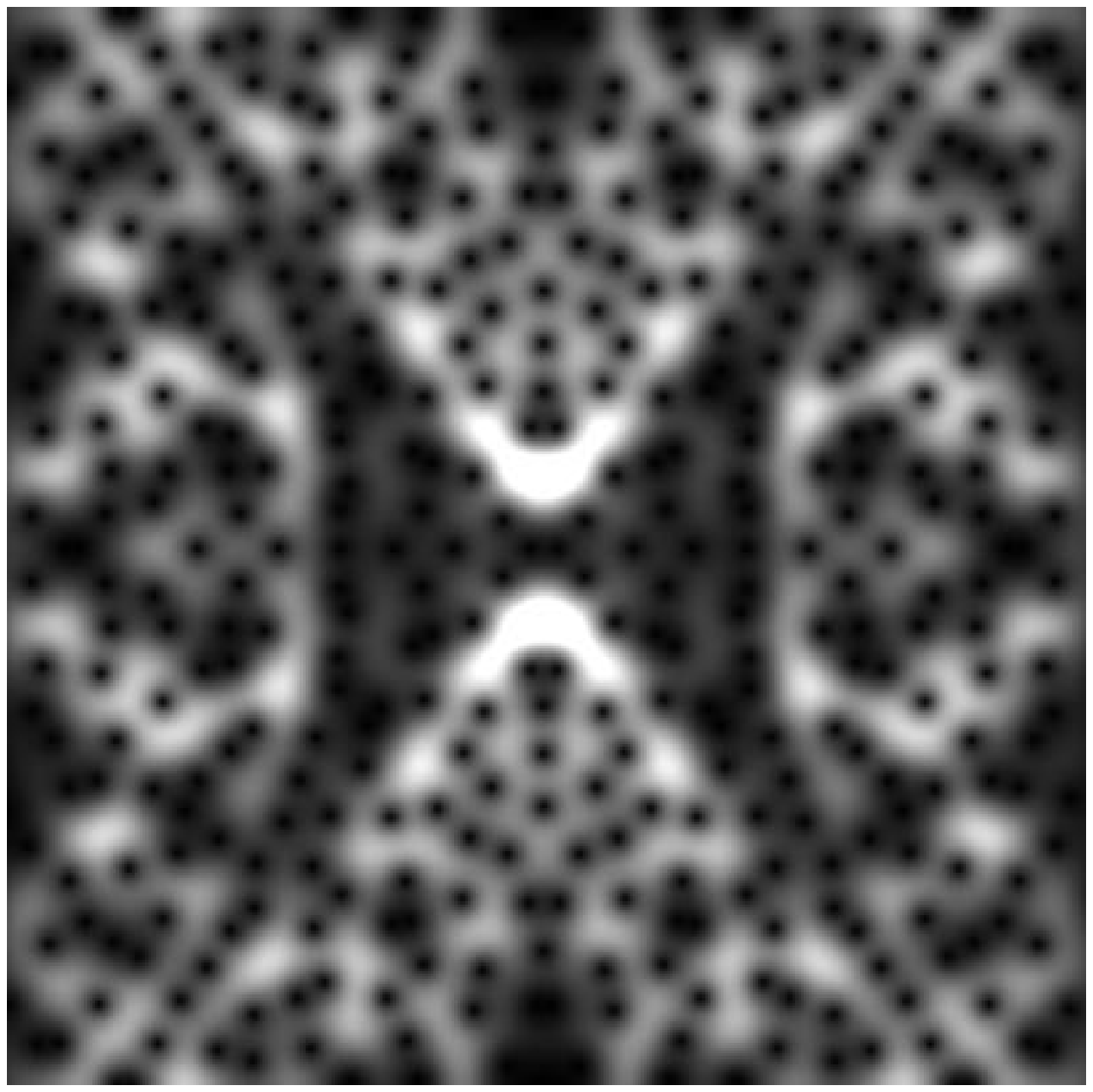}

\vspace*{-3.25in} \hspace*{4.6in}\includegraphics[width=2.5in] {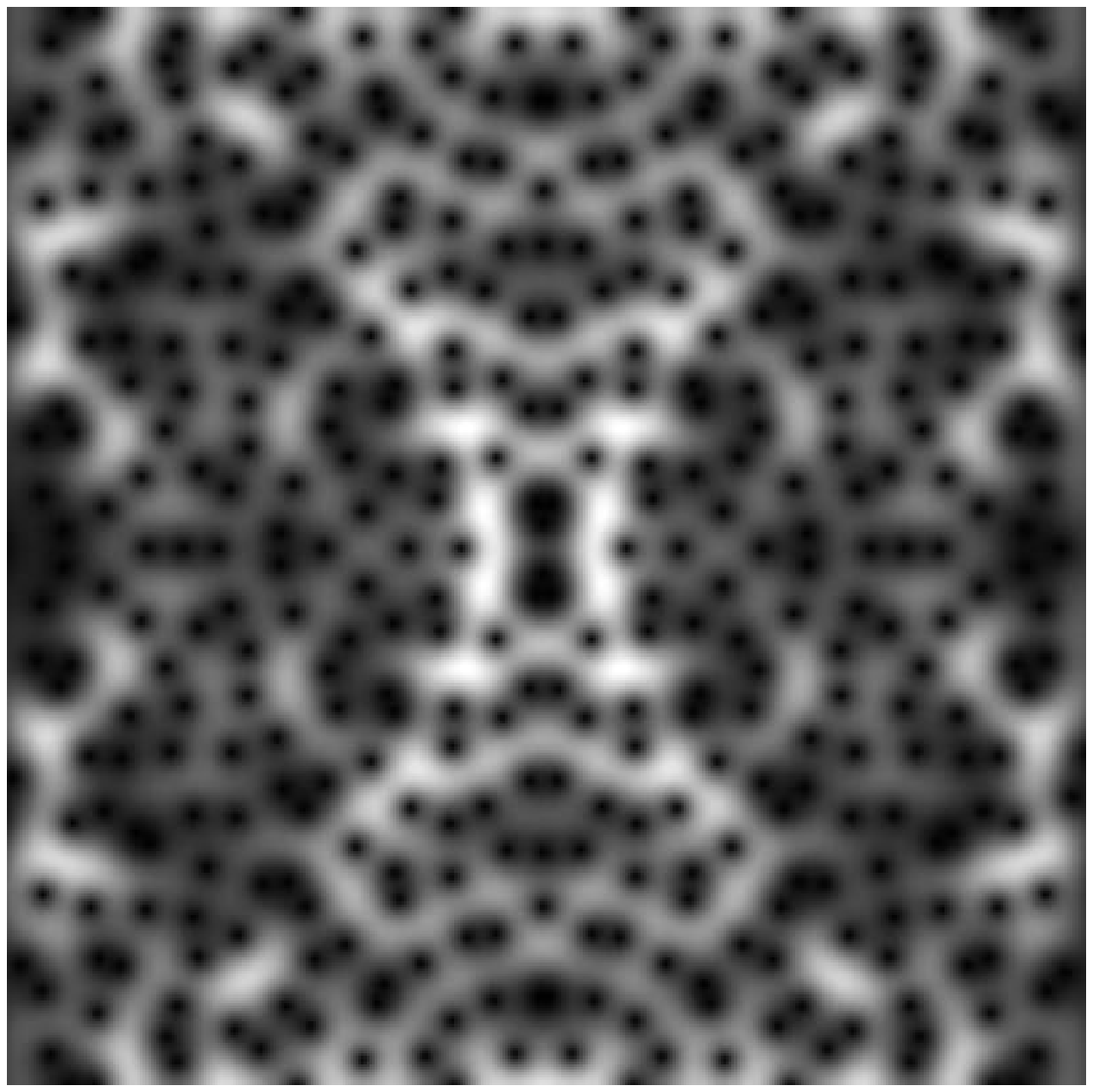}

\vspace*{-1.4in}\hspace*{-0.35in}\includegraphics[width=2.5in] {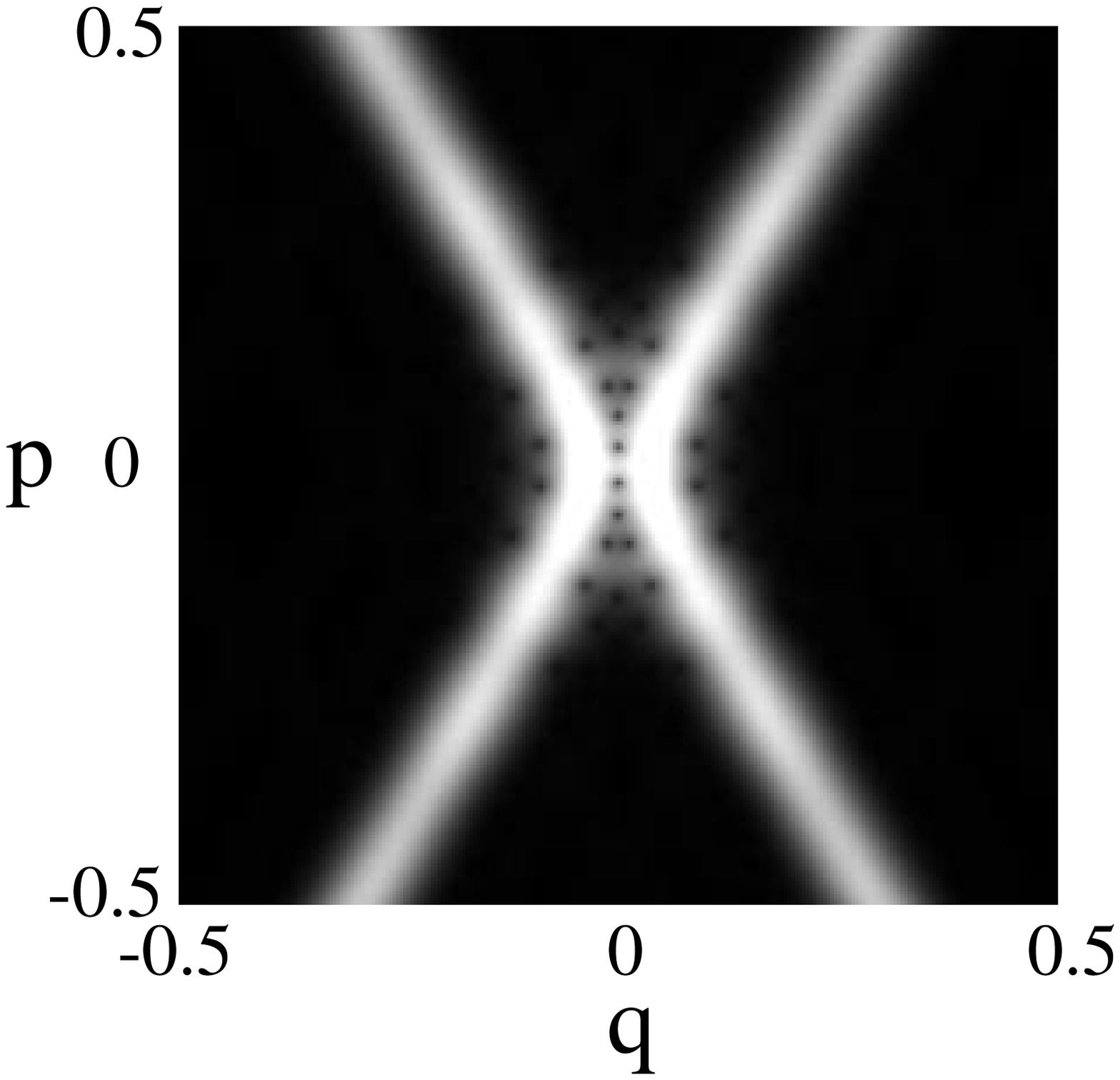}

\vspace*{-3.25in} \hspace*{1.3in}\includegraphics[width=2.5in] {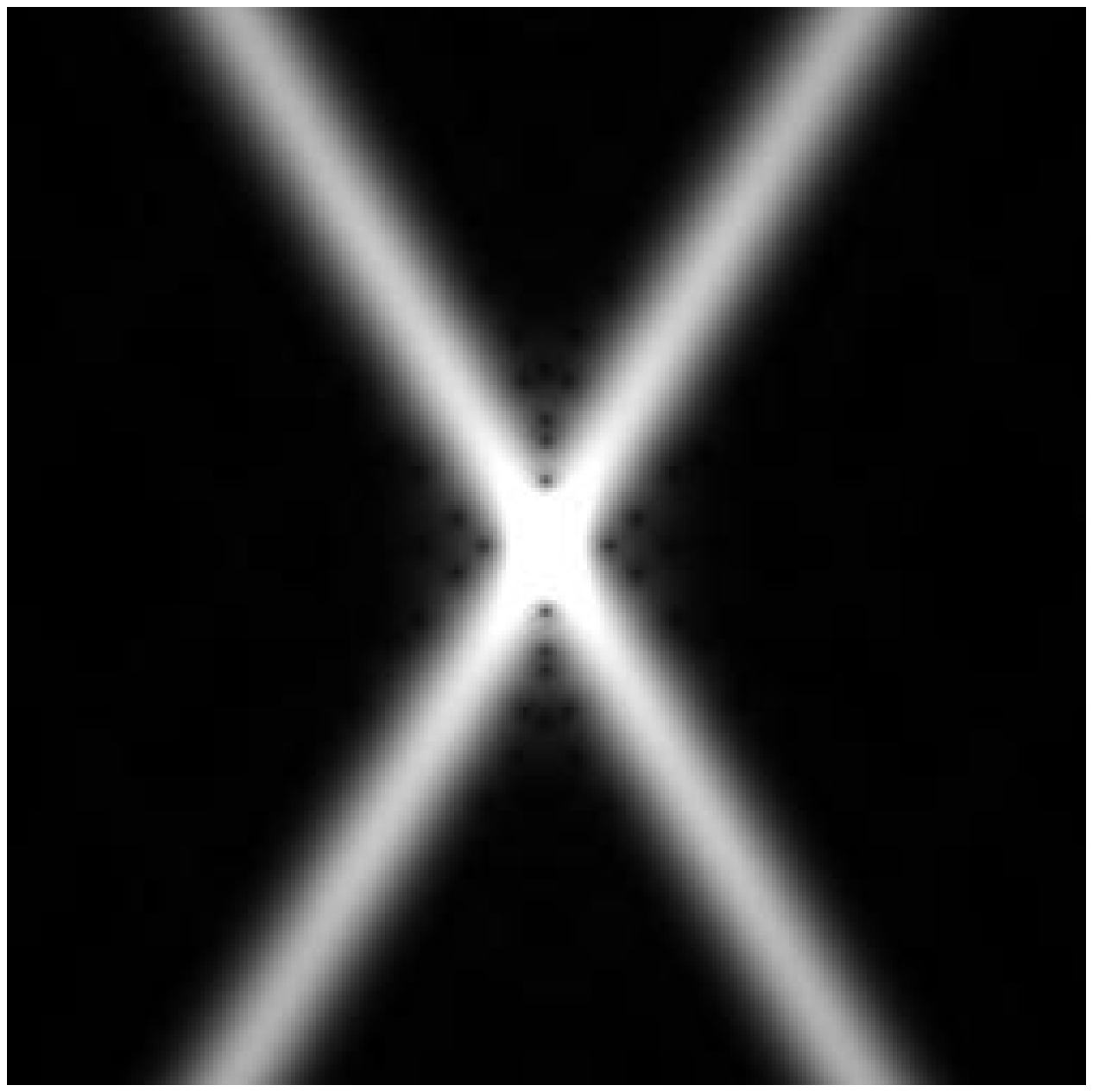}

\vspace*{-3.25in} \hspace*{2.95in}\includegraphics[width=2.5in] {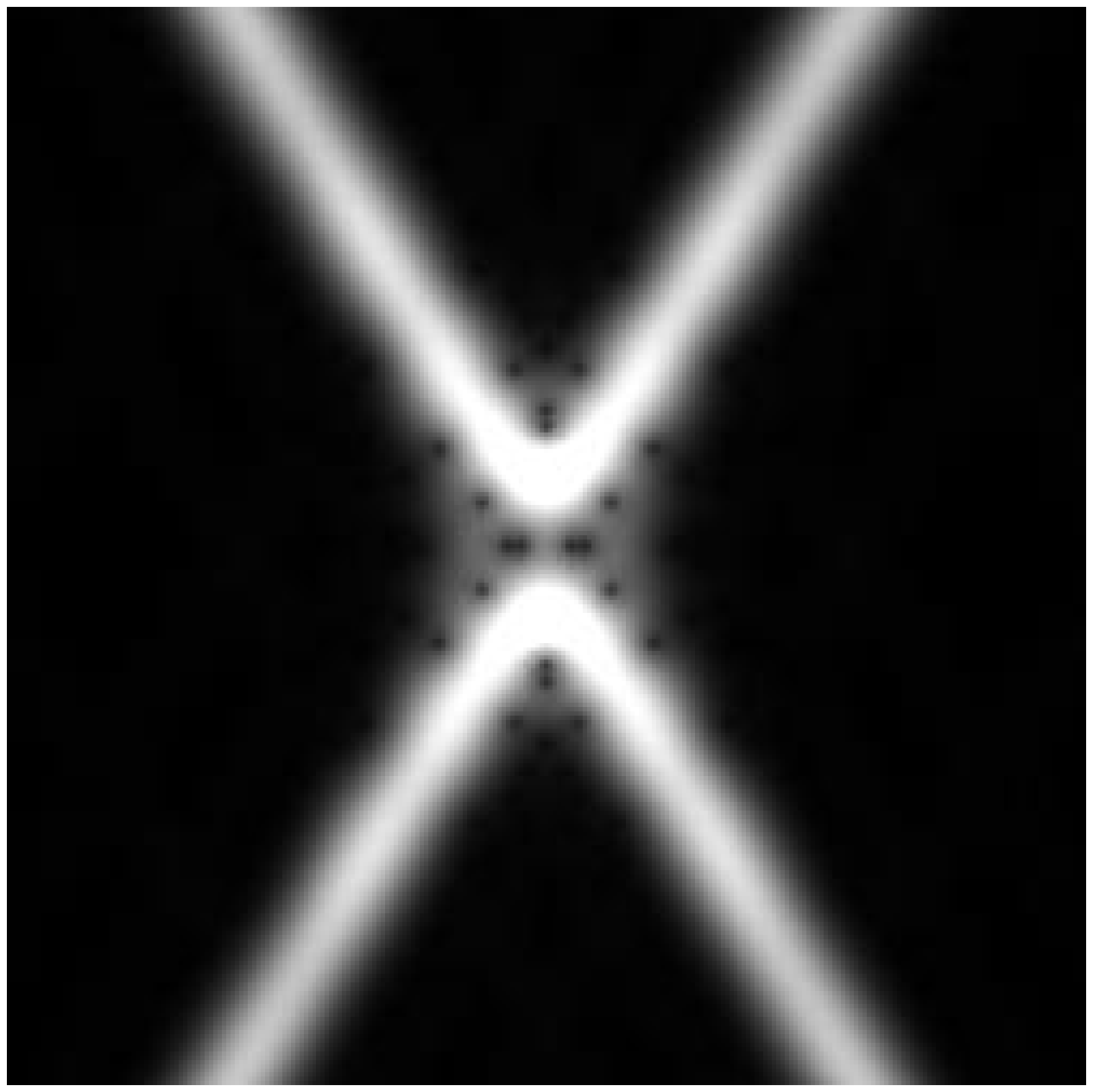}

\vspace*{-3.25in} \hspace*{4.6in}\includegraphics[width=2.5in] {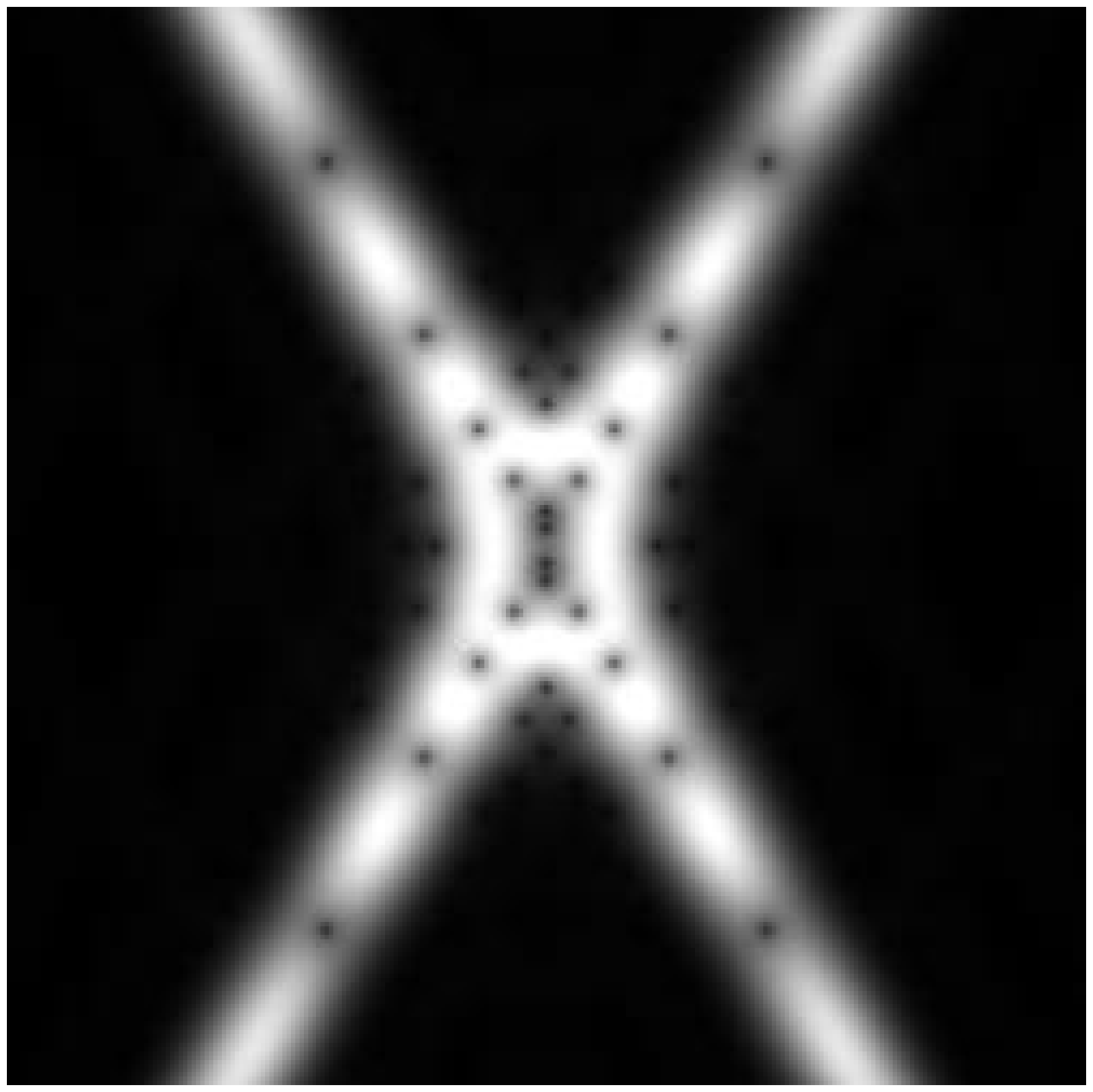}

\vspace*{-1.2cm}

\caption{ The Husimi functions of some scarred eigenstates of the perturbed
cat map. Starting from the left, these have the Hilbert-space dimensions
and eigenangles $(N=404, \theta = -1.690)$, $(N=400, \theta = 0.050)$,
$(N=400, \theta = 1.547)$ and $(N=404, \theta = 2.995)$.
We note that eigenangles are respectively close to $-\pi/2,0,\pi/2$ and
$\pi$ and that the states bear a qualitative resemblance
to the leading scar function at corresponding values of $\theta$.}
\label{individuals}
\end{figure}

We begin by considering some individual wavefunctions.
We show in Figure~\ref{individuals} some examples of eigenfunctions
of the quantum perturbed cat map which reflect the structure of the 
leading scar state in Figure~\ref{Husplot} at the 
corresponding eigenangle. The enhancement of the eigenstates
along the stable and unstable manifolds as seen here is a well-known
characteristic of scarred eigenfunctions. However, the dependence of 
the detailed pattern of this enhancement on eigenangle and the fact 
that it can persist into the antiscarred region, is less obvious.
Although we do not show examples here, we also expect to find 
eigenfunctions with enhanced overlaps with subsequent scar states
with $i=1,2,\cdots$, whose Husimi functions vary across the spectrum
as indicated in Figure~\ref{Husplot}. The 
Schnirelman theorem \cite{Schnir} guarantees that such structures 
cannot dominate individual wavefunctions in the semiclassical limit,
However we note that the scar states shrink in phase space as 
$\hbar\to 0$ and may dominate in a sufficiently small neighbourhood.
We also note that even when scar states do not dominate a given 
wavefunction, locally or otherwise, they should 
collectively form an efficient basis for calculating and characterising
such wavefunctions \cite{VE00}.

Let us now turn to overlap statistics.
We have seen that if we use the scar states $\ket{\chi_i(\theta)}$ 
as probe states, the statistics of the resulting overlaps
\begin{equation}
y_i = \braket{\chi_i (\theta)}{\psi_n},
\end{equation}
should be described by the Gaussian distribution $P(y_i;\theta)$ given in
Eq. (\ref{gaudis}). Furthermore, overlaps for different values of $i$
should be statistically independent. We now test this supposition using 
the perturbed cat-map model.

\begin{figure}[h]
\vspace*{-1.in}\hspace*{1.3in}\includegraphics[width=4.5in] {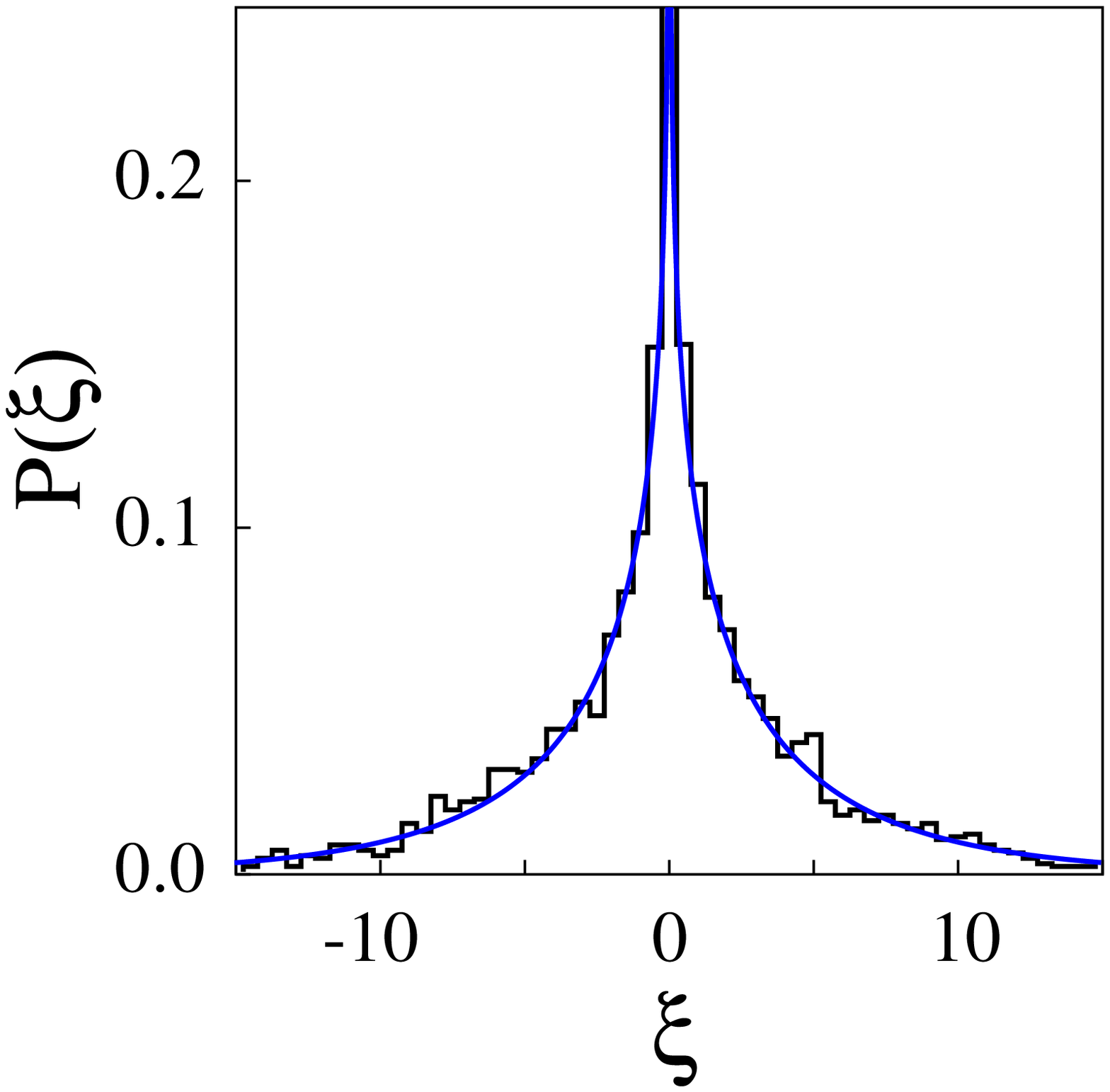}
\vspace*{-1.2in}
\caption{ The distribution of $\xi = y_0 \, y_1$, where
$y_i=\braket{\chi_i(0)}{\psi_n}$. The histogram shows the results of a
numerical calculation for a perturbed cat map in which the 
eigenstates $\ket{\psi_n}$ in the interval $-0.05\pi < \theta < 0.05\pi$
used. The solid line is the theoretical expectation
Eq.(\ref{Pxi}). The symmetry of this result supports the assertion of
statistical independence for $y_0$ and $y_1$.}
\label{indS}
\end{figure}

To verify statistical independence of distinct components $y_i$,
we consider the variable
\[
\xi = y_i \,y_j   \qquad \mbox{with}\quad i \ne j.
\]
The distribution for this variable can be calculated as a special case of
(\ref{pofy}), to give
\begin{equation}
P(\xi;\theta) = \frac{1}{\pi \sqrt{s_i(\theta)\, s_j(\theta)}}
K_0 \left(\frac{|\xi|}{\sqrt{s_i(\theta)\, s_j(\theta)}}\right),
\label{Pxi}
\end{equation}
where $K_0(z)$ is a modified Bessel function.
The independence of $y_i$ and $y_j$ is reflected in the fact that
this is an even function of $\xi$. Figure~\ref{indS} verifies that
distribution describes the joint  statistics of
$y_0$ and $y_1$ in the case of a perturbed cat map and supports the
assertion that these variables are statistically independent. Note 
that numerical
limitations restrict us to relatively modest truncations of the
harmonic-oscillator basis in this case --- the statistics shown have
been computed for scar states with $M=28$ and 
$\beta=\infty$ --- nevertheless the agreement is quite good.

\begin{figure}[h]
\vspace*{0.in} \hspace*{0.4in}\includegraphics[width=3.in] {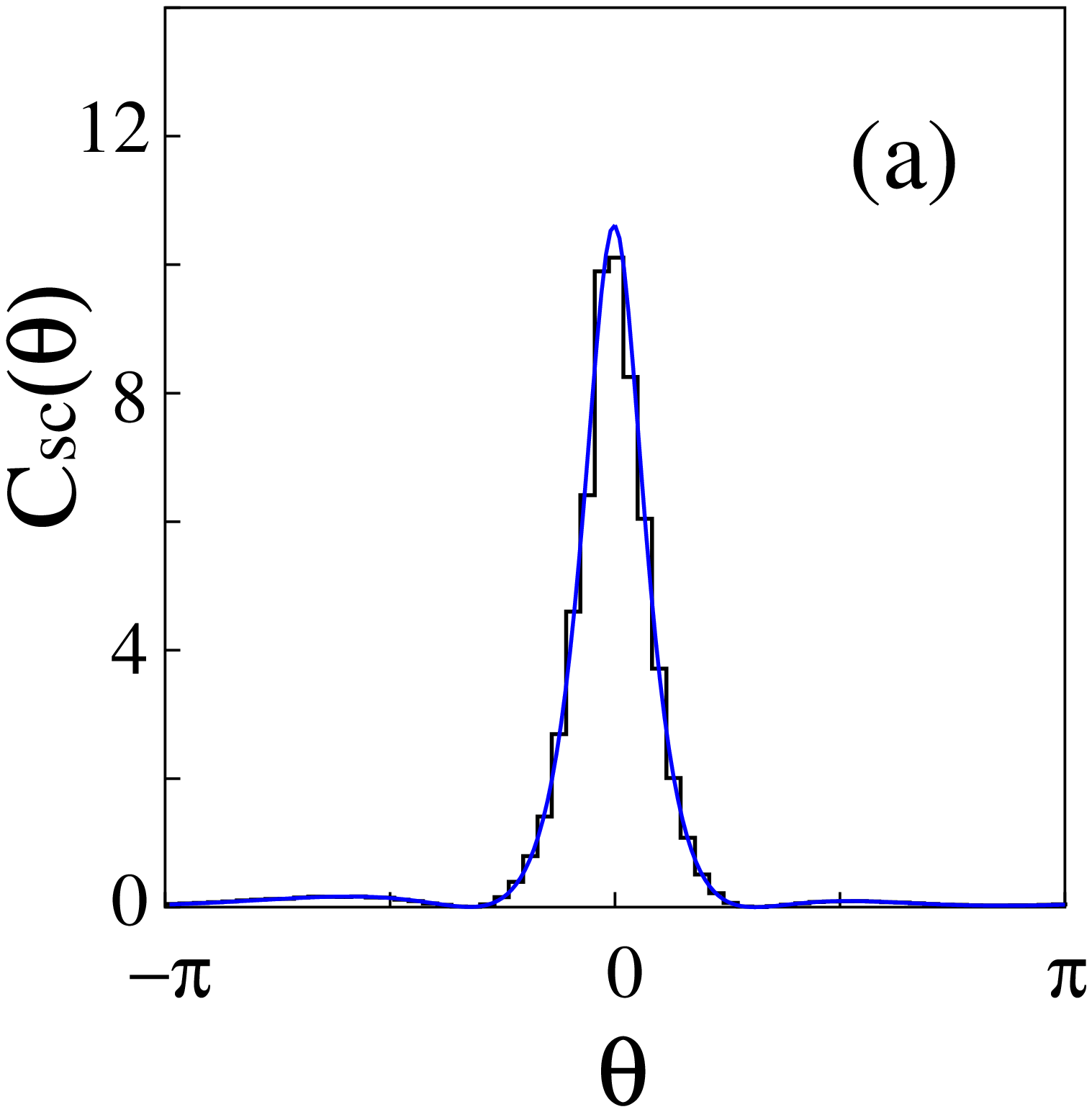}

\vspace*{-3.9in} \hspace*{3.1in}\includegraphics[width=3.in] {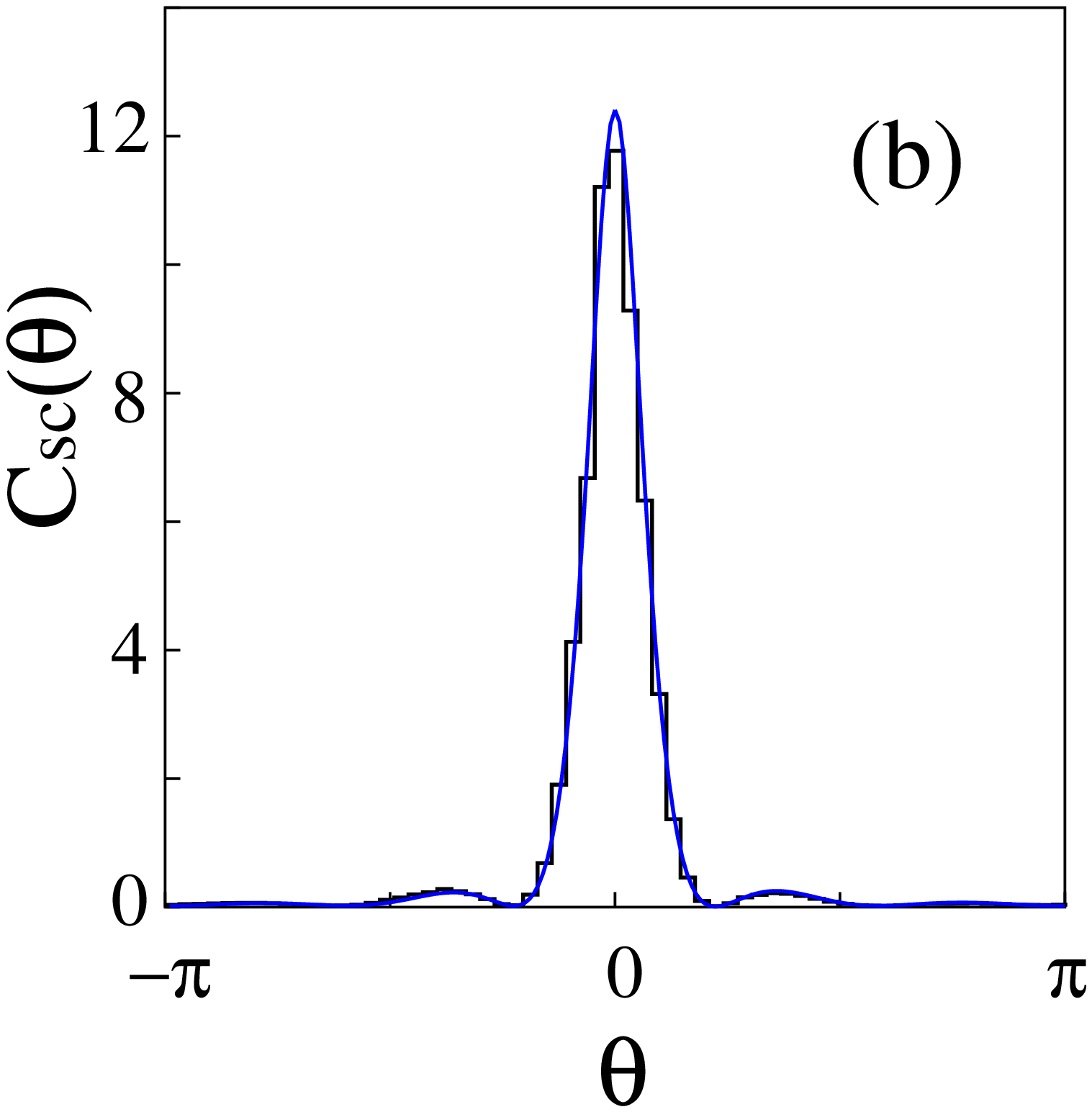}

\vspace*{-1.6in} \hspace*{0.4in}\includegraphics[width=3.in] {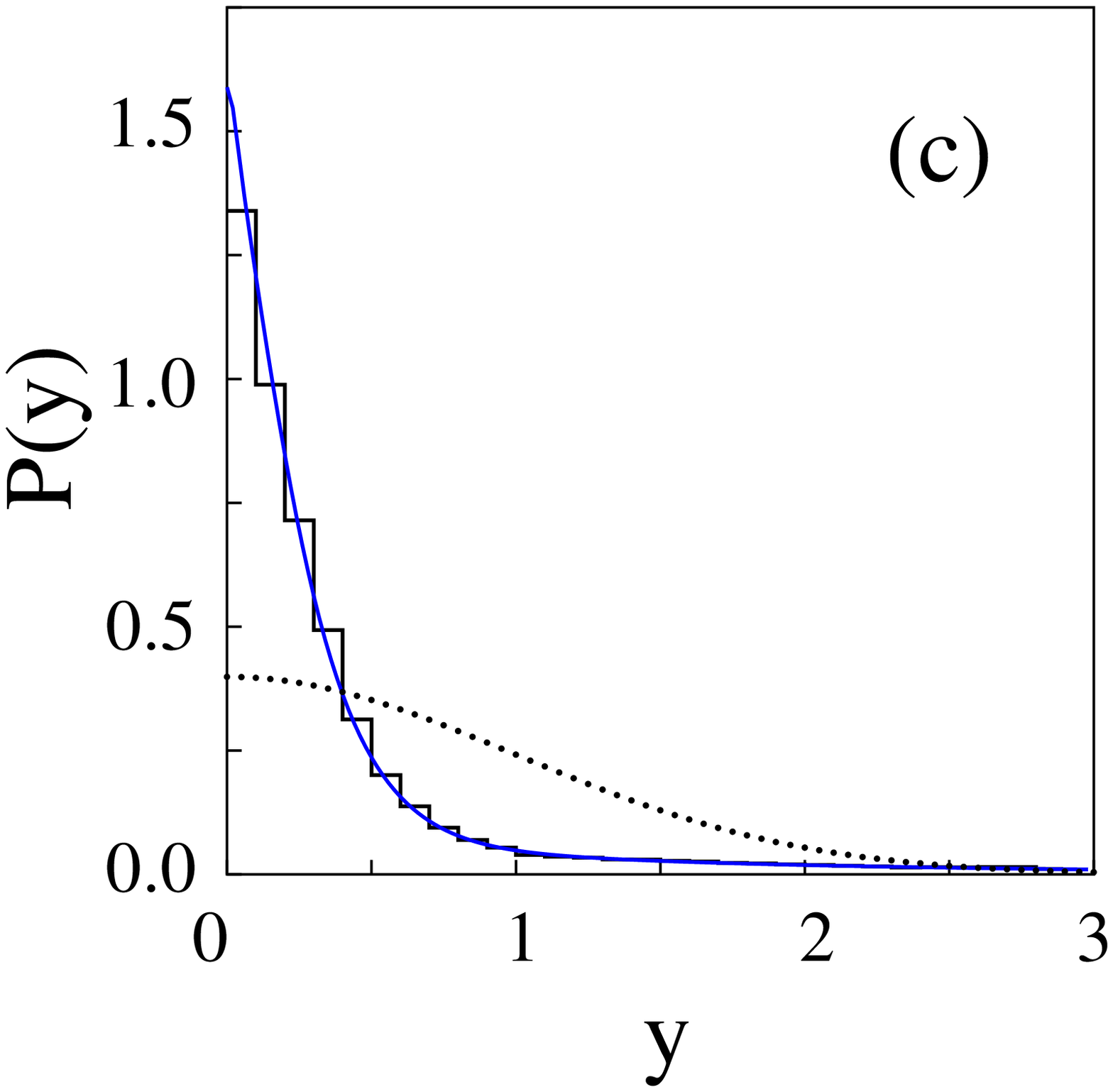}

\vspace*{-3.9in} \hspace*{3.1in}\includegraphics[width=3.in] {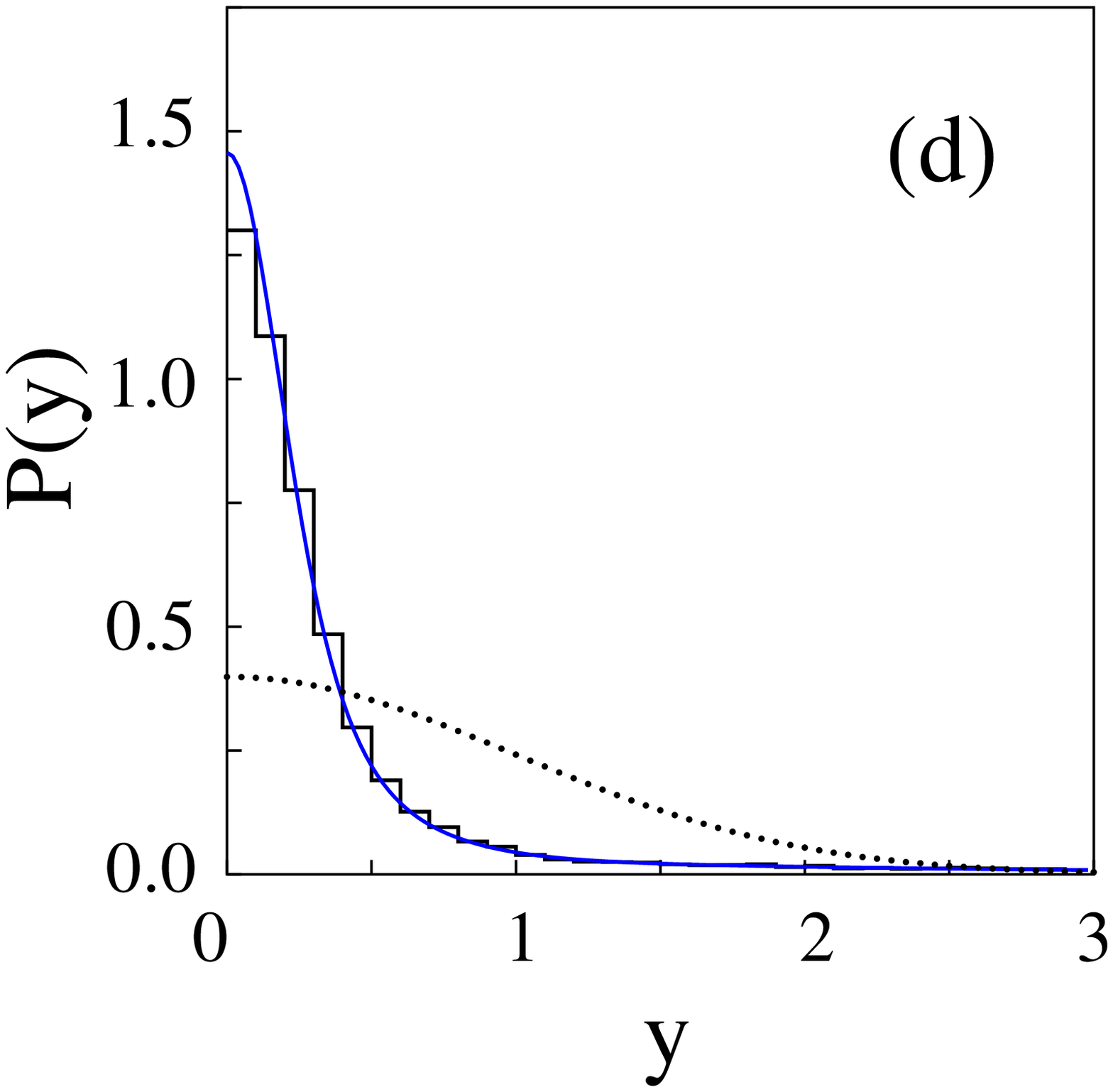}

\vspace*{-0.8in}
\caption{ The variance calculated for the scar state $\ket{\chi_0(0)}$
according to (\ref{Csc0}) is compared with results for the perturbed
cat map as a function of $\theta$. Cases (a) and (b) are respectively
for effective truncation dimensions $M=8$ and $M=28$ with
$\beta=\infty$ in the scarred region.
 The accumulated overlap distributions $P(y)$ are shown in (c) and (d), respectively,
 for the same truncation dimensions.
 The solid line comes from the present theory and
the dotted line is the RMT prediction.
}
\label{scarvariance}
\end{figure}

We also consider the statistics of a single component, concentrating on
the statistics of $y_0$, the overlap with the leading scar state. We
first construct the probe state $\ket{\chi_0(0)}$ for the maximally
scarred part of the spectrum and examine its overlap statistics as a
function of $\theta$. The variance at any other part of the spectrum
is then calculated for this fixed state using
\begin{eqnarray}
C_\sc (\theta) &=& \sum_t  e^{i\theta t}
			\braopket{\chi_0(0)}{ \Uh_\lin^t}{\chi_0(0)}.
\label{Csc0}
\end{eqnarray}
This is analogous to the envelope for the universal test state considered
in \cite{KAHE}.
A comparison with numerical results for the perturbed cat map is shown
in Figure~\ref{scarvariance} for two truncation dimensions $M$ of the
matrix $\tilde{\C}(\theta)$. We see that the overlaps are concentrated more in
the scarred region and that the height increases as $M$ is increased,
consistent
with the observation that $s_0(0)$ diverges logarithmically with $M$.
Note that, as for any fixed state, a spectral average of the variance
must average to unity since $\expect{C_\sc} =
\braopket{\chi_0(0)}{ \Uh_\lin^{t=0}}{\chi_0(0)}=1$. Any growth in the
scarred region must therefore be accompanied by a decay elsewhere in
the spectrum and this is also visible in  Figure~\ref{scarvariance}.
This does not contradict the fact that $s_0(\theta)$ remains large
over the whole spectrum because the leading scar state varies with
$\theta$ whereas in this calculation we have used a fixed state
$\ket{\chi_0(0)}$

We also show in Figure~\ref{scarvariance}(c) and (d) the distribution of
$y=\braket{\chi_0(0)}{\psi_n}$ about this mean behaviour,
averaged over eigenangle $\theta$. The distribution
\begin{equation}
p(y)=\frac{1}{2 \pi} \int \, d\theta \,\,p(y;\theta)=
       \frac{1}{2 \pi} \int \, d\theta 
\frac{1}{\sqrt{2 \pi C_\sc(\theta)}} \,\,e^{y^2/2C_\sc(\theta)},
\label{accPy}
\end{equation}
shown as a solid line describes quite accurately the wavefunction 
statistics, shown as a histogram. Note the marked difference
between these results and the RMT prediction indicated by 
a dotted line. Similar reults are found for the statistics of
$\braket{\chi_0(0)}{\psi_n}$ with $i>0$, although in the 
deviation from RMT is then less stark.

\section{Wavefunction statistics in billiard systems}

The discussion so far has been in terms of quantum maps, but we now 
address the physically more interesting question of time-independent 
systems, and in particular billiard systems.
In this section we describe how the the joint-probability 
distribution may be applied to such systems, for which an analogy 
with maps can be made through the boundary-integral formalism. In
more generic problems such a connection is made using the
transfer-operator formalism of Bogomolny \cite{BO92}.

\subsection{Boundary eigenfunctions}

We consider eigenfunctions of the Helmholtz equation
\begin{equation}
(\nabla^2 + k_n^2) \Psi_n({\bf x}) =0,
\end{equation}
satisfying the Dirichlet condition $\Psi_n({\bf x})=0$
on the boundary and we assume that the domain is such that the 
corresponding billiard dynamics is chaotic. In order to apply the 
results of the preceding sections it is convenient to reformulate 
this problem in terms of maps, and a natural means of doing this
is to use the boundary-integral method, which provides a quantum analog
of the classical Poincar\'e-Birkhoff mapping.

The boundary-integral method has been extensively described elsewhere
so here we just describe the key features, primarily to establish 
notation. For a more detailed description, and in particular for 
other investigations of scarring in the boundary representations,
we refer to \cite{BOA92,Prosen,TUVO,KS,SVS,BFSS}.
The essential point is that we replace the Helmholtz equation
in the interior of the billiard by the integral equation
\begin{equation}
\psi_n (s) =  \int \, K(s,s';k_n) \psi_n (s') \d s'
\label{mapB}
\end{equation}
over a coordinate $s$ on the boundary, where
\begin{equation}
\psi_n (s) = 
\left.\frac{\partial \Psi_n({\bf x})}{\partial n} 
				\right| _{{\bf x}={\bf x}(s)}
\end{equation}
serves as a boundary eigenfunction. The kernel is
\begin{equation}
K(s,s';k_n)= \left.\frac{\partial G({\bf x},{\bf x}(s');k_n)}{\partial n}  
\right|_{{\bf x}={\bf x}(s)}
\end{equation}
where 
\begin{equation}
G({\bf x},{\bf x'};k_n)= \frac{i}{2} H_0^{(1)}(k_n |{\bf x}-{\bf x'}|).
\label{green}
\end{equation}
is the Green function for the Helmholtz equation.

The integral equation (\ref{mapB}) will play the role of a quantum 
map for us. Even though the map is not unitary \cite{BOA92}, we will 
assert that a straightforward generalisation of the envelope matrix 
derived in Appendix~A constrains the variances of overlaps between
harmonic probe states and eigenfunctions. The important point is 
that iterates of the map concatenate semiclassically in the same
way that iterates of a unitary map do and while there are some
local modifications of amplitude on classical length scales due to
breaking of unitarity, these do not appreciably affect the 
statistics when phase-space localised probe states are used. In 
particular, we assert that the statistics of $\psi_n (s)$ near a 
periodic orbit are described by the joint-probability distribution 
with the structure described in previous sections, at least in the 
semiclassical limit. In order to write this distribution down we 
need to define analogues for some of the ingredients we used in the 
previous sections.

In describing the Poincar\'e-Birkhoff mapping we use canonical 
coordinates $(s,v)$ where $s$ is an arc-length coordinate
at which a trajectory collides with the boundary and
$v=\sin\alpha$, where $\alpha$ is the angle of incidence.
We consider wavefunction statistics in the neighbourhood of a point
with coordinates $(s_0,v_0)$ where a periodic orbit of length 
$\ell$ intersects the boundary. As a spectral parameter we
use $\theta=k\ell-\mu\pi/2$, where  $\mu$ is the Maslov index 
of the periodic orbit --- with this definition we will find that
that full scarring occurs when $\theta=0\;\mod\; 2\pi$.
As probe states we use eigenstates of a harmonic oscillator centred 
at $(s_0,v_0)$ on the boundary section. These are written as boundary 
wavefunctions of the form
\begin{equation}\label{defphil}
\phi_l(s)= \frac{1}{\sqrt{ \sigma 2^l l! \sqrt{\pi}}}
\exp\left[-\frac{(s-s_0)^2}{2 \sigma^2} +ikv_0(s-s_0)\right] 
H_l \left(\frac{s-s_0}{\sigma}\right),
\end{equation}
where $H_l (z)$ is a Hermite polynomial. As in \cite{BI01} we let
$\sigma$, which provides us with an aspect ratio in phase space of
the ellipse defined by the harmonic oscillator, to scale with $k$ as
\[
\sigma = \sqrt{\frac{\ell_0}{k}}
\]
where $\ell_0$ is a characteristic length of the billiard.
In practice we choose $\ell_0$ so that $Q=0$ (that is, so that the invariant
manifolds are orthogonal in the corresponding metric).
We scale the overlap variables $\x=(x_0,x_1,\cdots)$ as
\begin{equation}
  x_l \simeq  \sqrt{k} \braket{\phi_l }{ \psi_n}
=\sqrt{k} \int_{0}^{L} ds\,\, \phi_l^*(s) \psi_n(s),
\end{equation}
where $L$ is the boundary arc length. These scalings are such that
the statistics of the components $x_l$ do not vary over the spectrum.
We also note that $kL$ plays the same role
as the dimension $N$ in the case of quantum map.

\begin{figure}[h]
\vspace*{-2.5in} \hspace*{1.in}\includegraphics[width=4.5in] {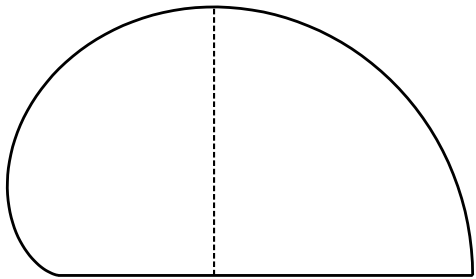}

\vspace*{-1.5in}
\caption{The cardioid billiard and the periodic orbit
used in numerical calculation.}
\label{shopo}
\end{figure}

For numerical purposes we consider the cardioid billiard, which
is defined in polar coordinates by
\begin{equation}
r=1+\cos \theta .
\end{equation}
We consider just the odd-parity states and therefore restrict ourselves
to the half-cardioid illustrated in Figure~\ref{shopo}.
For simplicity we treat the shortest periodic orbit, also shown in
Figure~\ref{shopo}, which has  stability exponent $\rho \simeq 0.693$
and length $\ell=3 \sqrt{3}/2$. We examine wavefunction
statistics where this intersects the boundary section at $s_0 \simeq 2.00$
and $v_0=0$.

\begin{figure}[h]
\vspace*{-1.6in} \hspace*{0.2in}\includegraphics[width=5.5in] {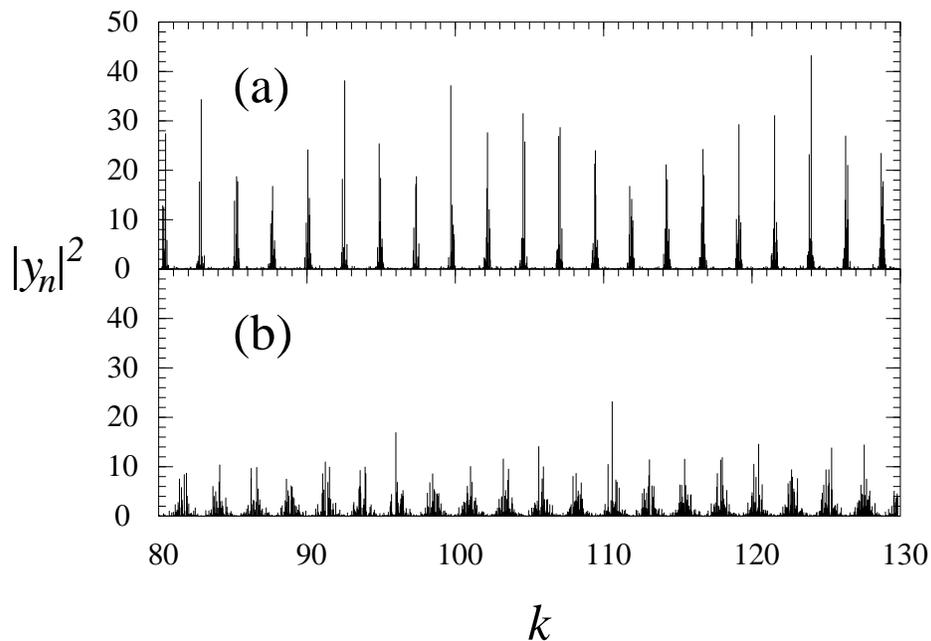}

\vspace*{-1.7in}

\caption{ The overlap probabilities $y_n^2 = k|\braket{\chi_0}{ \psi_n}|^2$
for the leading scar state are
shown as a function of $k$. In (a) the state $\chi_0$ is constructed
for the scarred part of the spectrum with $\theta=0$ and in (b) it
is constructed for the antiscarred part, with $\theta=\pi$. In each case the
scar state has been calculated using a truncation dimension of
$M=12$. }
\label{billiardfig}
\end{figure}

We calculate scar states as described in the previous sections,
with the obvious modifications appropriate to billiards, and denote
the corresponding boundary functions by $\chi_i(s;\theta)$.
Let us begin by considering the overlap probability
\begin{equation}
y_n^2= k|\braket{\chi_0(0)}{ \psi_n}|^2,
\label{yn2}
\end{equation}
for the leading scar state calculated for the scarred part of the 
spectrum. The scaling with $k$ here is designed so that the average
value does not change across the spectrum.
This overlap probability is calculated for individual eigenfunctions and the 
results shown in Figure~\ref{billiardfig}(a) as a function of $k$.
As expected we find strong peaks in typical values localised around
values of $k$ for which $\theta=k\ell-\mu\pi/2=0\;\mod\; 2\pi$, with 
a compensating suppression between peaks. The sharp contrast between the
peaks and valleys in this figure is a reflection of the  extent to which 
$\chi_0(s;0)$ maximises the deviation of statistics from RMT. We also show,
in Figure~\ref{billiardfig}(b) the overlap probabilities for the
leading scar state $\chi_0(s;\pi)$ constructed for the {\it antiscarred}
part of the spectrum where $\theta=\pi\;\mod\ 2\pi$. Once again we find
periodic peaks in the average value of the overlap probability, albeit
with a somewhat reduced contrast. Note also, however, that the peaks
in Figure~\ref{billiardfig}(b) occur when there are troughs in
 Figure~\ref{billiardfig}(a) and vice versa. In other words, we have 
defined a probe state for which significantly larger than average overlaps
occur in parts of the spectrum which are normally associated with 
antiscarring. Although we do not show them explicitly, each of these 
sequences can be defined also for subsequent scar states 
$\chi_i(s;\theta)$ with $i>0$; these subsequent sequences peak in the 
same parts of the spectrum but are statistically independent.

\begin{figure}[h]
\vspace*{-0.7in} \hspace*{0.4in}\includegraphics[width=3.in] {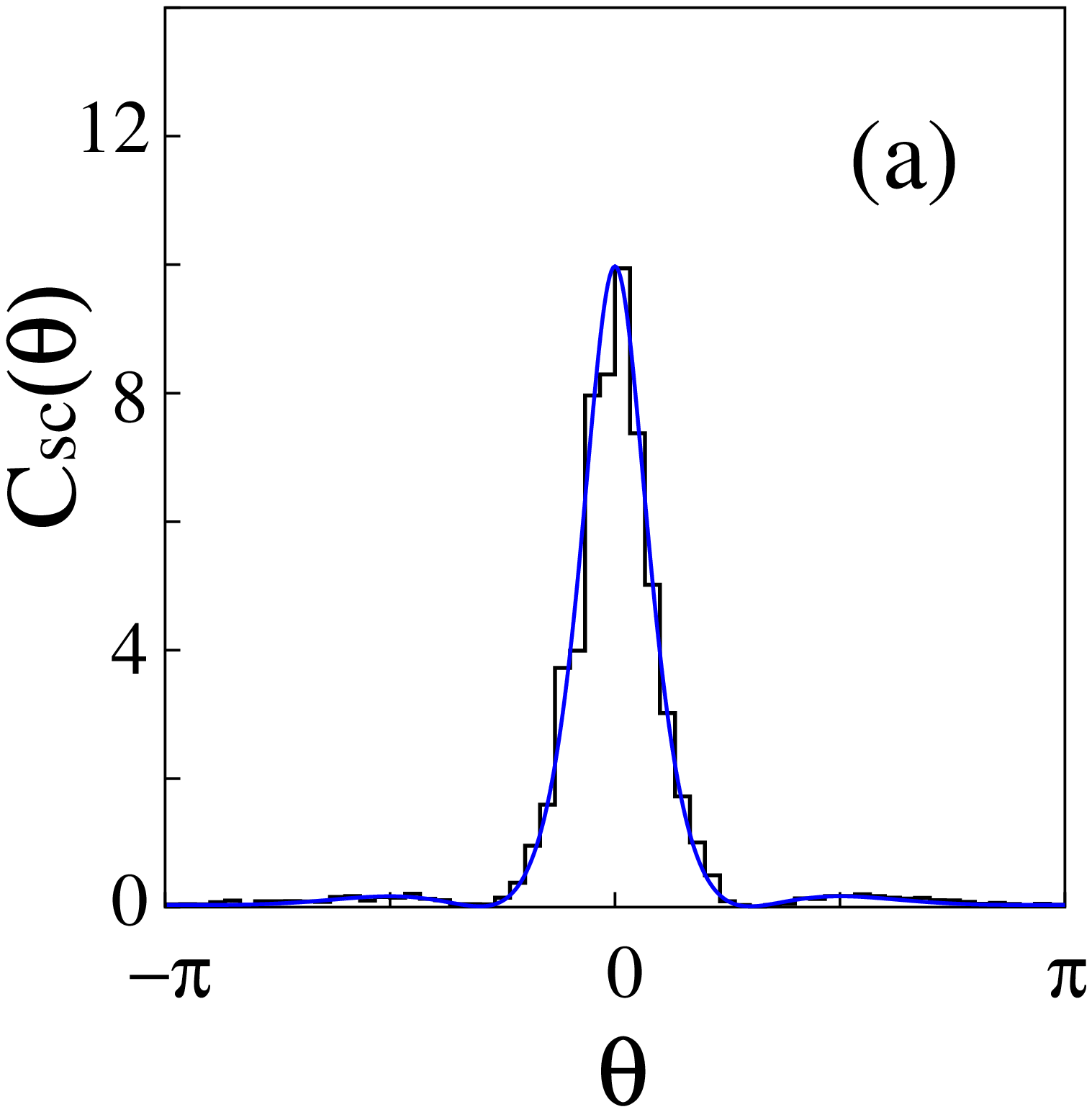}

\vspace*{-3.9in} \hspace*{3.1in}\includegraphics[width=3.in] {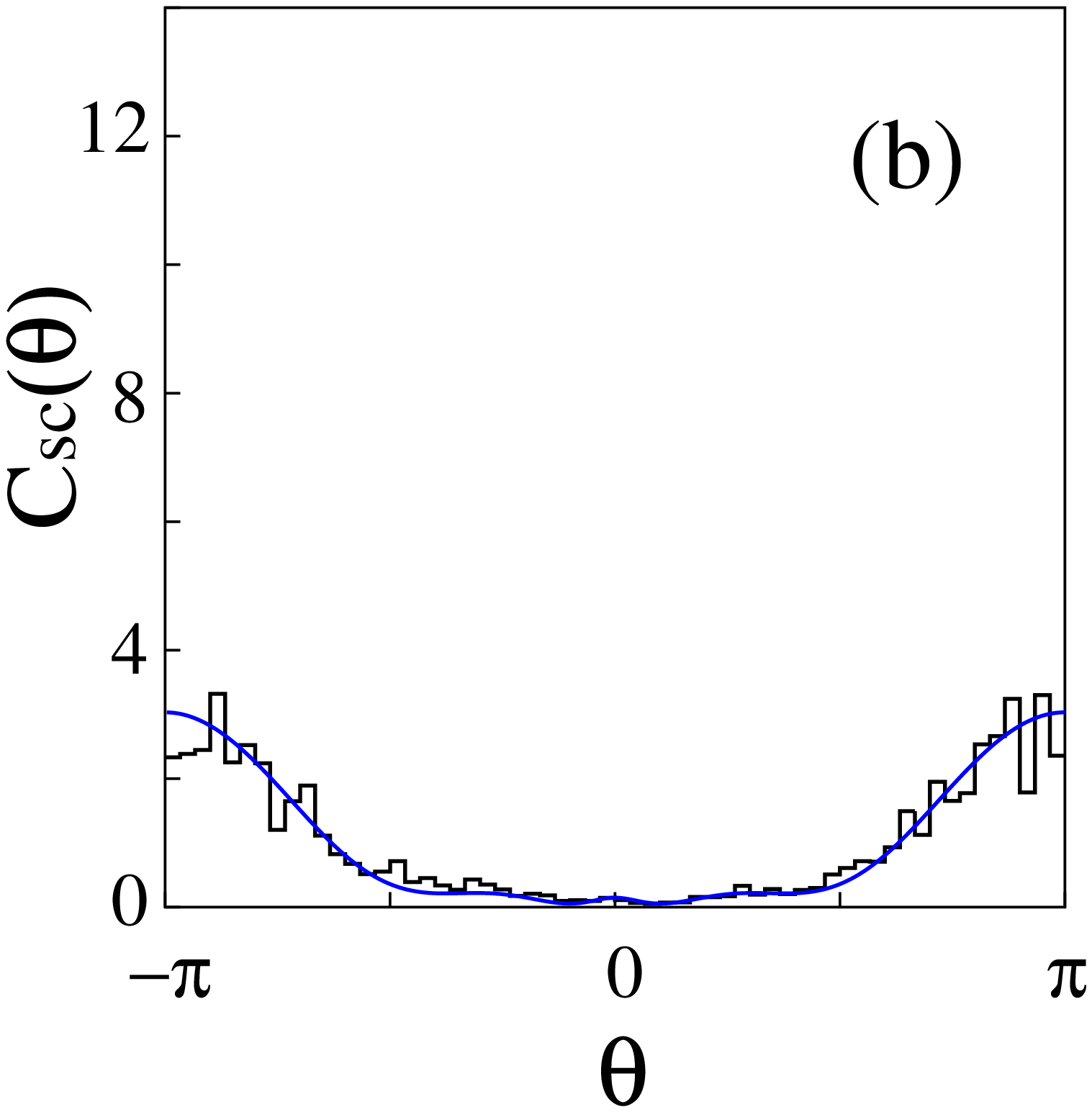}

\vspace*{-1.6in} \hspace*{0.4in}\includegraphics[width=3.in] {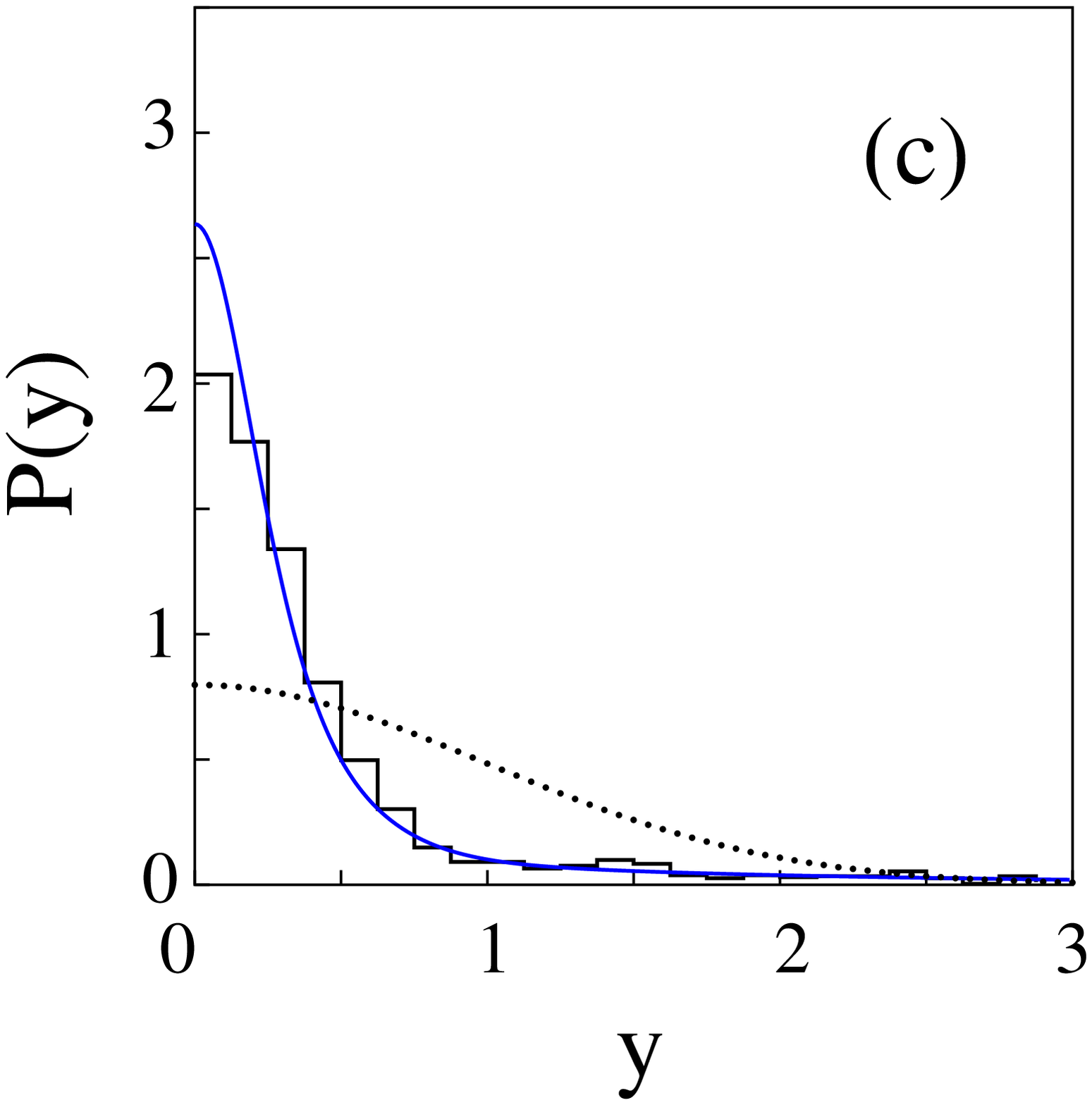}

\vspace*{-3.9in} \hspace*{3.1in}\includegraphics[width=3.in] {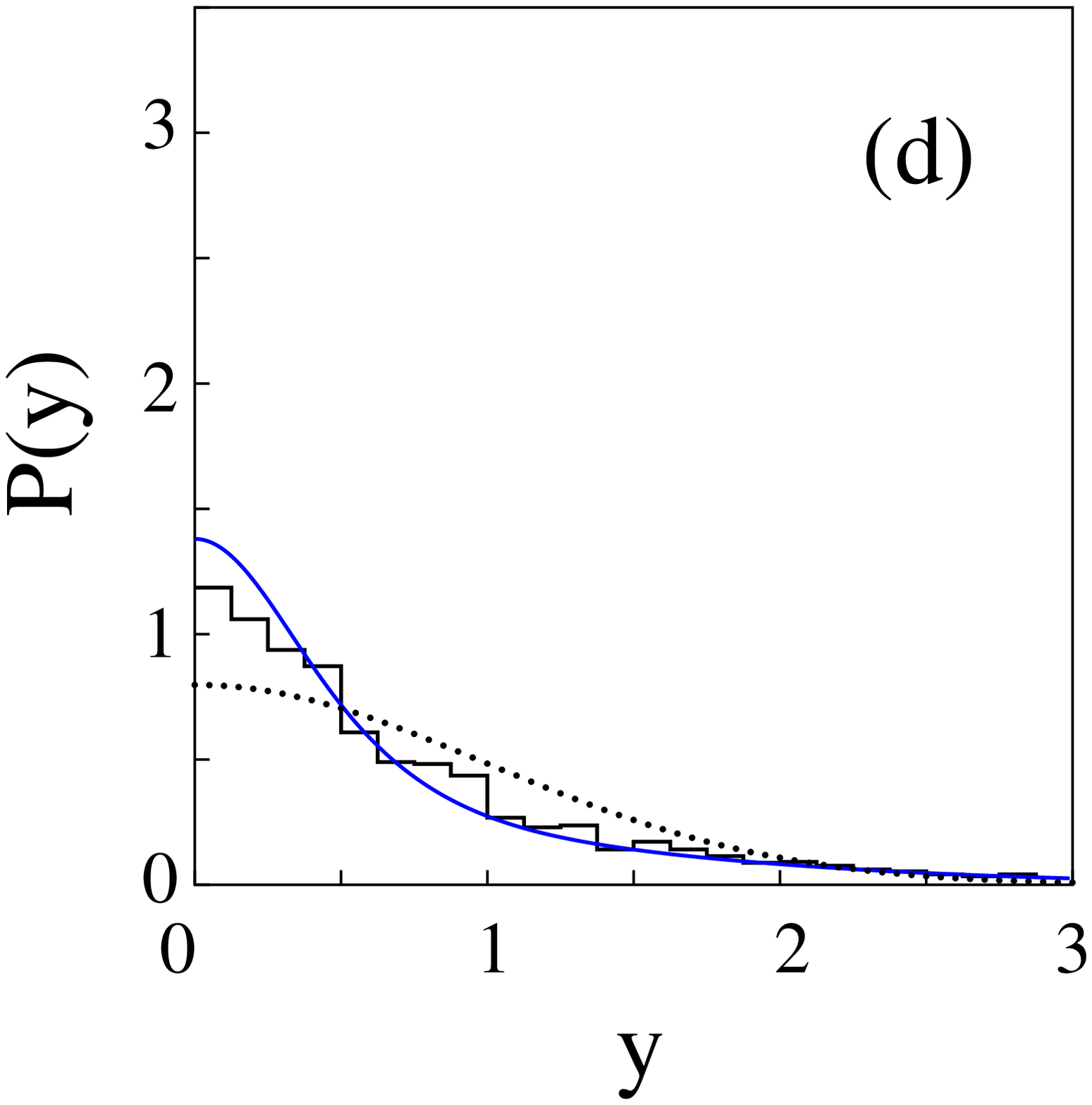}

\vspace*{-0.8in}
\caption{ The partial variance $\expect{y^2}$ is shown as a 
function of $\theta$. In (a) and (b) we respectively
show results for the scar state of the scarred region ($y^2 = 
k|\braket{\chi_0(0)}{\psi_n}|^2$) and of the antiscarred region 
($y^2 = k|\braket{\chi_0(\pi)}{\psi_n}|^2$). The solid lines show
the envelope functions $C_{\sc,0}(\theta )$ and $C_{\sc,\pi}(\theta )$
calculated in analogy with (\ref{Csc0}).
In (c) and (d) we show the corresponding accumulated overlap
distributions $P(y)$ describing the distribution of individual
overlaps about the averages depicted in (a) and (b).
The solid line comes from the present theory and
the dotted line is the RMT prediction. That is, in (c) we compare
a histogram of $y = k|\braket{\chi_0(0)}{\psi_n}|$ with theory
and in (d) we do the same for $y = k|\braket{\chi_0(\pi)}{\psi_n}|$.
}
\label{billenv}
\end{figure}

To make a more quantitative comparison with the theory of the
scar state we collate the overlap probabilities as a function
of $\theta$ and compare with the prediction of (\ref{Csc0}) (and its generalisation to the antiscarred region) in
Figures~\ref{billenv}(a) and (b). These describe envelopes for
Figure~\ref{billiardfig} and confirm that the detailed predictions
for the variances using the envelop matrix $\C(\theta)$ work
quantitatively in this billiard system. A comparison of the statistics
of the distribution of individual states about this average in (c)
and (d) also supports the hypothesis of Gaussian
statistics in this context. Calculations of correlation and the statistics
of higher scar states work similarly to the case of quantum maps, as
investigated in the previous section, and we do not give further numerical
details in the boundary representation here. Instead we turn to
a discussion of how we may formulate these statistical properties
in terms of full wavefunctions in the interior of the billiard.

\subsection{Full eigenfunctions}

\begin{figure}[h]

\vspace*{-1.9in}\hspace*{-1.in}\includegraphics[width=5.in]{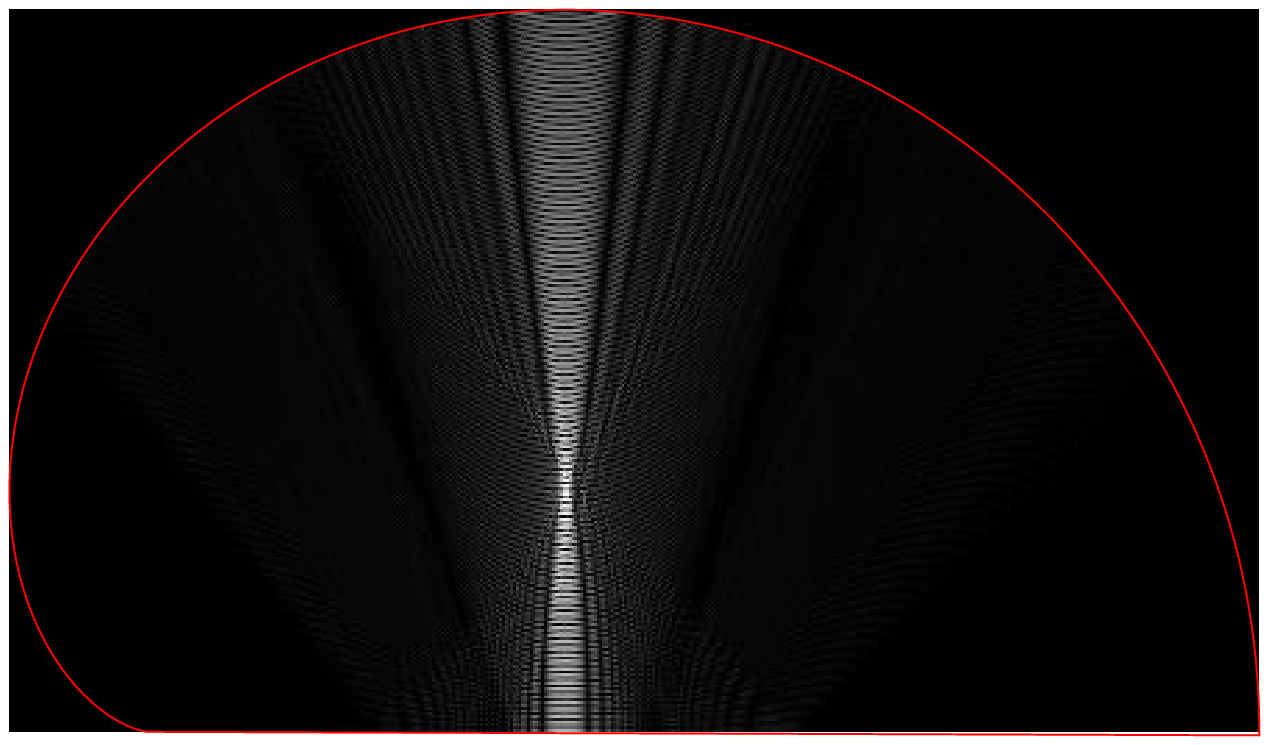}

\vspace*{-6.5in} \hspace*{2.6in}\includegraphics[width=5.in] {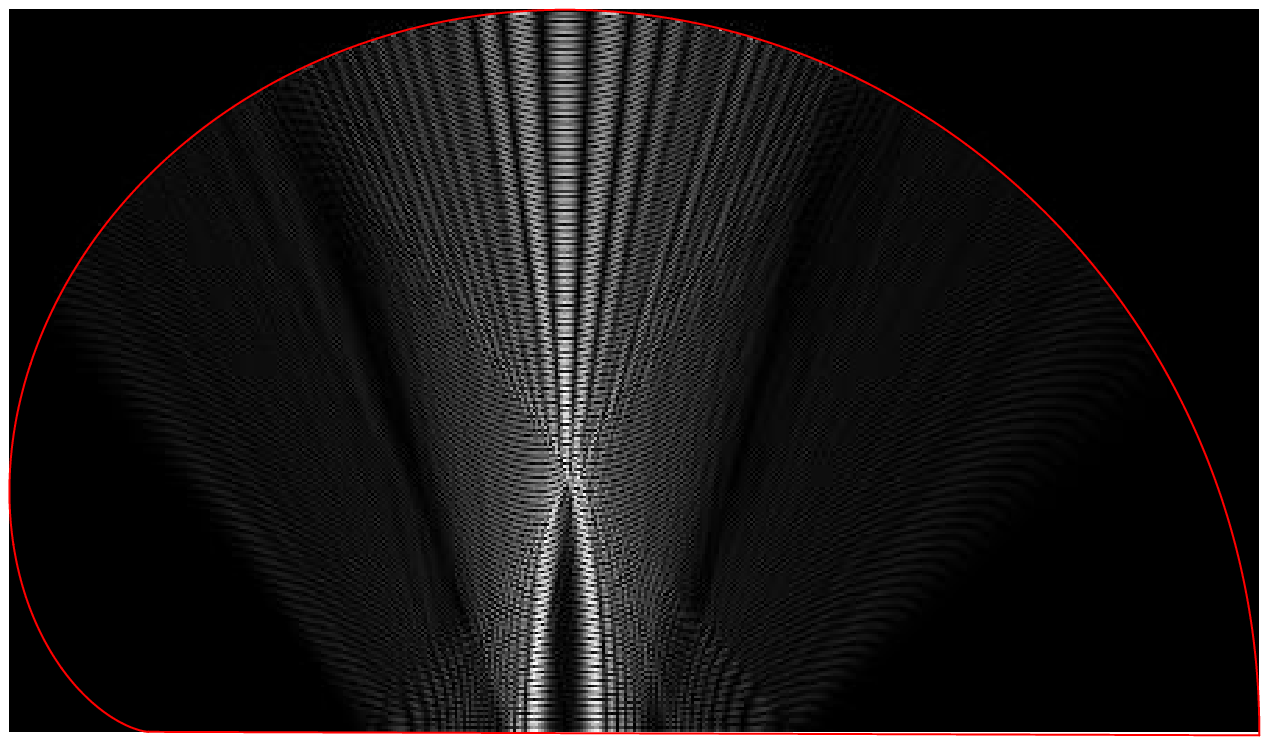}

\vspace*{-4.5in} \hspace*{-1.in}\includegraphics[width=5.in] {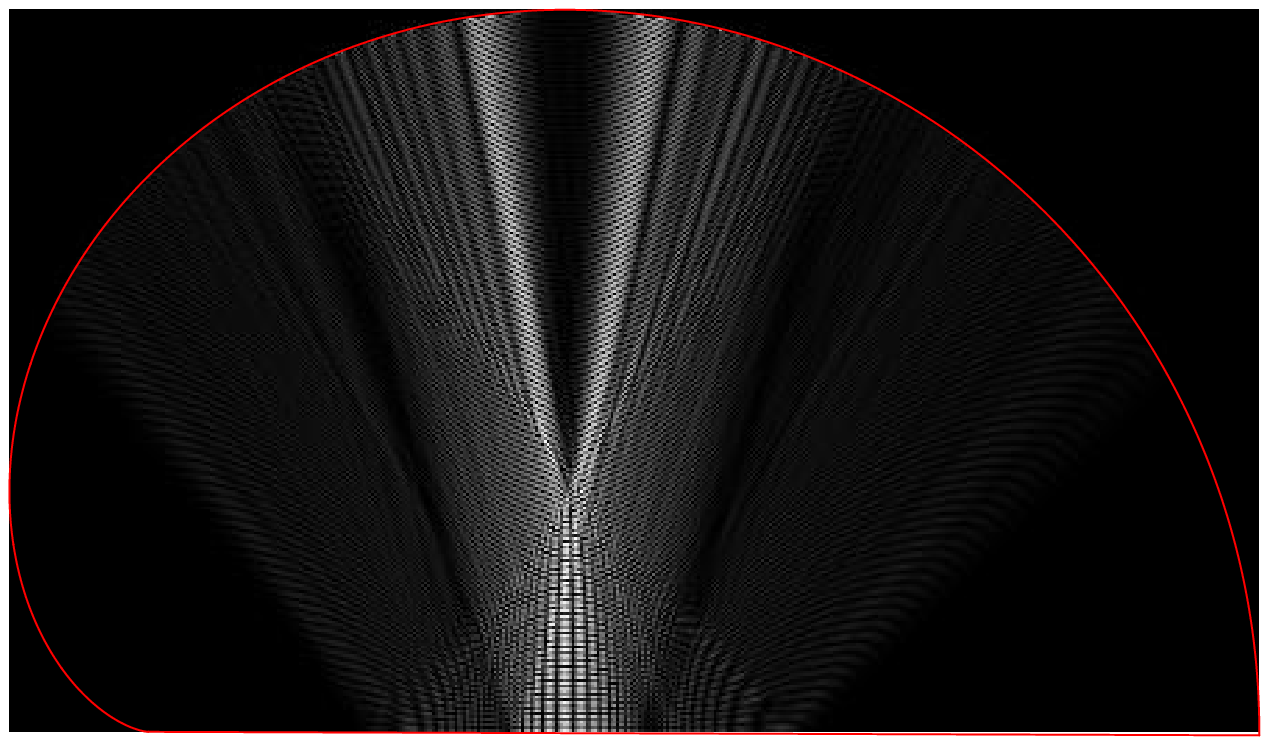}

\vspace*{-6.5in} \hspace*{2.6in}\includegraphics[width=5.in] {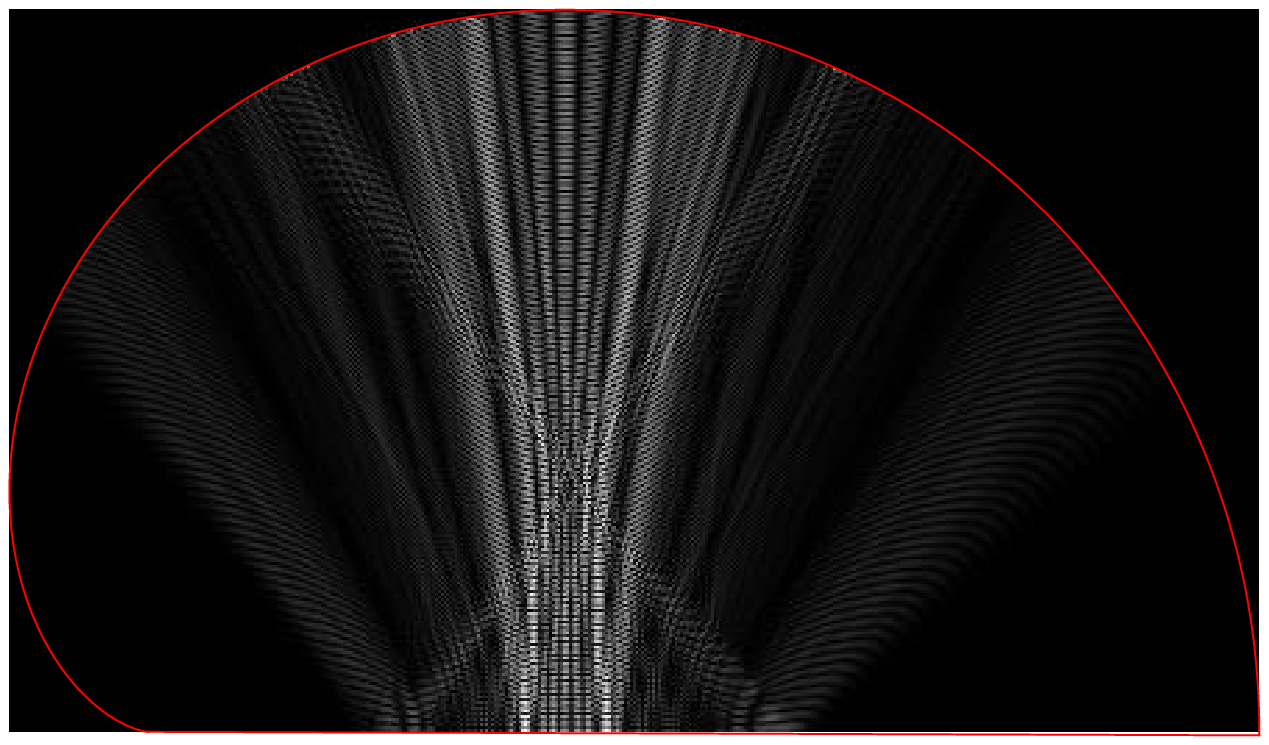}

\vspace*{-2.5in}

\caption{Examples of leading scarfunctions $\Phi_0^{scar} (\x;k)$ calculated
for (clockwise from top left) $k=999.40$, $998.80$, $1000.01$ and $1000.61$.
These correspond respectively to $\theta=0$, $\pi/2$, $\pi$ and $3\pi/2$.}
\label{scarfunctions}
\end{figure}

The scar states for billiards have so far been given in the
boundary-integral formalism. We now propose corresponding probe
states which are appropriate to full wavefunctions in the interior
of the billiard. Given an eigenfunction $\psi_n(s)$ of the
boundary-integral equation, the full eigenfunction $\Psi_n(x)$ is
obtained from it using Green's identity, giving
\begin{equation}
\Psi_n({\bf x}) = \frac{1}{2} \int
			G({\bf x},{\bf x}(s');k_n) \,\psi_n (s')\, \d s'.
\label{ptP}
\end{equation}
We define probe states in the interior of the billiard in analogy with
this equation. That is, given a probe state $\phi(s)$ in the boundary
representation, whether it be a harmonic-oscillator eigenstate,
a scar state or otherwise, we define a corresponding probe state in the
interior using
\begin{equation}
\Phi({\bf x}) = \frac{1}{2} \int
			G({\bf x},{\bf x}(s');k_n) \,\phi (s')\, \d s'.
\label{ptPhi}
\end{equation}
This function automatically satisfies the Helmholtz equation in the
interior but, because $\phi (s)$ does not satisfy the boundary integral
equation, does not satisfy the Dirichlet boundary conditions placed on
$\Psi_n(\x)$. Nevertheless it provides a perfectly well defined
function in the interior of the billiard and we may use it
to characterise the wavefunction near the periodic orbit.

We provide an argument in Appendix~\ref{billiardapp} which indicates
that, given a probe state which is localised in phase space around
a point $(s_0,v_0)$, then the boundary overlap $\braket{\phi}{\psi_n}$
and the full overlap  $\braket{\Phi}{\Psi_n}$ approximately coincide,
\begin{equation}
\braket{\Phi}{\Psi_n} \simeq g(s_0,p_0) \braket{\phi}{\psi_n},
\label{appen}
\end{equation}
up to a geometrical factor $g(s_0,p_0)$ which depends on the location
the probe state in phase space but is independent of $\psi_n$.
Following appropriate normalisation, the predictions for the statistics
of the boundary overlaps may therefore be carried over to the overlaps
defined in the interior.

\begin{figure}[h]

\vspace*{-1.8in}\hspace*{-1.in}\includegraphics[width=5.in]{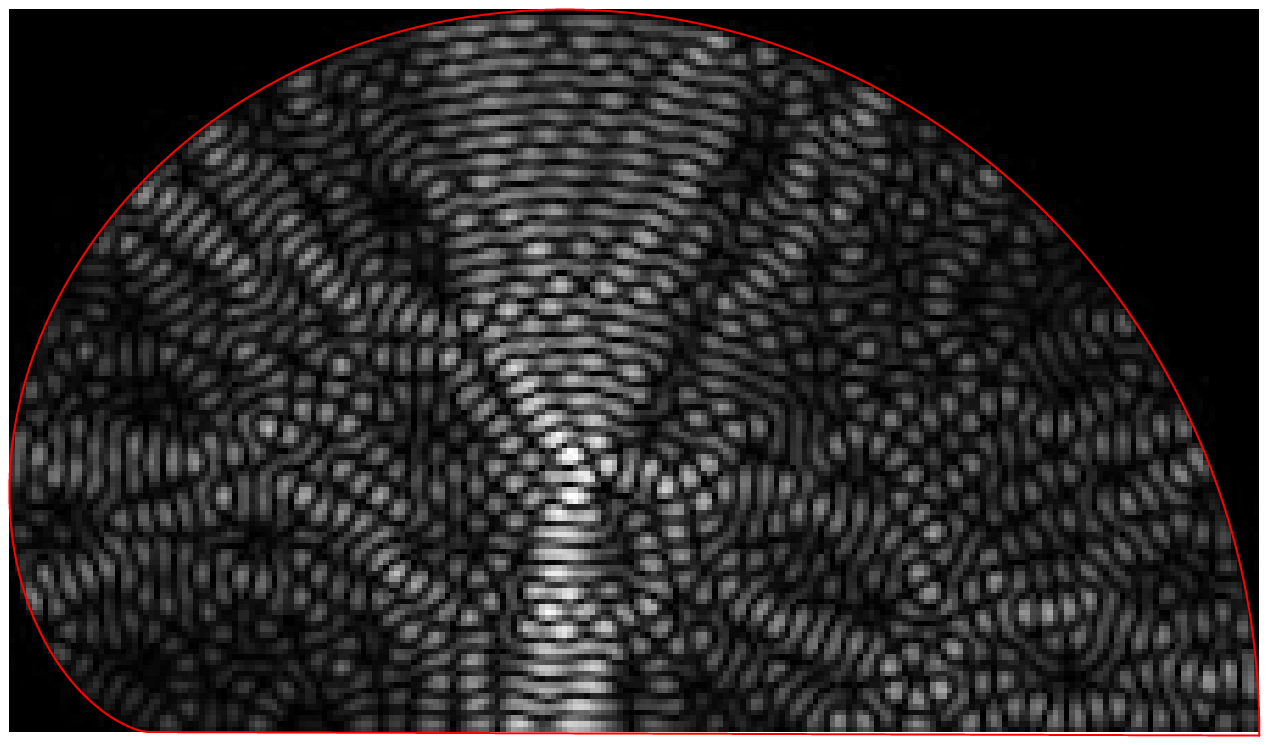}

\vspace*{-6.52in} \hspace*{2.6in}\includegraphics[width=5.in]{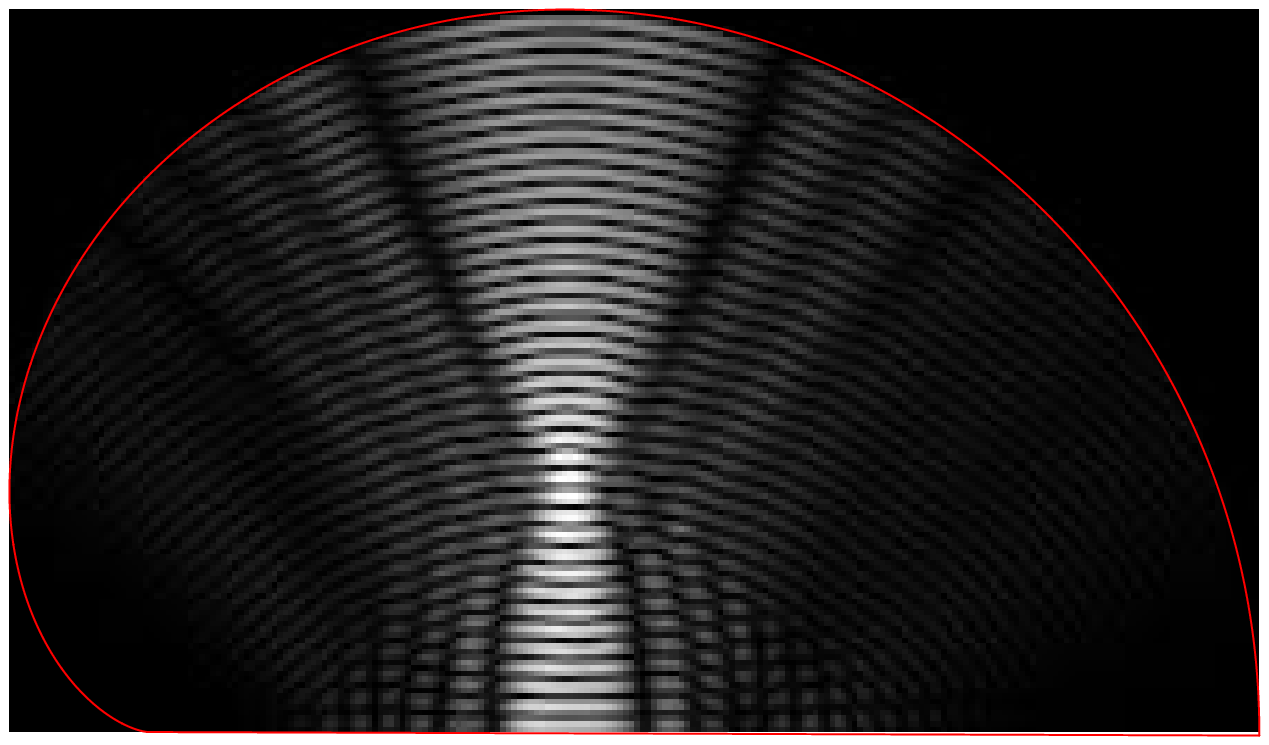}

\vspace*{-4.5in} \hspace*{-1.in}\includegraphics[width=5.in]{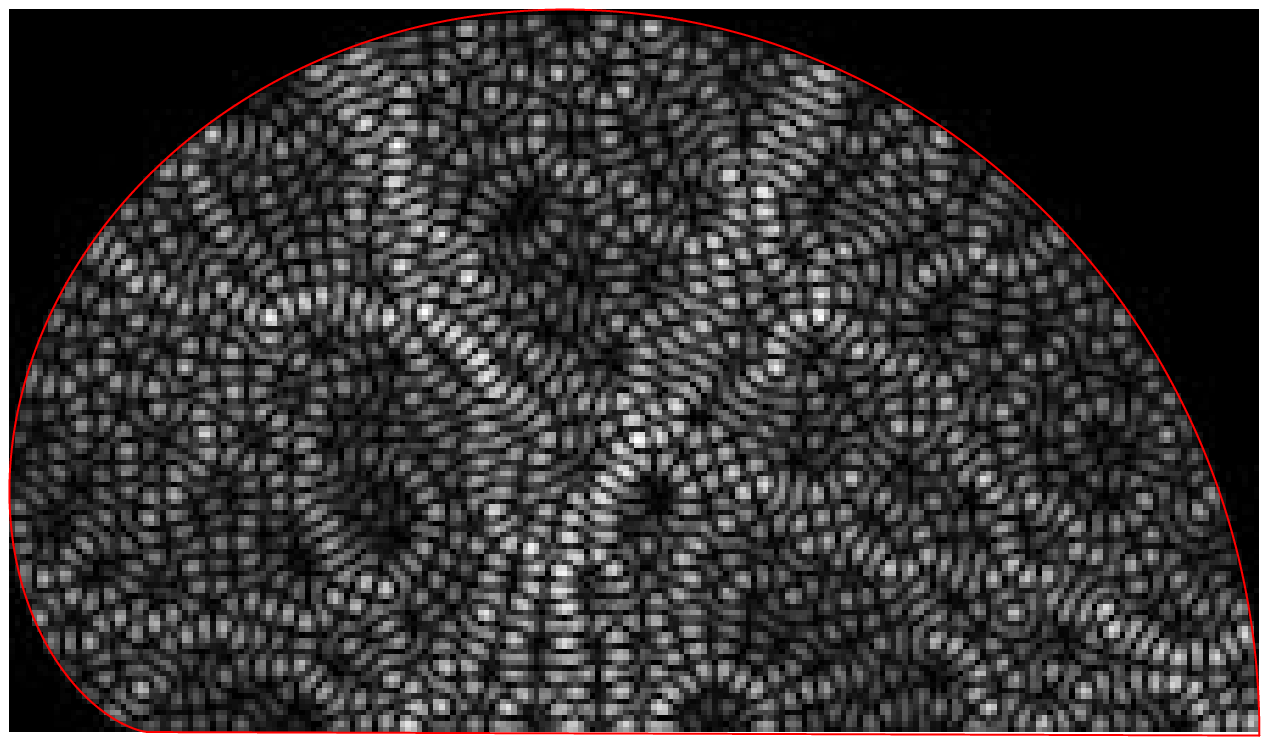}

\vspace*{-6.52in} \hspace*{2.6in}\includegraphics[width=5.in]{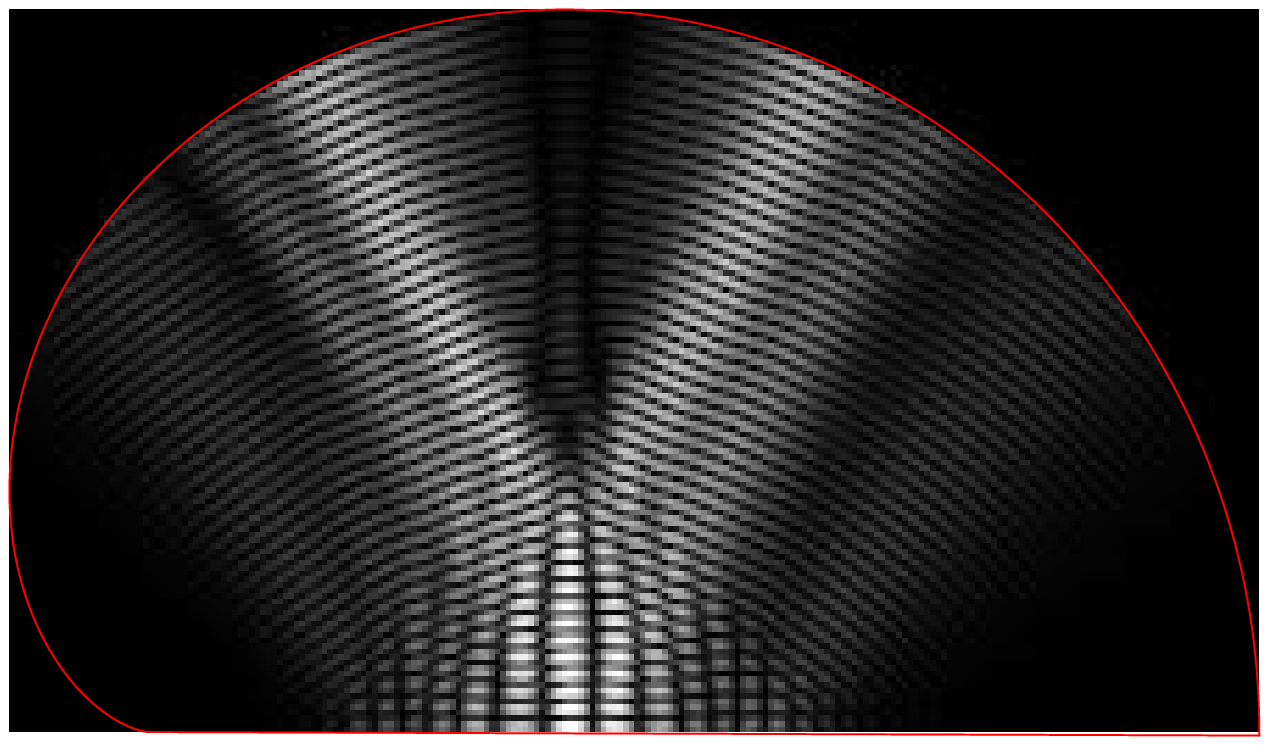}

\vspace*{-4.5in} \hspace*{-1.in}\includegraphics[width=5.in]{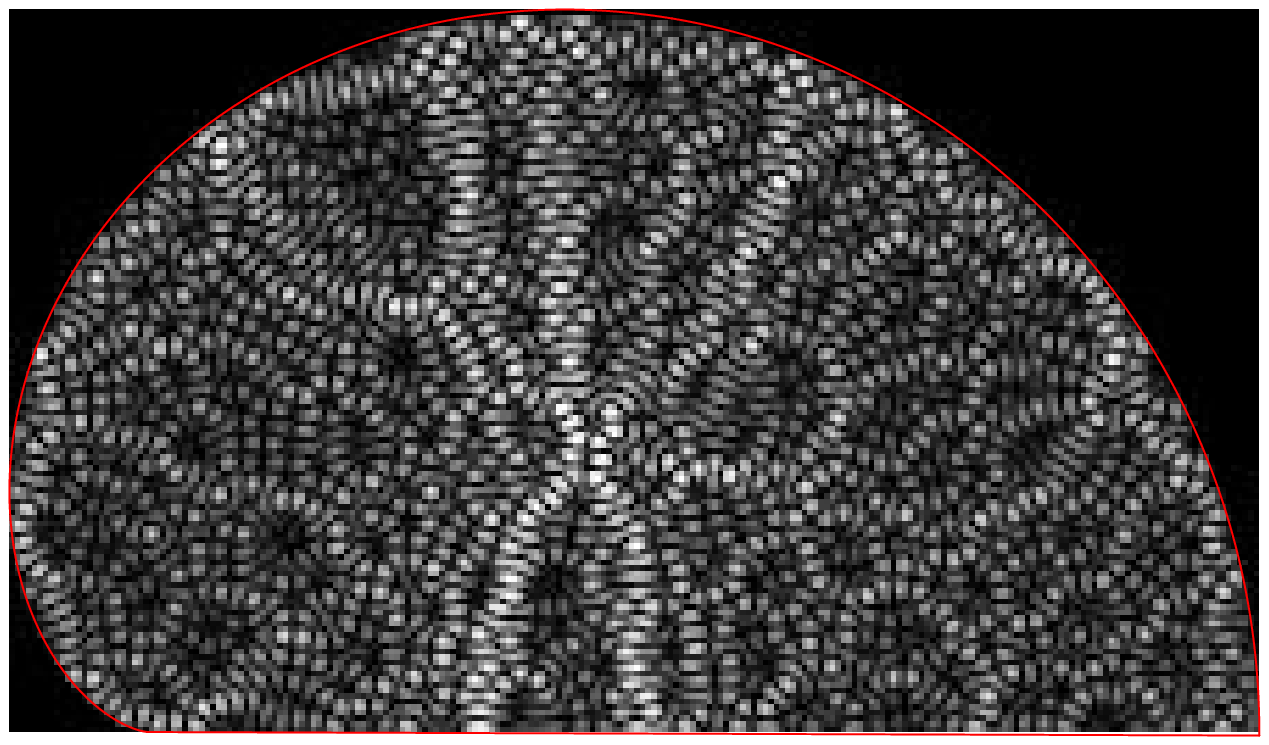}

\vspace*{-6.52in} \hspace*{2.6in}\includegraphics[width=5.in]{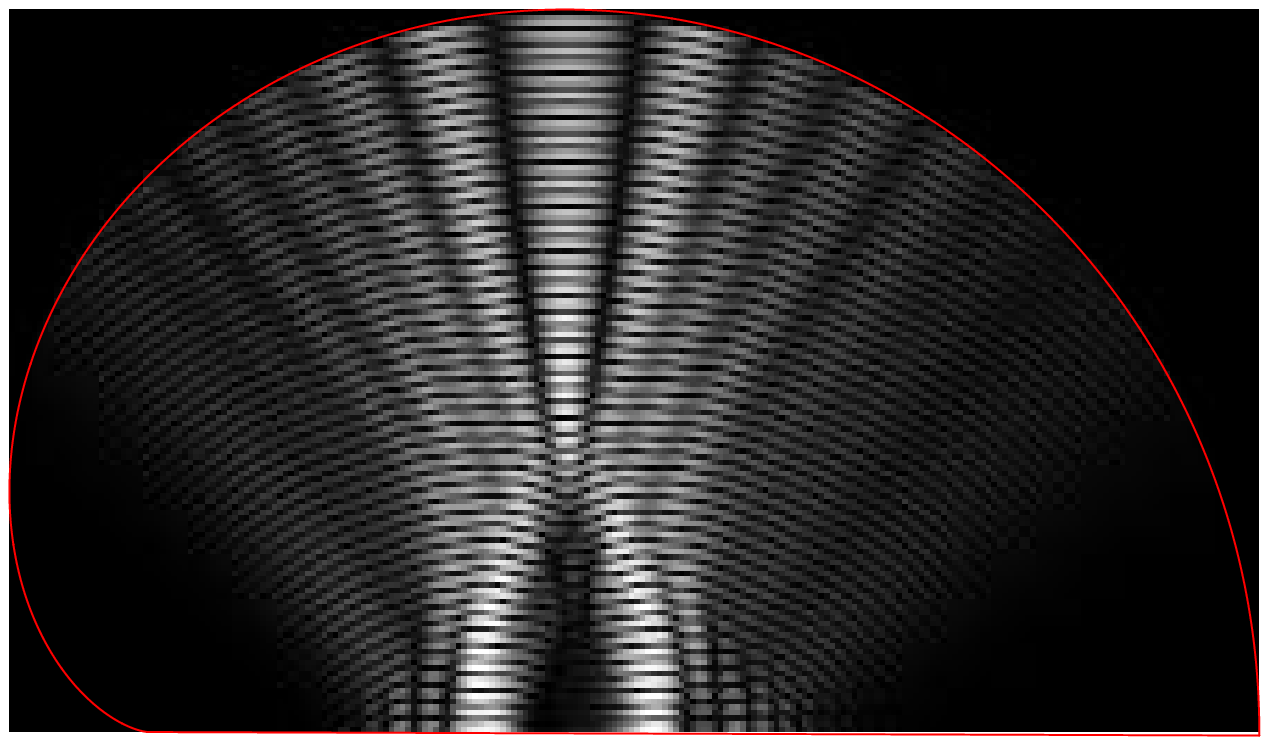}

\vspace*{-2.2in}

\caption{At left we show scarred eigenfunctions with, from top to
bottom, $k_n=92.54$, $107.48$ and $118.66$. These correspond
respectively to $\theta$ near $0$ (in the scarred region), $-\pi/2$ and
$\pi/2$. These may be compared with the corresponding scarfunctions
$\Phi_0^{scar} (\x;k_n)$, shown on the right. The overlap probabilities
are respectively $38.2$, $21.0$ and $21.5$
times larger than the RMT average.
Details of the scar state can readily be detected in the
eigenfunction in the scarred region. Although the correspondence
is less obvious for the other eigenfunctions, features such as numbers
and positions of nodes of the scarfunction can sometimes be seen
in the eigenfunctions.}
\label{wavefunctions}
\end{figure}

In particular, we may conclude that in a given part of the spectrum
the interior functions constructed from the leading
boundary scar states $\chi_i(s;\theta)$, $i=0,1,2,\cdots$, should have
especially large overlaps with the billiard eigenfunctions.
Let us denote by $\Phi_i^{scar} (\x;k)$ the interior function
obtained from a boundary scar state $\chi_i(s;\theta)$.
We show examples of these interior functions in Figure~\ref{scarfunctions}
corresponding to values $\theta=0$, $\pi/2$, $\pi$ and $3\pi/2$ of 
the spectral parameter $\theta$ and $k$ near $1000$. We remark that
these states are somewhat extended spatially around the periodic orbit 
in configuration space, a reflection  of the fact that phase space 
representations of the scar states extend quite far along the stable 
and unstable manifolds. We also remark that, 
near intersections of the periodic orbit with the boundary, these 
scarfunctions approximately satisfy the Dirichlet boundary conditions
placed on the full solution. This is a reflection of the fact that they
are derived from boundary states which are approximate eigensolutions 
of the boundary integral operator sufficiently near the periodic orbit
in phase space.

These functions provide us with 
templates for the behaviour of heavily scarred eigenfunctions near 
periodic orbits in different parts of the spectrum.
We compare in  Figure~\ref{wavefunctions} some cases of individual
scarred eigenfunctions with the corresponding leading scarfunction
$\Phi_0^{scar} (\x;k_n)$. In some cases, such as the eigenfunction with
$k_n=92.54$, one can identify characteristics of the corresponding
scarfunction (such as  the number of nodes seen along a periodic orbit).
In other scarred eigenfunctions, especially those in the antiscarred
region where the corresponding scarfunctions are not well-localised,
a detailed correspondence is less evident visually, although there is
nevertheless a large overlap with the appropriate $\Phi_0^{scar} (\x;k_n)$.
We emphasise that, although not shown explicitly here,
for each $k_n$ these interior functions are the leading members of
a sequence $\Phi_i^{scar} (\x;k_n)$, $i=0,1,2,\cdots$ and each of these
functions will tend to have especially large overlaps with $\psi_n$.

By tracking the scar states as a function of $\theta=k\ell-\mu\pi/2$
we thus define continuous families of interior functions
which approximately vanish on the boundary near bounces of the periodic 
orbit. While our notation suggests boundary functions
$\chi_i(s;\theta)$ which are $2\pi$-periodic in $\theta$, we should note
that as we increase  $\theta=k\ell-\mu\pi/2$ by $2\pi$, the effective
value of $\hbar$ decreases and there is a slow modulation of 
$\chi_i(s;\theta)$ with $k$. (While the coefficients
of $\chi_i(s;\theta)$ in a basis of the harmonic states $\phi_l(s)$
defined in (\ref{defphil}) are asymptotically periodic in $\theta$
as $M\to\infty$, the states 
$\phi_l(s)$ themselves shrink as $k$ increases). To emphasise this
evolution, we label the interior functions $\Phi_i^{scar} (\x;k)$
with $k$ rather than $\theta$. As $k$ increases by $2\pi/\ell$
the form of  $\Phi_i^{scar} (\x;k)$ transverse to the periodic orbit
is approximately periodic but an extra node appears in the
longitudinal direction. Over many such quasiperiods, there is a gradual
shrinkage in the transverse length scale and an increase in the number
of nodes along the orbit. Throughout this evolution,
$\Phi_i^{scar} (\x;k)$ exactly satisfies the Helmholtz equation and
approximately satisfies the Dirichlet boundary conditions near the
periodic orbit.

We remark finally that although we do not show so explicitly here,
these considerations should also allow us to characterise scarring
in general time-dependent systems. Starting with a formal representation
of eigenstates on a Poincar\'e section as in the Bogomolny transfer
method, we can define an analogous extension of probe states from
the section to the full space and therefore provide statistical
predictions for overlap statistics which can be calculated concretely
from the full wave function.

\section{Conclusion}\label{conclusion}

We have shown that the joint-probability distribution
$P(\x;\theta)$ describing wavefunction statistics in a
harmonic-oscillator basis near a periodic orbit leads to the
definition of a set of probe states which describe in a natural way
the morphology of scarring in a given part of the spectrum.
That is, in a given part of the spectrum (parametrised by $\theta$),
we define a set of probe states $\ket{\chi_i(\theta)}$ with $i=0,1,\cdots$
such that the overlaps $\braket{\chi_i(\theta)}{\psi_n}$
are expected to be especially large, on average, and statistically
independent. In particular these probe states offer a template for
the structure of scarred wavefunctions and tell us how the shape of
scarring changes across the spectrum.

The scar states, in offering a basis set for which we expect enhanced
overlap with chaotic eigenfunctions, may in particular
prove useful in explicit computation as outlined in \cite{VE01}. 
We note also the
fact that we can obtain such enhanced overlap probabilities in parts 
of the spectrum normally associated with antiscarring, where
we expect a suppression of overlap with a Gaussian basis state
on the periodic orbit. We have illustrated our results with
explicit calculation for a quantum map model and have also outlined 
briefly how these ideas may be applied to billiard systems.

\vspace*{1cm}
\noindent {\bf Acknowledgements}\\
\noindent This paper is supported by the EPSRC under the 
Fast Stream scheme.

\appendix

\section{Appendix: The matrix elements of the linear evolution operator}
\label{appgetA}

\noindent We here give explicit formulas for the calculation of the
the  linear correlation matrix ${\bf A}(t)$ in a harmonic oscillator
basis. A detailed derivation of these results has been given in 
\cite{CR02} and here we state the key features.

To begin we characterise
linear classical evolution near a periodic orbit using the stability 
exponent $\rho$ and the parameter $Q$ described in the main text.
We then define angles $\psi(t)$ and $\phi(t)$ using
\begin{equation}
\cosh \rho t + iQ\sinh\rho t = \sec\psi(t) \,e^{i\phi(t)},
\end{equation}
and we choose these angles to lie in the ranges 
$0<\psi(t)<\pi/2$ and $-\psi(t)<\phi(t)<\psi(t)$ respectively.

Then we may calculate the matrix elements $A_{lk}(t)$ for $l=k+2n$
and $n\ge 0$ using the polar form
\begin{equation}\label{givecorr}
A_{k+2n,k}(t) =  G_{kn}(\psi(t)) \;
e^{i(k+1/2)(\phi(t)-\mu t\pi)+in\phi(t)+ in\pi/2},
\end{equation}
where $\mu$ is the Maslov index of the periodic orbit concerned and
the amplitude is
\[
G_{kn}(\psi) = \sqrt{\frac{k!}{(k+2n)!}}\;\frac{(2n)!}{2^nn!}\;\;
\sin^n\psi \,\sqrt{\cos\psi} \; C_k^{n+1/2}(\cos\psi),
\]
where $C_k^\alpha(x)$ denotes a Gegenbauer polynomial.
We assume $n\geq 0$ and $t>0$ in (\ref{givecorr}) and 
use $A_{lk}^*(-t)=A_{lk}(t)=A_{kl}(t)$
to calculate the correlation function when $t<0$ or $l<k$.
We note finally that $A_{kl}(t)=0$ if $k$ and $l$ do not have 
the same parity.

\section{Appendix: interior and boundary overlaps}
\label{billiardapp}
\noindent
In this appendix we show that if a boundary probe function $\phi(s)$
is appropriately localised near a point $(s_0,v_0)$ in phase space
then there is a relationship of the form given in (\ref{appen}) between
the boundary overlap $\braket{\phi}{\psi_n}$ 
the full overlap $\braket{\Phi}{\Psi_n}$ of the corresponding
interior functions, defined by (\ref{ptP}) and (\ref{ptPhi}).
Substituting these defining relations into the interior 
integral for $\braket{\Phi}{\Psi_n}$ we find
\begin{equation}
\braket{\Phi}{\Psi_n} = \int_0^L \d s \int_0^L \d s'\;
f(s,s') \, \phi^*(s) \psi_n(s'),
\label{original}
\end{equation}
where 
\begin{equation}\label{deff}
 f(s,s') = \frac{1}{4}\int_D  \,G^*({\bf x},{\bf x}(s);k) \,
		G({\bf x}, {\bf x}(s');k)  \, \d\x 
\end{equation}
and $D$ denotes the billiard domain. We consider in particular the case
where the probe state is of the form $\phi(s)=e^{ik v_0 s} h(s)$ 
where we suppose that $h(s)$ is supported in a neighbourhood of 
$s_0$ that is small in comparison with the length scales of the 
billiard but
varies on a scale longer than the length scale $1/k$ typical of 
wavefunctions. This is the 
case for example for the basis probe states $\phi_l(s)$, which vary and
decay on a length scale of order $1/\sqrt{k}$. It is also the case for any
finite combination of them such as the truncated scar states we use.
We claim that $f(s,s')$  
then behaves essentially as a delta-function when integrated against 
$\phi(s)$, giving
\[
\int  \phi^*(s) f(s,s') \d s \simeq g(s',v_0) \phi^*(s')
\]
where  $g(s',v_0)$ is determined from the geometry of the billiard
and varies with $s'$ on a classical length scale. Since $\phi(s)$ 
is localised to a neighbourhood of a point $s_0$ that is smaller 
than classical length scales then we may further replace $g(s',v_0)$
by $g(s_0,v_0)$ in the overlap integral and approximate
\[
\braket{\Phi}{\Psi_n} \simeq g(s_0,v_0) \braket{\phi}{\psi_n} .
\]
The overlaps $\braket{\Phi}{\Psi_n}$ and $\braket{\phi}{\psi_n}$
then coincide up to a factor which depends on the position in phase space
of the probe state but is independent of $\psi_n$. This means in particular
that predictions for the statistics of $\braket{\phi}{\psi_n}$
carry over to $\braket{\Phi}{\Psi_n}$ after appropriate normalisation.

We justify this assertion by using the asymptotic
representation of the Hankel function
\[
 H_0^{(1)}(k|{\bf x}-\x(s)|) \simeq 
        \sqrt{\frac{2}{\pi k|{\bf x}-\x(s)|}}
			e^{i k|{\bf x}-\x(s)|-i \pi/4 } 
\]
for the Green functions in the integral (\ref{deff}), giving
\begin{equation}\label{approx1}
f(s,s') \simeq \frac{1}{8\pi k}\int_D 
\frac{e^{ik(|{\bf x}-\x(s')|-|{\bf x}-\x(s)|)}}
{\sqrt{|{\bf x}-\x(s)||{\bf x}-\x(s')| }}\;\d\x.
\end{equation}
Fixing $s$ and $s'$ temporarily, we are free to choose the origin 
of coordinates to be the midpoint between 
$\x(s)$ and $\x(s')$, so that $\x(s)={\bf a}/2$ and 
$\x(s')=-{\bf a}/2$. For large $k$ the rapid oscillation in the 
integrand when $s$ and $s'$ are distinct means that $f(s,s')$ is 
peaked around $s\approx s'$. To 
evaluate the integral near this peak, we consider contributions from 
points for which $r=|\x|\gg a=s-s'$. We then approximate the 
phase of the integrand by
\[
k\left(|{\bf x}-\x(s')|-|{\bf x}-\x(s)|\right) = ka\sin\theta
				+O\left(\frac{ka^3}{r^2}\right),
\]
where $(r,\theta)$ are polar coordinates for $\x$ relative the
midpoint of $\x(s)$ and $\x(s')$, which we approximate as lying 
on the boundary, and the normal to the boundary there. This substitution
for the phase is valid as long as $a\ll (R^2/k)^{1/3}$, where $R$ is the 
length scale of the billiard. We then 
approximate the integral in (\ref{approx1}) by 
\begin{eqnarray*}
f(s,s') &\simeq& \frac{1}{8\pi k} \int_D  \frac{e^{ika\sin\theta}}
{r}\;r\d r \d\theta\\[3pt]
&=& \frac{1}{8\pi k}\int_{-\pi/2}^{\pi/2} e^{ika\sin\theta} L(\theta)\,\d\theta
\end{eqnarray*}
where $L(\theta)$ is the length inside the billiard of a line 
passing through the origin at polar angle $\theta$. For convex billiards,
 $r=L(\theta)$ simply defines the shape of the boundary in polar 
coordinates.

As a function of $(s,s')$, this integral is effectively supported 
on a length scale  $1/k$ near $s\approx s'$ or $a\approx 0$. 
When integrated
against a probe state which varies over a longer scale such as $1/\sqrt{k}$,
we may therefore treat it as a $\delta$-like pulse. In fact, we may 
approximate
\begin{eqnarray*}
\int \phi^*(s) f(s,s') \d s  &=& \int h^*(s) e^{-ikv_0 s}f(s,s') \,\d s \\[3pt]
&\simeq& \frac{1}{8\pi k} e^{-ikv_0 s'} h^*(s')
		\int_{-\infty}^\infty  \left[
\int_{-\pi/2}^{\pi/2} e^{ika(\sin\theta-v_0)} 
					L(\theta)\,\d\theta\right]\d a\\[3pt]
&\simeq&  \frac{1}{8\pi k} \phi^*(s')
		\int_{-\pi/2}^{\pi/2}\left[
	\int_{-\infty}^\infty e^{ika(\sin\theta-v_0)} \d a
\right]
	L(\theta)\,\d\theta\\[3pt]
&\simeq&  \frac{1}{8\pi k} \phi^*(s')
		\int_{-\pi/2}^{\pi/2}
       \left[2\pi\delta(k\left(\sin\theta-v_0)\right)\right]
	L(\theta)\,\d\theta\\[3pt]
&=&  \frac{1}{4 k^2 \cos\theta_0} L(\theta_0)\phi^*(s'),
\end{eqnarray*}
where $v_0=\sin\theta_0$. Here, $L(\theta_0)$ is the length of a chord which 
leaves the point $s'$ at an angle $\theta_0$ to the normal.
It depends on $s'$ and we therefore
replace $L(\theta_0)$ by $L(s',\theta_0)$ in our notation to emphasise
that it changes as we change the centre $s\approx s'$ of the 
peaked function $f(s,s')$.

If $h(s)$ is sufficiently slowly varying we may therefore
approximate the expression for the inner product (\ref{original}) by
\[
\braket{\Phi}{\Psi_n} \simeq \frac{1}{4 k^2\cos\theta_0 }
			\int_0^L L(s',\theta_0)
			\phi^*(s') \psi_n(s') \,\d s'.
\]
If, however, $h(s)$ is at the same time sufficiently localised
around $s_0$, then we can replace this in turn by
\[
\braket{\Phi}{\Psi_n} \simeq \frac{L(s_0,\theta_0)}{4 k^2 \cos\theta_0}
			\int_0^L \phi^*(s') \psi_n(s') \,\d s'
 = \frac{L(s_0,\theta_0)}{4 k^2 \cos\theta_0} \,\braket{\phi}{\psi_n}
\]
This verifies our assertion that $\braket{\Phi}{\Psi_n}$ and 
$\braket{\phi}{\psi_n}$ approximately coincide up to a geometrical 
factor
\[
g(s_0,v_0) =  \frac{L(s_0,\theta_0)}{4 k^2 \cos\theta_0} 
\]
 which depends on $k$ and the location of the probe
function $\phi(s)$ but is independent of $\psi_n(s)$.

It should be noted that for scarring around a general periodic orbit
we will need to consider probe functions which are localised not 
around a single point $(s_0,v_0)$ on the boundary section but 
around a series of points 
$(s_t,v_t)$, $t=0,1,\cdots,T$ representing the bounces of a periodic 
orbit on the boundary. In this case we consider probe states of the form
$\phi(s) = \sum_t\phi_t(s)$ where each component is localised in the section
around a single point $(s_t,v_t)$. The interior overlap 
$\braket{\Phi}{\Psi_n}$ is then a sum of terms of the form given above, each 
with a geometrical factor $g(s_t,v_t)$ which depends on the the angle 
of incidence of the periodic orbit at the corresponding bounce and 
the length of the orbit segment to the next bounce. We can 
recover a simple scaling such as given in (\ref{appen}) if we
scale each of the components $\phi_t$ of $\phi$ by the corresponding
geometrical factor before computing the overlap (or alternatively
by redefining the boundary inner product). A detailed treatment 
of this procedure will presumably lead to a formulation in which the
boundary integral operator is replaced, semiclassically at least,
by a unitary evolution (see \cite{BOA92}, but do not pursue that option here.
Instead we note that our explicit calculations have been for a 
two-bounce orbit normally incident on the boundary and in that 
case the geometrical factor is the same for both bounces and no 
such adjustment is necessary.

\end{document}